\documentclass[onecolumn]{aa}
\usepackage{lscape}
\usepackage{graphicx}
\usepackage{longtable}
\usepackage{float}

\usepackage{rotating}
\usepackage{graphicx}
\usepackage{booktabs}

\begin{document}


\title {
Gas phase Elemental abundances in Molecular cloudS (GEMS)
V. Methanol in Taurus\thanks{Based on observations carried out with the IRAM NOEMA interferometer. IRAM is supported by INSU/CNRS (France), MPG (Germany) and IGN (Spain)}
}


\author{S. Spezzano\inst{1}  \and A. Fuente\inst{2}  \and P. Caselli\inst{1}  \and A. Vasyunin\inst{3}  \and D. Navarro-Almaida\inst{2} \and M. Rodr\'iguez-Baras\inst{2}  \and A. Punanova\inst{3}  \and C. Vastel\inst{4}  \and V. Wakelam\inst{5}}

\institute{Max Planck Institut f\"ur Extraterrestrische Physik, Giessenbachstrasse 1, 85748 Garching, Germany \and Observatorio Astron\'omico Nacional(OAN), 28014 Madrid, Spain \and Ural Federal University, 620002 Yekaterinburg, Russia \and IRAP, Universit\'e de Toulouse, CNRS, UPS, CNES, 31400 Toulouse, France \and Laboratoire d’astrophysique de Bordeaux, Univ. Bordeaux, CNRS, B18N, all\'ee Geoffroy Saint-Hilaire, 33615 Pessac, France}

\abstract {Methanol, one of the simplest complex organic molecules in the Interstellar Medium (ISM), has been shown to be present and extended in cold environments such as starless cores. Studying the physical conditions at which CH$_3$OH starts its efficient formation is important to understand the development of molecular complexity in star-forming regions.} 
{We aim at studying methanol emission across several starless cores and investigate the physical conditions at which methanol starts to be efficiently formed, as well as how the physical structure of the cores and their surrounding environment affect its distribution.}
{Methanol and C$^{18}$O emission lines at 3 mm have been observed with the IRAM 30m telescope within the large program "Gas phase Elemental abundances in Molecular CloudS" (GEMS) towards 66 positions across 12 starless cores in the Taurus Molecular Cloud. A non-LTE radiative transfer code was used to compute the column densities in all positions. We then used state-of-the-art chemical models to reproduce our observations.} 
{We have computed N(CH$_3$OH)/N(C$^{18}$O) column density ratios for all the observed offsets, and two different behaviours can be recognised: the cores where the ratio peaks at the dust peak, and the cores where the ratio peaks with a slight offset with respect to the dust peak ($\sim $10000 AU). We suggest that the cause of this behaviour is the irradiation on the cores due to protostars nearby which accelerate energetic particles along their outflows. The chemical models, which do not take into account irradiation variations, can reproduce fairly well the overall observed column density of methanol, but cannot reproduce the two different radial profiles observed.}
{We confirm the substantial effect of the environment onto the distribution of methanol in starless cores. We suggest that the clumpy medium generated by  protostellar outflows might cause a more efficient penetration of the interstellar radiation field in the molecular cloud and have an impact on the distribution of methanol in starless cores. Additional experimental and theoretical work is needed to reproduce the distribution of methanol across starless cores.}

\titlerunning{Methanol GEMS in Taurus}
\maketitle

\section{Introduction}
\label{sect_intro}
Methanol, CH$_3$OH, and even more complex organic molecules (COMs) have been widely observed in star-forming regions. COMs are defined as organic molecules with $\geq$ 6 atoms \citep{vandishoeck09}, and their formation is considered to be the first step in the chemical complexity that will eventually be inherited by forming planets in the process of star and planetary system formation. In the past decade several papers have reported on the detection of methanol and other COMs towards starless cores showing that COMs are present and their emission is extended in cold and shielded environments \citep{bacmann12, bizzocchi14, vastel14, jimenez-serra16, scibelli20, scibelli21}. Starless cores provide the ingredients that will eventually build-up a planet, both the organic as well as the refractory material. Therefore, it is valuable to understand the chemistry that drives the build-up of chemical complexity in stellar nurseries such as starless cores. \\
COMs have been widely observed towards hot cores and hot corinos, both characterised by an active chemistry driven by ice sublimation at T$\geq$100 K (e.g. \cite{caselli93, cazaux03,caux11, jorgensen12, belloche14}). 
COMs are mostly formed on the surface of dust grain via: i) hydrogenation of carbon monoxide (producing small COMs like formaldehyde and methanol, \citealt{charnley95,watanabe02}); and ii) radical-radical (associative) reactions (producing larger COMs like methyl formate and dimethyl ether, \citealt{garrod&herbst06}). In order for radical-radical COM formation to happen, it is crucial to have a central heating source like a young stellar object (YSO) because the mobility of radicals on the surface appears at T$\geq$30 K.
Because of the lack of a central heating source, the detection of COMs toward starless cores challenged our understanding of COMs in the interstellar medium. Large COMs like acetaldehyde, dimethyl ether and methyl formate have been observed towards dark cloud cores and starless cores \citep{bacmann12,vastel14,jimenez-serra16,nagy19,scibelli20, scibelli21, jimenez-serra21}.

The high-level of molecular complexity revealed by these studies towards cold cores has been a challenge for chemical models \citep{oberg10}. Recent laboratory work has shown that COMs like acetaldehyde, vinyl alcohol, and also glyicine is possible in interstellar ice analogs with non-energetic processes \citep{chuang20, ioppolo21}. The formation of COMs on the grains depends on the efficiency for the large radicals to diffuse on the surfaces. Although it is an uncertain parameter, a larger diffusion efficiency will produce more COMs on the surfaces \citep{ruaud15, walsh16}. The formation in the ices also depends on the chemical network itself, which is probably not complete for the larger molecules. \cite{ruaud15} for instance have proposed new routes to form COMs at low temperature (and low diffusion efficiency) via reactions induced by the formation of complexes with the main ice components. New gas-phase pathways have also been proposed but still require the efficient production and desorption of methanol \citep{vasyunin13, balucani15}. If the efficiency on the surface is sufficient, remains the problem of desorbing these molecules strongly bound to the surfaces. The most promising non-thermal desorption mechanisms seem to be chemical reactive desorption from CO-rich surfaces \citep{minissale16, vasyunin17}, chemical explosions \citep{rawlings13, ivlev15}, and grain sputtering induced by cosmic-rays \citep{wakelam21}.\\ 
Methanol is one of the simplest COMs and the precursor of many of the O-bearing COMs, hence understanding the processes responsible for its release in the gas phase is of pivotal importance to constrain chemical models and reproduce the chemical complexity that we observe in starless cores. \\

In this paper we present our results on the emission of methanol towards 12 starless cores as well as towards the clouds where they are embedded. We study the dependency, or the lack thereof, between the emission of methanol and physical parameters such as visual extinction, volume density, as well as environment. The paper is structured as follows: Section~\ref{sect_observations} describes the observations, Section~\ref{sect_dataset} presents the dataset as well as the TMC-1 molecular cloud and the B213 filament, where the 12 observed  starless cores are located. The analysis of the observed data and its results are in Sections~\ref{sect_analysis} and \ref{sect_results}. The results of the chemical modelling are presented in Section~\ref{sec_modelling}. Our conclusions are discussed in Section~\ref{sect_conclusions}. More detailed information on each source is presented in the Appendice.\\

\section{Observations}
\label{sect_observations}
The molecular transitions used in this study are reported in Table~\ref{table:parameters}.
The 3\,mm methanol lines were observed within the "Gas phase Elemental abundances in Molecular CloudS" (GEMS) (PI: Asunci\'on Fuente) IRAM 30m Large Program
and the observing procedure and receiver setups were described by \citet{Fuente19}.  
The HPBW is varying with the frequency as HPBW($"$)=2460 /$\nu$ where $\nu$ is in GHz.
The observing mode was frequency switching with a frequency throw of 6 MHz well adapted to remove standing waves between the secondary mirror and the receivers. 
The Eight MIxer Receivers (EMIR) and the Fast Fourier Transform Spectrometers (FTS) with a spectral resolution of 49~kHz were used for these observations. 
The intensity scale is T$_{MB}$, which is related with T$_A^*$ by $T_{MB} =(F_{eff} / B_{eff})T_A^*$ (see Table B.1 in \citealp{Fuente19}). GEMS observations
in  TMC~1 were completed with C$^{18}$O 2$\rightarrow$1 observations carried out with the IRAM 30m telescope in a previous project.

\begin{landscape}
\begin{figure*}
 \centering
 \includegraphics [height=0.85\textwidth, keepaspectratio]{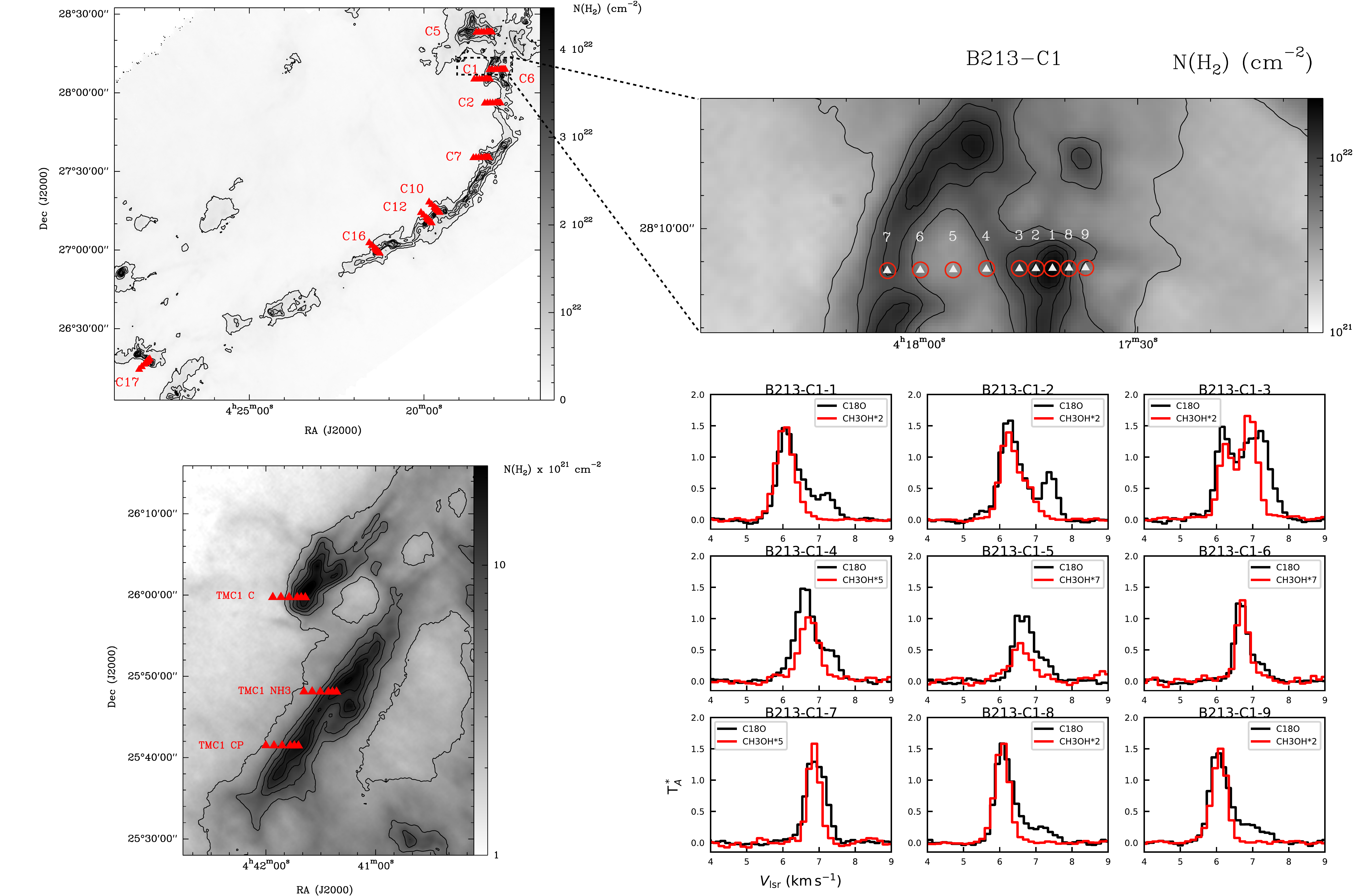}
 \caption[angle=90]{Left panels: H$_2$ column density map of the TMC-1 molecular cloud (lower panel) and the B213 filament (upper panel) in Taurus,computed from $Herschel$ and $Planck$ data \citep{palmeirim13, rodriguez21}. The offsets observed towards the starless cores in our dataset are marked with a triangle.\\
 Right panel: H$_2$ column density of B213-C1 (upper panel). The triangles mark the positions observed with the GEMS large project, and the red circles show the beam size of the IRAM 30m telescope at 97 GHz ($\sim$26$"$). The spectra of the 2$_{1,2}$-1$_{1,1}$ ($E_2$) transition of methanol (in red) overlaid with the $J$ = 1-0 transition of C$^{18}$O (in black) in all offsets towards B213-C1 are shown in the lower panel. 
}
  \label{fig:figure1}
    \end{figure*}
\end{landscape}

For TMC~1, we use observations of the CH$_3$OH  $J_{K_a,K_c}$ = 1$_{0,1}$ $\rightarrow$0$_{0,0}$ line carried out with the Yebes 40m radiotelescope in Q-Band \citep{Tercero20}
during March-April 2018.
The 40m telescope is equipped with HEMT receivers for the 2.2-50 GHz range, and a SIS receiver for the 85-116 GHz range. Single-dish 
observations in K-band (21-25 GHz) and Q-band (41-50 GHz) were performed simultaneously. The backends consisted of FFTS  covering a bandwidth of $\sim$2 GHz 
in band K and $\sim$9 GHz in band Q, with a spectral resolution of $\sim$38 kHz. 
Central frequencies were 23000~MHz and 44750~MHz for the K and Q band receivers, respectively.
The observing procedure was position-switching, and the OFF-positions are 
RA(J2000) = 04$^{\rm h}$42$^{\rm m}$24$^{\rm s}$.24 Dec(2000):25$^{\circ}$41$'$27$''$.6 for TMC~1-CP,
RA(J2000) = 04$^{\rm h}$42$^{\rm m}$29$^{\rm s}$.52  Dec(2000):25$^{\circ}$48$'$07$''$.2 for TMC~1-NH$_3$,
RA(J2000) = 04$^{\rm h}$42$^{\rm m}$32$^{\rm s}$.16  Dec(J2000):25$^{\circ}$59$'$42$''$.0 for TMC~1-C. 
These positions were checked to be empty of emission before the observations. 
The intensity scale is T$_{MB}$ with conversion factors of 4.1 Jy/K in band K 
(T$_{MB}$/T$_A^*$=1.3) and in 5.7 Jy/K in band Q (T$_{MB}$/T$_A^*$=2.1).
The HPBW of the telescope is 42$"$ at 7\,mm and 84$"$ at 1.3\,cm. 
As an example of our dataset, Figure~\ref{fig:figure1} shows to the left the H$_2$ column density map of the TMC-1 molecular cloud (upper panel) and the B213 filament (lower panel) computed from $Herschel$ and $Planck$ data \citep{palmeirim13, rodriguez21}, and on the right a zoom-in into the H$_2$ column density map of B213-C1 where the observed offsets are marked (upper panel) and the spectra of the 2$_{1,2}$-1$_{1,1}$ ($E_2$) transition of methanol are overlaid with the 1-0 transition of C$^{18}$O (lower panel). The spectra observed towards the other sources in our sample are shown in Figures ~\ref{fig:TMC-1C-co}-~\ref{fig:B213-C17-all}.

\newpage
\newpage

\begin{table}
\caption{Spectroscopic parameters of the observed lines}
\label{table:parameters}
\scalebox{1}{
\begin{tabular}{cccccc}
\hline\hline
Molecule & Transition & Rest frequency\tablefootmark{a} & E$_{up}$& A & n$^*$\tablefootmark{b}
\\
&      &(MHz)   & (K)  & ($\times$10$^{-5}$ s$^{-1}$)&cm$^{-3}$\\
\hline
CH$_3$OH   &  $J_{K_a,K_c}$ = 1$_{0,1}$-0$_{0,0}$ ($A^+$) &   48372.460(2)       &  2.3\tablefootmark{c}   & 3.5$\times$10$^{-7}$   &7$\times$10$^3$\\
CH$_3$OH   &  $J_{K_a,K_c}$ = 2$_{1,2}$-1$_{1,1}$ ($E_2$) &   96739.358(2)       &  12.53\tablefootmark{c}   & 2.6$\times$10$^{-4}$   &3$\times$10$^4$\\
CH$_3$OH   &  $J_{K_a,K_c}$ = 2$_{0,2}$-1$_{0,1}$ ($A^+$) &   96741.375(2)       &   6.96\tablefootmark{c}    & 3.4$\times$10$^{-4}$    &3$\times$10$^4$\\
CH$_3$OH   &  $J_{K_a,K_c}$ = 2$_{0,2}$-1$_{0,1}$ ($E_1$) &   96744.545(2)       &  20.08\tablefootmark{c}   & 3.4$\times$10$^{-4}$    &3$\times$10$^4$\\
CH$_3$OH   &  $J_{K_a,K_c}$ = 2$_{0,2}$-1$_{0,1}$ ($E_1$-$E_2$) &  108893.945(2)      &  13.10\tablefootmark{c}   & 1.47$\times$10$^{-5}$   &5$\times$10$^5$\tablefootmark{d}\\\
C$^{18}$O    &  $J$ = 1-0 &   109782.173(2)      &  5.27   & 6.3$\times$10$^{-8}$  &2$\times$10$^3$\\
C$^{18}$O    &  $J$ = 2-1 &   219560.354(2)       &  15.81  & 6.0$\times$10$^{-7}$  &2$\times$10$^4$\\

\hline
\end{tabular}
}
\tablefoot{
\tablefoottext{a}{Frequencies and uncertainties from the CDMS \citep{Mueller2005}}
\tablefoottext{b}{n$^*$ is the critical density, calculated at 10 K.}
\tablefoottext{c}{Energy relative to the ground 0$_{0,0}$, $A$ rotational state.}
\tablefoottext{d}{Calculated with the rate coefficient reported in \cite{bizzocchi14}}
}
\end{table}

\begin{table}
\caption{Source sample }
\label{table:sources}
\scalebox{1.0}{
\begin{tabular}{ccc}
\hline\hline
&Right Ascension & Declination\\ 
&(J2000)&(J2000) \\
\hline \hline
TMC-1 C &04:41:38.80&+25:59:42.0\\
TMC-1 CP&04:41:41.90 &+25:41:27.1 \\
TMC-1 NH$_3$& 04:41:21.30&+25:48:07.0\\
B213-1 &04:17:41.80&+28:08:47.0\\
B213-2 &04:17:50.60&+27:56:01.0\\
B213-5 &04:18:03.80&+28:23:06.0\\
B213-6 &04:18:08.40&+28:05:12.0\\
B213-7 &04:18:11.50&+27:35:15.0\\
B213-10 &04:19:37.60&+27:15:31.0\\
B213-12&04:19:51.70&+27:11:33.0\\
B213-16 &04:21:21.00&+27:00:09.0\\
B213-17 &04:27:54.00&+26:17:50.0\\
\hline
\end{tabular}
}
\end{table}

\section{Dataset}
\label{sect_dataset}

We have observed several lines of CH$_3$OH and two lines of C$^{18}$O towards the GEMS sources in Taurus with the aim of understanding how these two chemically related molecules behave towards several starless cores within the same environment. The GEMS dataset allows us to study the abundance of both molecules, as well as their ratios, towards different offsets within and outside the cores, and hence tracing gas with different physical properties such as volume density and temperature. Our source sample is listed in Table~\ref{table:sources}.\\
All transitions of methanol studied in this paper, except the $J_{K_a,K_c}$ = 1$_{0,1}$-0$_{0,0}$ ($A^+$) transition at 48 GHz, have been observed towards every offset in our source sample (although they were not always detected).
Both C$^{18}$O transitions have been observed towards every offsets in the starless cores in TMC-1, while only the $J$ = 1-0 transition has been observed towards the starless cores in the B213 filament.
Figures~\ref{fig:TMC-1C-co}-~\ref{fig:B213-C17-all} show the spectra of the 2$_{1,2}$-1$_{1,1}$ ($E_2$) transition of methanol overlaid with the 1-0 transition of C$^{18}$O towards all offsets in our dataset, with the exception of B213-C1, shown in Figure~\ref{fig:figure1}.

\subsection{TMC-1 cloud}
Taurus Molecular Cloud-1 (TMC-1) is a cold and dense cloud with a filamentary structure of about 5$\arcmin \times$40$\arcmin$ in the centre of the Taurus molecular cloud at a distance of 140.2$^{+1.3}_{-1.3}$ pc \citep{galli19}. 
The chemistry of TMC-1 has been the object of many studies in the literature (e.g. \citealt{ guelin82, pratap97, saito02, schnee07,suutarinen11}). 
The most peculiar feature of TMC-1 is the chemical differentiation between the southern part, the cyanopolyyne peak (TMC-1 CP), where carbon-chain molecules are more abundant, and the northern part, the ammonia peak (TMC-1 NH$_3$), where NH$_3$ and N$_2$H$^+$ are more abundant. The origin of this differentiation has been extensively discussed, and it can be explained by either a small variation in the chemical timescale ($\sim$ 10$^5$ yr), and/or by the different density (with the ammonia peak being denser) \citep{hirahara92, suzuki92}. Also a variation in the C/O ratio can reproduce an abundance gradient similar to that observed \citep{pratap97}.
We present here the observations of methanol and C$^{18}$O towards the cyanopolyyne and ammonia peaks (TMC-1 CP and NH$_3$). We also present observations towards the pre-stellar core TMC1-C \citep{schnee07, schnee10}.  Several observational evidences suggest that  this core is a later evolutionary stage than TMC1 (CP),  presenting higher deuteration fractions, more similar to those in L1544 \citep{crapsi05, navarro21}. The spectra of the 2$_{1,2}$-1$_{1,1}$ ($E_2$) transition of methanol overlaid with the 1-0 transition of C$^{18}$O towards TMC-1C, TMC-1 CP and TMC-1 NH$_3$ are shown in Figures~\ref{fig:TMC-1C-co}-~\ref{fig:TMC-1NH3-co}.

\subsection{B213 filament}
B213 (L1495) is one of the longest and most prominent filaments in the Taurus molecular cloud, as it extends for over 80$\arcmin$, and it is located towards the north-west of the cloud at a distance of 129.9$^{+0.4}_{-0.3}$ pc \citep{galli19}. The gas towards B213 presents a very complex velocity structure that has been interpreted as the result of colliding filaments \citep{duvert86}. B213 has been observed in CO isotopologues with large scale maps \citep{onishi96, goldsmith08}. The denser gas has been studied using ammonia, H$^{13}$CO$^+$, N$_2$H$^+$ and SO \citep{benson89, hacar13}. Furthermore, some dense cores in B213 contain young stellar objects \citep{rebull10}.
We present here the observations of methanol and C$^{18}$O towards nine starless cores in B213. The spectra of the 2$_{1,2}$-1$_{1,1}$ ($E_2$) transition of methanol overlaid with the 1-0 transition of C$^{18}$O towards the starless cores we observed in B213 are shown in Figure~\ref{fig:figure1} and Figures~\ref{fig:B213-C2-all}-~\ref{fig:B213-C17-all}.

\section{Analysis}
\label{sect_analysis}
We have observed the three rotational transitions belonging to the 2-1 transitions of methanol at 96 GHz (symmetry A, E1 and E2) as well as the 1-0 transition at 108 GHz (symmetry E2-E1), in all cores of our sample. In addition, we have observed the $J_{K_a,K_c}$ = 1$_{0,1}$-0$_{0,0}$ ($A^+$) transition of methanol at 48 GHz towards the three cores in TMC-1. The intensity ratios among the four lines at 3 mm in our dataset cannot be reproduced assuming LTE, implying that we have to use a non-LTE code if we want to derive a reliable column density of methanol in all positions within our dataset. Since methanol lines have a relatively high critical density (between 0.7 and 3 $\times$10$^4$ cm$^{-3}$, see Table~\ref{table:parameters}), the correspondent column density is very sensitive to variations in the H$_2$ volume density. The H$_2$ volume density towards all observed positions within the GEMS catalogue has been calculated by modelling the emission of CS, C$^{34}$S, and $^{13}$CS with the radiative transfer code RADEX \citep{vandertak07} in \cite{rodriguez21}. However, the lines of methanol and CS could come from different velocity components and/or different layers within the same cloud. The line profiles of C$^{18}$O and CH$_3$OH are different in many positions. Therefore, in order to be consistent and derive accurate column densities we need to use the volume density derived from CH$_3$OH data. 
We hence decided to use the Markov Chain Monte Carlo method (MCMC) together with the RADEX non-LTE radiative transfer code within the CASSIS software \citep{vastel15} in order to compute the methanol column density, the excitation temperature, and the gas volume density towards all positions in our dataset. To assess the goodness of the fit, the reduced $\chi^2$ was computed for each model. The reduced $\chi^2$ in our sample ranges from 1.5 to 4.5, with most of the models having a reduced $\chi^2$ $\sim$ 2. The results of the MCMC+RADEX analysis with CASSIS on methanol are reported in Table~\ref{table:results}.
We used the collisional rates for methanol reported in \cite{rabli10} at 10 K.  

We did not use the MCMC method for the C$^{18}$O because for most of the cores we observed only one transition (with the exception of the three cores in TMC-1), and the observed lines of C$^{18}$O are not very sensitive to variations in the volume densities. Furthermore, C$^{18}$O often has more velocity components than CH$_3$OH, and we are interested in deriving a column density for C$^{18}$O from the velocity component that corresponds to the one observed with CH$_3$OH.
We used two approaches to derive the column densities of C$^{18}$O in our dataset. In the offsets with a good match between the rest velocity and line width of the lines of methanol and C$^{18}$O, we fit the lines of C$^{18}$O and used RADEX to derive the column density, assuming the T$_{kin}$ and n$_{H_2}$ derived for CH$_3$OH. In these cases, the errors derived on the v$_{LSR}$ and $\delta$v from the line fit have been propagated to derive the error on the column density of C$^{18}$O, see for example the values for TMC-1C and B213-C6 reported in Table~\ref{table:results}.
For the offsets with a bad match between the rest velocity and line width of the lines of methanol and C$^{18}$O (see for example the spectra of TMC-1 NH$_3$ in Figure~\ref{fig:TMC-1NH3-co}), we have used the GUI of CASSIS and performed a single-line fit with RADEX of the portion of the spectrum of C$^{18}$O that is comparable with methanol in velocity and line-width. In order to do this we have fixed the v$_{LSR}$ and adjusted the line-width in order to be similar to the line-width of methanol, and at the same time model well the line shape of C$^{18}$O. The error on the column densities derived in this manner, and reported in Table~\ref{table:results}, has been estimated to be 15$\%$.
While in the case of CH$_3$OH our analysis shows that the molecules are not in LTE (as the resulting T$_{ex}$ is lower than the kinetic temperature), this is not the case for the C$^{18}$O 1-0 transition, where T$_{ex}$ $\sim$ T$_{kin}$.

The visual extinction tabulated in Table~\ref{table:results} has been calculated from the H$_2$ column density maps reported in \cite{rodriguez21}, using the formula A$_V$ [mag] = N(H$_2$)[cm$^{-2}$]/9.4$\times$10$^{20 }$[cm$^{-2}$ mag$^{-1}$] \citep{bohlin78}.\\

We do not report on the upper limits for the column density of methanol towards positions where the line was not detected because it is not possible to derive them in the same fashion as we did for the methanol column densities reported in Table~\ref{table:results}, and we would have to assume a value for the H$_2$ volume density. The uncertainty induced by this assumption would make the comparison of the upper limits on N(CH$_3$OH) with the N(CH$_3$OH) computed with the MCMC+RADEX method not meaningful.

\begin{longtable}{p{0.12\textwidth}|p{0.05\textwidth}|p{0.09\textwidth}p{0.07\textwidth}p{0.09\textwidth}p{0.07\textwidth}p{0.08\textwidth}|p{0.08\textwidth}p{0.07\textwidth}p{0.07\textwidth}}
\caption{Results from the MCMC and RADEX analysis.}
\label{table:results}
\\\hline\hline
&&&CH$_3$OH &&&&&C$^{18}$O\\
         &A$_{V}$\tablefootmark{a}& N$_{tot}$             & T$_{kin}$               & n$_{H_2}$                & v$_{LSR}$            &  $\delta$v           & N$_{tot}$    & v$_{LSR}$   & $\delta$v \\
         & mag              & [10$^{13}$cm$^{-2}$]  & [K]                   & [10$^4$cm$^{-3}$]       & [km s$^{-1}$]         & [km s$^{-1}$]         & [10$^{15}$cm$^{-2}$]  & [km s$^{-1}$]   & [km s$^{-1}$] \\
\hline
\textbf{TMC-1 C}&&&&&&&&&\\
1 &      19.9                  &  2.7$^{+0.4}_{-0.30}$ &  10.4$^{+0.2}_{-0.3}$ &  3.8$^{+1.0}_{-0.7}$  & 5.18  &  0.34  & 3.62(1)   & 5.21   & 0.46   \\&&&&&&&&\\
2 &  18.5                   &  2.7$^{+0.3}_{-0.3}$ &  10.3$^{+0.5}_{-0.12}$ &  2.5$^{+0.5}_{-0.5}$  & 5.14  &  0.35  & 3.3(1)   & 5.19   & 0.46   \\&&&&&&&&\\
3 &     13.3                  &  2.1$^{+0.5}_{-0.2}$ &  12.5$^{+0.3}_{-0.3}$ &  1.01$^{+0.2}_{-0.3}$  & 5.12  &  0.32  & 2.67(2)   & 5.16   & 0.46   \\&&&&&&&&\\
4 &     4.8                 &  1.4$^{+0.3}_{-0.5}$ &  11.3$^{+1.0}_{-0.6}$ &  0.15$^{+0.08}_{-0.02}$  & 5.09  &  0.39  &  1.54(1)  &  5.13  &  0.51  \\&&&&&&&&\\
\textbf{TMC-1 CP}&&&&&&&&&\\
1 &       18.2                 &  2.0$^{+0.6}_{-0.3}$ &  10.5$^{+0.9}_{-0.4}$ &  1.5$^{+0.3}_{-0.5}$  & 5.63  &  0.33  & 2.3$\pm$0.34   &\tablefootmark{b}  & 0.28\tablefootmark{c}   \\&&&&&&&&\\
1 &      18.2                    &  1.2$^{+0.1}_{-0.1}$ &  10.7$^{+0.2}_{-0.4}$ &  3.87$^{+0.30}_{-0.51}$  & 6.07 &  0.32  & 0.9$\pm$0.13   & \tablefootmark{b}  & 0.3\tablefootmark{c}   \\&&&&&&&&\\
2 &    16.7                   &  1.3$^{+0.3}_{-0.2}$ &  12.1$^{+0.3}_{-0.5}$ &  1.6$^{+0.2}_{-0.5}$  & 5.66  &  0.38  &  2.2$\pm$0.33  & \tablefootmark{b}   & 0.28\tablefootmark{c}   \\&&&&&&&&\\
2 &     16.7                  &  1.1$^{+0.2}_{-0.1}$ &  10.3$^{+0.2}_{-0.2}$ &  3.5$^{+0.3}_{-0.8}$  & 6.06  &  0.37  & 1.0$\pm$0.15   &  \tablefootmark{b} & 0.28\tablefootmark{c}   \\&&&&&&&&\\
3 &     13.7                  &  0.6$^{+0.1}_{-0.1}$ &  10.7$^{+0.2}_{-0.3}$ &  4.2$^{+0.9}_{-0.6}$  & 5.68  &  0.36  & 1.8$\pm$0.27   & \tablefootmark{b}  & 0.28\tablefootmark{c}   \\&&&&&&&&\\
3 &       13.7               &  0.7$^{+0.1}_{-0.1}$ &  10.6$^{+0.2}_{-0.5}$ &  4.2$^{+0.3}_{-0.4}$  & 6.01  &  0.36  &  1.2$\pm$0.18  & \tablefootmark{b}   & 0.3\tablefootmark{c}   \\&&&&&&&&\\
4 &       7.3               &  2.3$^{+1.9}_{-1.0}$ &  10.3$^{+0.4}_{-0.2}$ &  0.16$^{+0.05}_{-0.04}$  & 5.67  &  0.33  & 1.2$\pm$0.18   & \tablefootmark{b}  & 0.3\tablefootmark{c}   \\&&&&&&&&\\
4 &           7.3           &  0.3$^{+0.1}_{-0.1}$ &  12.2$^{+0.3}_{-0.7}$ &  0.34$^{+0.07}_{-0.05}$  & 6.02  &  0.33  &  0.7$\pm$0.10  & \tablefootmark{b}  & 0.3\tablefootmark{c}   \\&&&&&&&&\\
\textbf{TMC-1 NH$_3$}&&&&&&&&&\\
1 &        17.0                &  1.7$^{+1.5}_{-0.3}$ &  11.3$^{+0.4}_{-0.4}$ &  1.5$^{+0.8}_{-1.0}$  & 5.59  &  0.43  &  0.7$\pm$0.10  &\tablefootmark{b}   & 0.4\tablefootmark{c}   \\&&&&&&&&\\
1 &          17.0              &  1.0$^{+0.2}_{-0.2}$ &  10.6$^{+0.2}_{-0.4}$ &  2.8$^{+0.7}_{-0.8}$  & 6.13  &  0.38  &  0.45$\pm$0.07  &\tablefootmark{b}   & 0.4\tablefootmark{c}   \\&&&&&&&&\\
2 &      15.6                 &  3.0$^{+0.7}_{-0.6}$ &  11.0$^{+0.4}_{-0.7}$ &  0.7$^{+0.2}_{-0.1}$  & 5.60  &  0.37  &   0.8$\pm$0.12 &  \tablefootmark{b}  &  0.4\tablefootmark{c}  \\&&&&&&&&\\
2 &       15.6                &  3.8$^{+1.4}_{-0.7}$ &  10.9$^{+0.9}_{-0.5}$ &  0.4$^{+0.1}_{-0.1}$  & 6.01  &  0.39  & 0.5$\pm$0.07   & \tablefootmark{b}   &  0.4\tablefootmark{c}  \\&&&&&&&&\\
3 &      12.9                 &  8.8$^{+0.6}_{-0.7}$ &  13.8$^{+0.5}_{-1.5}$ &  0.35$^{+0.05}_{-0.04}$  & 5.75  &  0.44  &  1.1$\pm$0.16  &\tablefootmark{b}   &  0.4\tablefootmark{c}  \\&&&&&&&&\\
3 &        12.9               &  1.7$^{+0.5}_{-0.8}$ &  10.2$^{+0.2}_{-0.1}$ &  0.29$^{+0.19}_{-0.07}$  & 6.01  &  0.35  &  0.7$\pm$0.10  & \tablefootmark{b}  &  0.4\tablefootmark{c}  \\&&&&&&&&\\
4 &      10.0               &  3.0$^{+0.7}_{-0.4}$ &  10.6$^{+1.5}_{-0.4}$ &  0.26$^{+0.08}_{-0.06}$  & 5.84  &  0.44  & 0.9$\pm$0.13   &  \tablefootmark{b}  &  0.45\tablefootmark{c}  \\&&&&&&&&\\
\textbf{B213-1}&&&&&&&&&\\
1           &        26.9               &  6.9$^{+0.5}_{-0.5}$ &  10.1$^{+0.1}_{-0.1}$ &  1.1$^{+0.1}_{-0.1}$  & 5.97  &  0.50  & 1.3$\pm$0.19  &\tablefootmark{b}   & 0.6\tablefootmark{c}   \\&&&&&&&&\\
2           &        13.6               &  5.8$^{+0.4}_{-0.6}$ &  10.4$^{+0.2}_{-0.2}$ &  1.1$^{+0.2}_{-0.1}$  & 6.19  &  0.50  &  1.5$\pm$0.22  & \tablefootmark{b}   & 0.6\tablefootmark{c}   \\&&&&&&&&\\
3           &         13.4              &  4.6$^{+1.3}_{-1.1}$ &  10.1$^{+0.1}_{-0.1}$ &  1.0$^{+0.6}_{-0.3}$  & 6.21  &  0.48  &  1.2$\pm$0.18  &\tablefootmark{b}   &  0.6\tablefootmark{c}  \\&&&&&&&&\\
3           &          13.4             &  3.2$^{+1.7}_{-2.1}$ &  10.1$^{+0.1}_{-0.1}$ &  1.7$^{+0.7}_{-0.7}$  & 6.83  &  0.43  & 1.2$\pm$0.18   & \tablefootmark{b}  &  0.6\tablefootmark{c}  \\&&&&&&&&\\
4           &          6.0             &  1.3$^{+0.5}_{-0.3}$ &  10.2$^{+0.2}_{-0.1}$ &  1.1$^{+0.7}_{-0.4}$  & 6.62  &  0.66  & 1.3$\pm$0.19   &\tablefootmark{b}   & 0.6\tablefootmark{c}   \\&&&&&&&&\\
5           &           4.2            &  0.4$^{+0.3}_{-0.2}$ &  10.3$^{+0.1}_{-0.1}$ &  1.0$^{+0.8}_{-0.4}$  & 6.53  &  0.47  & 0.9$\pm$0.13   & \tablefootmark{b}   & 0.6\tablefootmark{c}   \\&&&&&&&&\\
6           &            5.2           &  0.8$^{+0.4}_{-0.2}$ &  10.3$^{+0.4}_{-0.3}$ &  1.1$^{+0.4}_{-0.4}$  & 6.65  &  0.43  &  0.71(1)  &  6.63  &  0.47  \\&&&&&&&&\\
7           &          10.5             &  0.7$^{+0.2}_{-0.2}$ &  10.3$^{+0.15}_{-0.16}$ &  4.4$^{+1.5}_{-2.1}$  & 6.80  &  0.48  &  1.17(1)  & 6.84   & 0.64   \\&&&&&&&&\\
8           &        10.3               &  3.1$^{+0.3}_{-0.2}$ &  10.1$^{+0.12}_{-0.08}$ &  5.6$^{+1.5}_{-1.2}$  & 5.97  &  0.49  & 1.24(3)   & 6.06   & 0.61 \\&&&&&&&&\\
9           &         5.5              &  2.8$^{+0.2}_{-0.2}$ &  10.7$^{+0.16}_{-0.19}$ &  6.5$^{+0.8}_{-1.5}$  & 5.98  &  0.49  & 1.26(3)  & 6.02   & 0.66 \\&&&&&&&&\\
\textbf{B213-2}&&&&&&&&&\\
1           &         20.9              &  3.5$^{+2.2}_{-1.7}$ &  10.5$^{+0.7}_{-0.4}$ &  2.4$^{+1.0}_{-1.0}$  & 7.29  &  0.55  &  0.5$\pm$0.07  & \tablefootmark{b}  & 0.5\tablefootmark{c}   \\&&&&&&&&\\
2           &        10.6               &  0.7$^{+0.2}_{-0.2}$ &  10.3$^{+0.1}_{-0.2}$ &  4.4$^{+0.4}_{-1.3}$  & 7.02 &  0.46  & 0.6$\pm$0.09   & \tablefootmark{b}  & 0.6\tablefootmark{c}   \\&&&&&&&&\\
8           &          16.5             &  1.5$^{+1.1}_{-0.4}$ &  10.9$^{+0.2}_{-0.3}$ &  2.4$^{+1.1}_{-1.0}$  & 6.97  &  0.40  & 0.8$\pm$0.12   &  \tablefootmark{b} &  0.5\tablefootmark{c}  \\&&&&&&&&\\
9           &               9.8        &  1.4$^{+0.2}_{-0.2}$ &  10.2$^{+0.2}_{-0.1}$ &  6.2$^{+0.7}_{-2.6}$  & 6.85  &  0.33 &  0.9$\pm$0.05   &  \tablefootmark{b}  &  0.5\tablefootmark{c}  \\&&&&&&&&\\
\textbf{B213-5}&&&&&&&&&\\
1           &           23.6            &  7.8$^{+0.48}_{-0.9}$ &  10.2$^{+0.1}_{-0.1}$ &  1.1$^{+0.2}_{-0.1}$  & 6.26  &  0.50  & 1.5$\pm$0.22   & \tablefootmark{b} & 0.7\tablefootmark{c}   \\&&&&&&&&\\
2           &          13.0             &  3.0$^{+0.5}_{-0.5}$ &  10.2$^{+0.3}_{-0.1}$ &  1.2$^{+0.3}_{-0.1}$  & 6.44  &  0.70  & 1.0$\pm$0.15   & \tablefootmark{b}   &  0.7\tablefootmark{c}  \\&&&&&&&&\\
3           &        10.9               &  1.0$^{+0.3}_{-0.1}$ &  10.1$^{+0.06}_{-0.07}$ &  2.7$^{+0.5}_{-1.5}$  & 6.73  &  0.68  & 1.0$\pm$0.15   & \tablefootmark{b}   &  0.7\tablefootmark{c}   \\&&&&&&&&\\
4           &        13.3               &  0.3$^{+0.1}_{-0.1}$ &  10.1$^{+0.1}_{-0.1}$ &  1.0$^{+0.1}_{-0.1}$  & 6.37  &  0.44  & 0.57(1)   & 6.3   &  0.37  \\&&&&&&&&\\
4           &        13.3               &  0.8$^{+0.2}_{-0.2}$ &  10.1$^{+0.1}_{-0.1}$ &  0.96$^{+0.2}_{-0.1}$  & 7.07  &  0.37  & 4.72(1)   & 7.11   &  0.55   \\&&&&&&&&\\
5           &         14.0              &  0.2$^{+0.1}_{-0.1}$ &  10.0$^{+0.1}_{-0.1}$ &  0.6$^{+0.6}_{-0.1}$  & 6.28  &  0.42  & 0.63(1)   & 6.33   &  0.35   \\&&&&&&&&\\
5           &             14.0          &  0.4$^{+0.1}_{-0.1}$ &  10.4$^{+0.1}_{-0.1}$ &  1.9$^{+0.1}_{-0.2}$  & 7.09  &  0.37  & 4.88(2)   & 7.13   &  0.55   \\&&&&&&&&\\
6           &           14.5            &  1.2$^{+0.7}_{-0.4}$ &  10.4$^{+0.17}_{-0.15}$ &  4.8$^{+2.0}_{-1.2}$  & 7.25  &  0.55  & 6.14(2)   & 7.15   & 0.60 \\&&&&&&&&\\
7           &               18.9        &  1.8$^{+1.7}_{-1.1}$ &  10.7$^{+0.7}_{-0.3}$ &  5.5$^{+0.8}_{-0.9}$  & 7.3  &  0.51  & 7.75(2)   & 7.21   & 0.59   \\&&&&&&&&\\
8           &               12.9        &  2.9$^{+0.3}_{-0.4}$ &  10.1$^{+0.1}_{-0.1}$ &  1.2$^{+0.2}_{-0.1}$  & 6.36  &  0.50  & 1.2$\pm$0.18   & \tablefootmark{b}  &  0.6\tablefootmark{c}   \\&&&&&&&&\\
9           &             10.9          &  0.8$^{+0.3}_{-0.1}$ &  10.3$^{+0.1}_{-0.1}$ &  1.3$^{+0.4}_{-0.6}$  & 6.06  &  0.48  & 0.77(1)   & 6.12   &  0.52 \\&&&&&&&&\\
9           &            10.9           &  0.3$^{+0.2}_{-0.1}$ &  10.2$^{+0.1}_{-0.1}$ &  1.0$^{+0.6}_{-0.4}$  & 7.00  &  0.37  & 3.06(1)   & 7.11   & 0.55    \\&&&&&&&&\\
\textbf{B213-6}&&&&&&&&&\\
1           &      22.2                 &  3.5$^{+0.5}_{-0.4}$ &  10.5$^{+0.3}_{-0.2}$ &  2.5$^{+1.3}_{-0.6}$  & 6.85  &  0.53  & 0.69(1)   & 6.77   & 0.68   \\&&&&&&&&\\
2           &           15            &  2.4$^{+0.3}_{-0.3}$ &  10.6$^{+0.1}_{-0.1}$ &  4.8$^{+0.6}_{-1.0}$  & 6.77  &  0.40  & 0.83(1)   &  6.79  &  0.46 \\&&&&&&&&\\
3           &           5.6            &  0.7$^{+0.5}_{-0.1}$ &  10.1$^{+0.1}_{-0.1}$ &  3.3$^{+1.4}_{-2.3}$  & 6.80  &  0.39  &  1.11(1)  &  6.84  & 0.44   \\&&&&&&&&\\
8           &           18.2            &  4.0$^{+0.9}_{-0.5}$ &  10.1$^{+0.1}_{-0.1}$ &  1.7$^{+0.6}_{-0.6}$  & 6.94  &  0.37  &  0.73(1)  & 6.92   &  0.49  \\&&&&&&&&\\
9           &          11.3             &  2.5$^{+2.5}_{-2.1}$ &  10.7$^{+0.6}_{-0.6}$ &  5.2$^{+3.3}_{-3.0}$  & 7.06  &  0.40  & 0.76(2)   &  6.99  &  0.44  \\&&&&&&&&\\
\textbf{B213-7}&&&&&&&&&\\
1           &         20.2              &  9.5$^{+2.7}_{-1.4}$ &  10.2$^{+0.1}_{-0.1}$ &  0.48$^{+0.15}_{-0.13}$  & 6.91  &  0.44 & 2.08(1)  &  6.87  & 0.41   \\&&&&&&&&\\
2           &           14.6            &  2.7$^{+0.3}_{-0.3}$ &  10.4$^{+0.1}_{-0.2}$ &  7.4$^{+0.5}_{-3.0}$  & 6.92  &  0.41  & 2.38(2)  & 6.91   &  0.47  \\&&&&&&&&\\
3           &          7.2             &  6$^{+1}_{-1}$ &  10.1$^{+0.2}_{-0.1}$ &  4.5$^{+1.7}_{-1.2}$  & 6.96  &  0.43  & 1.87(2)  &  6.96  & 0.55   \\&&&&&&&&\\
8           &          16.6             &  2.5$^{+0.3}_{-0.2}$ &  10.3$^{+0.3}_{-0.2}$ &  8.9$^{+0.7}_{-1.5}$  & 6.91  &  0.47  & 1.82(2)  & 6.84   & 0.43   \\&&&&&&&&\\
9           &            8.7           &  7$^{+1}_{-4}$ &  11.0$^{+0.1}_{-0.8}$ &  0.31$^{+0.67}_{-0.08}$  & 6.80  &  0.34  & 1.51(1)  & 6.88   &  0.52  \\&&&&&&&&\\
\textbf{B213-10}&&&&&&&&&\\
1           &         20.7              &  1.3$^{+0.2}_{-0.3}$ &  10.2$^{+0.2}_{-0.1}$ &  1.2$^{+0.7}_{-0.3}$  & 6.76  &  0.33  & 1.17(2)   & 6.77   & 0.45   \\&&&&&&&&\\
2           &            17.5           &  5.1$^{+1.4}_{-1.2}$ &  10.5$^{+0.2}_{-0.2}$ &  0.27$^{+0.13}_{-0.08}$  & 6.69  &  0.27  &  1.34(2)  & 6.68   & 0.47    \\&&&&&&&&\\
3           &            11.8           &  4.9$^{+0.8}_{-1.0}$ &  10.3$^{+0.2}_{-0.1}$ &  0.27$^{+0.11}_{-0.05}$  & 6.60  &  0.31  &  1.36(2)  & 6.57   &  0.47  \\&&&&&&&&\\
4           &          6.1             &  1.2$^{+1.0}_{-0.4}$ &  10.1$^{+0.1}_{-0.1}$ &  0.72$^{+0.45}_{-0.39}$  & 6.58  &  0.32  &  0.90(1)  & 6.59   &  0.45  \\&&&&&&&&\\
7           &        17.6               &  2$^{+1}_{-1}$ &  10.6$^{+0.2}_{-0.4}$ &  0.58$^{+0.62}_{-0.21}$  & 6.82  &  0.32  & 1.11(1)   & 6.83   & 0.42   \\&&&&&&&&\\
8           &            14.5           &  0.8$^{+0.2}_{-0.1}$ &  10.1$^{+0.1}_{-0.1}$ &  5.5$^{+0.6}_{-2.0}$  & 6.89  &  0.31  & 1.31(1)   & 6.86   & 0.40 \\&&&&&&&&\\
9           &            9.1           &  0.6$^{+0.1}_{-0.1}$ &  10.1$^{+0.1}_{-0.8}$ &  4.9$^{+0.4}_{-0.7}$  & 6.73  &  0.29  & 1.07(2)   & 6.74   & 0.42   \\&&&&&&&&\\
\textbf{B213-12}&&&&&&&&&\\
1           &          22.1             &  1.4$^{+0.4}_{-0.3}$ &  10.2$^{+0.1}_{-0.1}$ &  0.83$^{+0.3}_{-0.2}$  & 6.65  &  0.39  & 0.5$\pm$0.07   & \tablefootmark{b}   & 0.5\tablefootmark{c}   \\&&&&&&&&\\
2           &        17.6               &  2$^{+1}_{-1}$ &  10.8$^{+0.2}_{-0.2}$ &  0.3$^{+0.2}_{-0.1}$  & 6.69  &  0.39  &  0.45$\pm$0.07  & \tablefootmark{b}   & 0.5\tablefootmark{c}   \\&&&&&&&&\\
3           &       11.5                & 6$^{+1}_{-4}$ &  10.1$^{+0.1}_{-0.1}$ &  0.13$^{+0.78}_{-0.24}$  & 6.69  &  0.37  & 0.45$\pm$0.07   & \tablefootmark{b}   &  0.5\tablefootmark{c}  \\&&&&&&&&\\
8           &           8.7            &  2.0$^{+2.0}_{-0.8}$ &  10.5$^{+0.21}_{-0.12}$ &  0.41$^{+0.29}_{-0.19}$  & 6.51  &  0.32  &  0.7$\pm$0.10  &\tablefootmark{b}   & 0.5\tablefootmark{c}   \\&&&&&&&&\\
\textbf{B213-16}&&&&&&&&&\\
1           &             24.8          &  1.8$^{+0.1}_{-0.11}$ &  10.1$^{+0.09}_{-0.07}$ &  5.1$^{+0.6}_{-1.0}$  & 6.68  &  0.34  & 1.3$\pm$0.19   & \tablefootmark{b}  & 0.4\tablefootmark{c}   \\&&&&&&&&\\
2           &          13.3             &  1.2$^{+0.1}_{-0.1}$ &  10.3$^{+0.11}_{-0.10}$ &  4.3$^{+0.7}_{-1.1}$  & 6.61  &  0.44  & 1.5$\pm$0.22   & \tablefootmark{b}   & 0.5\tablefootmark{c}   \\&&&&&&&&\\
3           &         10.5              &  3.3$^{+0.1}_{-0.1}$ &  10.2$^{+0.14}_{-0.15}$ &  0.37$^{+0.1}_{-0.1}$  & 6.48  &  0.43  &  1.5$\pm$0.22  &  \tablefootmark{b}  & 0.5\tablefootmark{c}   \\&&&&&&&&\\
4           &        6.9               &  3.3$^{+0.2}_{-0.1}$ &  10.2$^{+0.07}_{-0.07}$ &  0.16$^{+0.1}_{-0.1}$  & 6.35  &  0.40  & 1.2$\pm$0.18   &\tablefootmark{b}   & 0.4\tablefootmark{c}   \\&&&&&&&&\\
7           &        22.6               &  1.8$^{+0.1}_{-0.1}$ &  10.34$^{+0.12}_{-0.17}$ &  4.5$^{+0.7}_{-0.9}$  & 6.71  &  0.33  &  1.2$\pm$0.18  & \tablefootmark{b}  & 0.35\tablefootmark{c}   \\&&&&&&&&\\
8           &          9.6             &  1.3$^{+0.3}_{-0.2}$ &  10.5$^{+0.17}_{-0.20}$ &  2.4$^{+0.8}_{-1.0}$  & 6.73  &  0.31  & 1.2$\pm$0.18   &  \tablefootmark{b}  & 0.35\tablefootmark{c}   \\&&&&&&&&\\
9           &          5.2             &  1.5$^{+0.9}_{-0.6}$ &  10.3$^{+0.46}_{-0.22}$ &  0.32$^{+0.39}_{-0.15}$  & 6.56  &  0.41  & 1.0$\pm$0.15   & \tablefootmark{b}   & 0.35\tablefootmark{c}   \\&&&&&&&&\\
\textbf{B213-17}&&&&&&&&&\\
1           &         6.4              &  0.4$^{+0.3}_{-0.2}$ &  10.4$^{+0.3}_{-0.3}$ &  1.4$^{+1.2}_{-0.7}$  & 7.24  &  0.42  & 0.8$\pm$0.12   &   \tablefootmark{b}  & 0.4\tablefootmark{c} \\&&&&&&&&\\
7           &         12.0              &  4.1$^{+2.8}_{-3.7}$ &  10.3$^{+0.2}_{-0.2}$ &  7.4$^{+0.9}_{-2.1}$  & 7.39  &  0.42  & 0.87(9)   &   7.22   & 0.52 \\&&&&&&&&\\
8           &          19.7             &  5.6$^{+3.0}_{-2.2}$ &  10.5$^{+0.4}_{-0.3}$ & 4.5$^{+1.5}_{-2.4}$  & 7.31  &  0.42   & 0.83(1)   &  7.21   & 0.48 \\&&&&&&&&\\
9           &           10.2            &  0.8$^{+0.6}_{-0.3}$ &  10.2$^{+0.2}_{-0.1}$ &  3.7$^{+1.1}_{-3.0}$  & 7.07  &  0.41  & 0.66(1)   &  7.13    & 0.51\\&&&&&&&&\\
\hline

\end{longtable}
\tablefoot{
\tablefoottext{a}{Derived from $Herschel$ and $Planck$ data \citep{palmeirim13, rodriguez21}.}
\tablefoottext{b}{Fixed to the correspondent $v_{LRS}$ of CH$_3$OH. }
\tablefoottext{c}{Fixed to a value that matches best the $\delta$v of CH$_3$OH and the line shape of C$^{18}$O.}
v$_{LSR}$ and $\delta$v  are reported without uncertainty because the statistical uncertainty is much lower than the spectral resolution.

}

\section{Results}
\label{sect_results}
\begin{figure*}
\centering
 \includegraphics [width=0.5\textwidth]{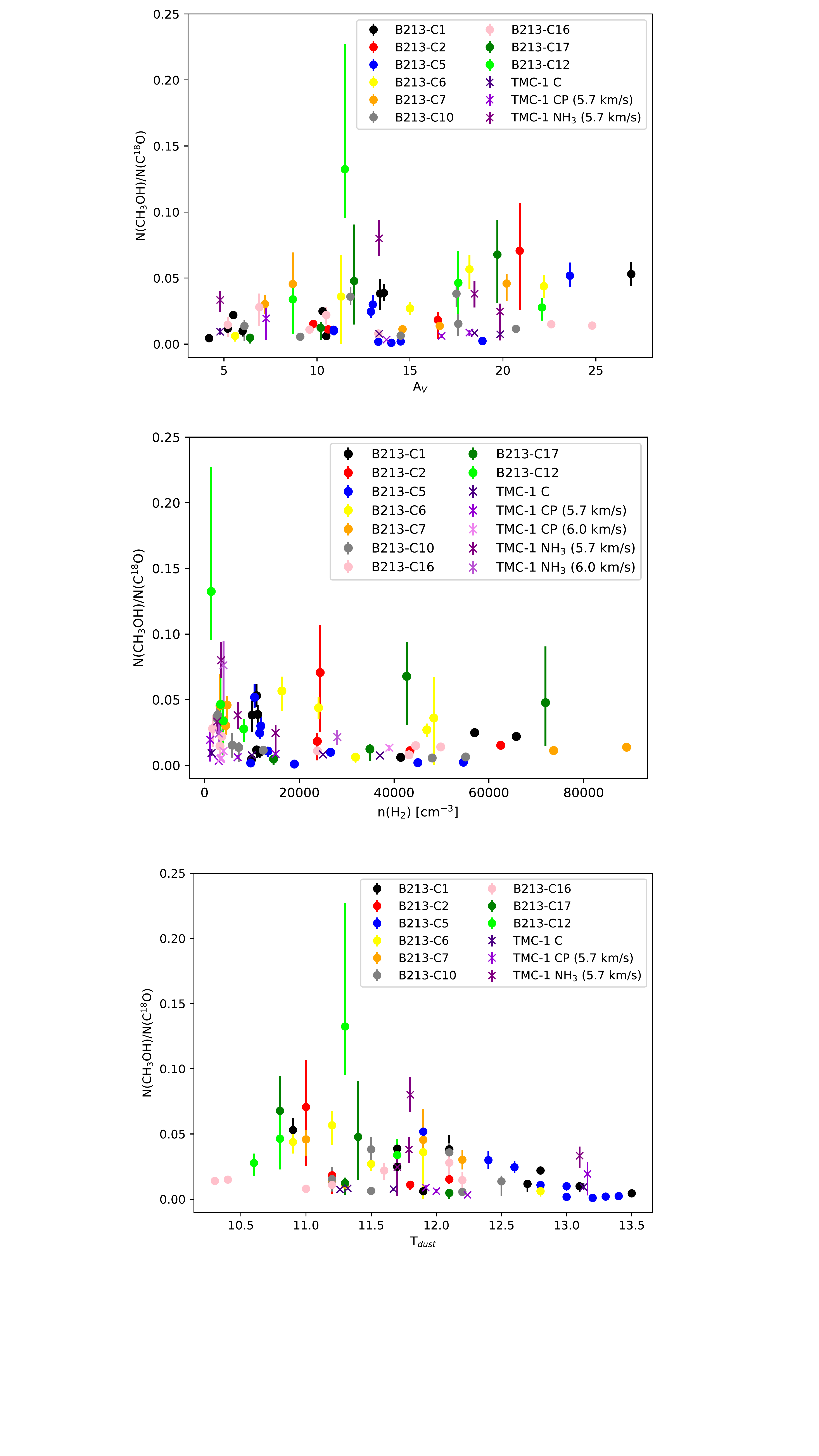}
 \caption{Methanol and C$^{18}$O column density ratios computed in all positions within our sample plotted against the: visual extinctions derived from $Herschel$ and $Planck$ data \citep{palmeirim13, rodriguez21} (upper panel); volume density derived from the MCMC+RADEX analysis of the methanol lines (central panel);  T$_{dust}$ derived from $Herschel$ data in \cite{rodriguez21} (lower panel).
}
  \label{fig:correlations}
\end{figure*}

Figure~\ref{fig:correlations} shows the column density ratio of methanol and C$^{18}$O in all observed positions in our sample  plotted against the visual extinction and the dust temperature derived from $Herschel$ and $Planck$ data \citep{palmeirim13, rodriguez21}, as well as against the volume density derived from the methanol MCMC+RADEX analysis, presented here in Section~\ref{sect_analysis}. 
Figure~\ref{fig:correlations} shows no correlation among N(CH$_3$OH)/N(C$^{18}$O) and A$_V$, T$_{dust}$, or n$_{H_2}$. This also holds if we consider B213 and TMC-1 separately or if we consider solely the column density of methanol instead of the column density ratio with respect to C$^{18}$O. In Figure~\ref{fig:correlations}, the points belonging to B213 and TMC-1 are shown as full circles and as crosses, respectively. In the case of multiple velocity components reported for methanol and C$^{18}$O in Table~\ref{table:results}, we assumed that most of the dust emission comes from the fiber with a larger methanol column density (supposedly the densest). As a consequence, only the velocity component with the largest methanol column density is plotted against A$_V$ and T$_{dust}$ in Figure~\ref{fig:correlations}. In particular, in the case of TMC-1 CP and NH$_3$, we plot the velocity component at 5.7 km/s. 
As we aim at studying methanol in different positions across different cores, comparing solely the column density of methanol would give us a biased picture because of the different densities of the positions involved. On the other hand, comparing methanol abundances with respect to molecular hydrogen is also not an optimal solution given the presence of different velocity components in several spectra, corresponding to different fibers on the line of sight. We decided to use the column density ratio of methanol and carbon monoxide because they are chemically related, as the depletion of carbon monoxide on the surface of dust grains is necessary to the formation of methanol. Using the rare isotopologue C$^{18}$O we can limit, if not exclude, the risk of large optical depth.    

\begin{figure}
\centering
 \includegraphics [width=0.5\textwidth]{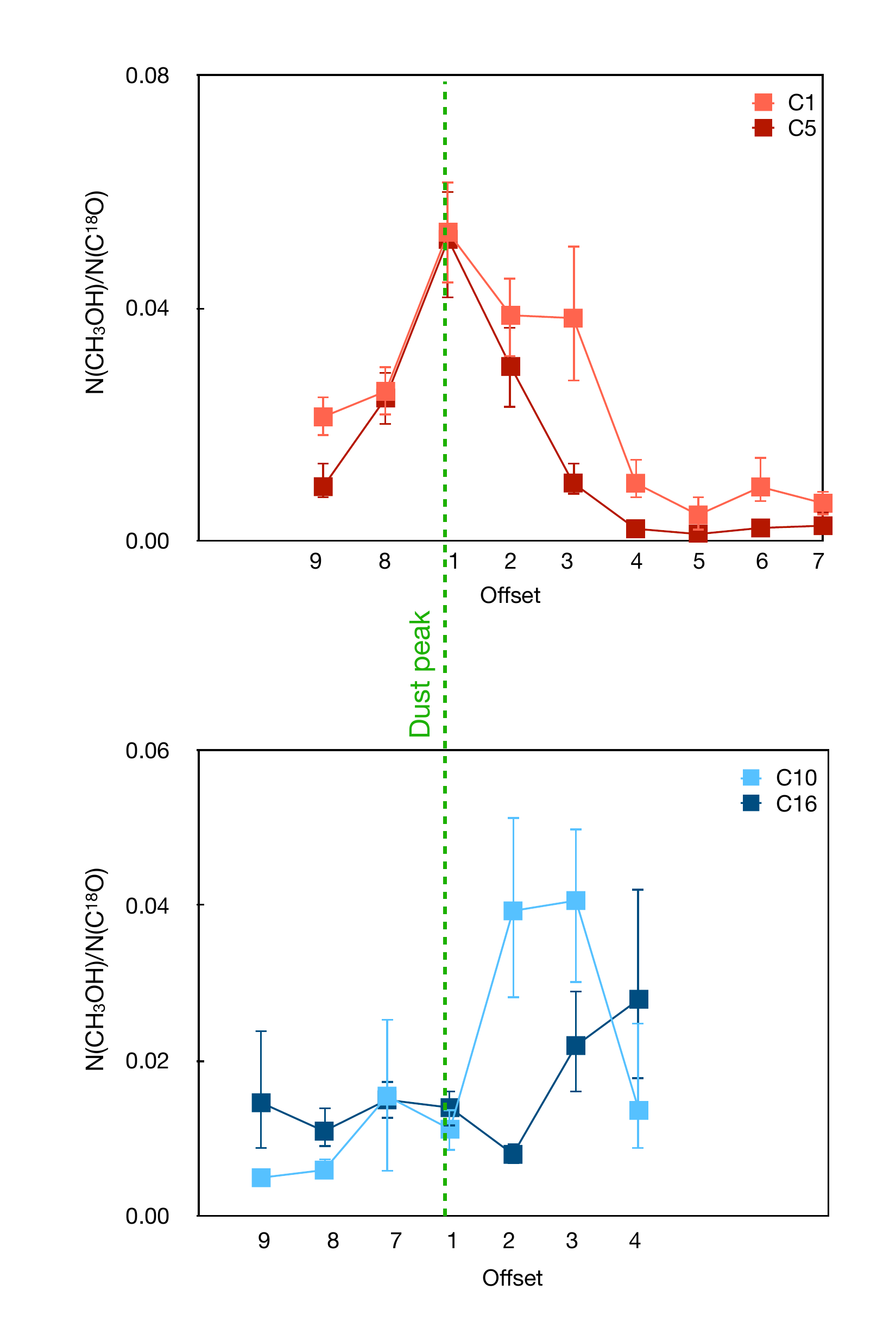}
 \caption{Upper panel: methanol and C$^{18}$O column density ratio for two cores in the North of B213, C1 and C5. Lower panel: methanol and C$^{18}$O column density ratio for two cores in the South of B213, C10 and C16. The column density ratios are plotted with respect to the observed offsets, and offset 1 always refers to the dust peak (also shown as a dashed green line). The positions of the offsets in the cores are shown in Figures~\ref{fig:B213-C1-results}, \ref{fig:B213-C5-results}, \ref{fig:B213-C10-results}, and \ref{fig:B213-C16-results}.
}
  \label{fig:summary}
\end{figure}

A possible reason for the lack of definite trends in Figure~\ref{fig:correlations} might be that methanol emits from the outer layers of starless cores (e.g. in L1544 as shown in \citealt{bizzocchi14, vastel14}). Furthermore, the emission of methanol is not homogeneous around pre-stellar and starless cores, and its distribution has been shown to depend on the density structure around the core, and hence the illumination onto the core due to the interstellar radiation field \citep{spezzano16, spezzano20}. In \cite{spezzano16}, for example, the emission maps of methanol and cyclopropenylidene towards the inner 2.5$"\times$2.5$"$ of the pre-stellar core L1544 show that both molecules have asymmetric distribution around the core, with methanol peaking towards the North-East and cyclopropenylidene towards the South-West. It is important to note that the visual extinction at the methanol and the $c$-C$_3$H$_2$ emission peak in L1544 is the same, within error-bars. This shows that the H$_2$ column density of the medium computed on the line-of-sight is not a good enough indicator to discriminate among the regions where the illumination is effective enough to keep more Carbon in its atomic form and hence available to form hydrocarbons like $c$-C$_3$H$_2$, and the regions where there is enough shielding to allow carbon to be locked in to carbon monoxide, and eventually form methanol. In addition to the density structure around the core, also larger scale environmental effects have shown to influence the distribution of methanol around starless and pre-stellar cores. In \cite{spezzano20}, methanol was mapped towards 4 pre-stellar and 2 starless cores, and it was shown that the methanol peak was influenced by the presence of nearby massive stars in the two cores mapped in Ophiucus (OphD and HMM-1).\\
The GEMS dataset offers the opportunity to study the emission of methanol towards 12 starless core, a larger sample with respect to previous studies. Furthermore, the GEMS dataset is not limited to the cores, giving us information on the transition from core to cloud.
In order to spot trends that might be related to the density structure surrounding the core, in Figures~\ref{fig:TMc1C-results} to \ref{fig:B213-C17-results} we plot the column density ratios ordered by offset position across the observed cut within the core for each source, while the offsets are shown on top of the N(H$_2$) column density maps of the cores in the upper panel of the Figure to facilitate the comparison. The individual results are discussed in detail in Appendix~\ref{sec_app_dataset}. Generally, we can recognise two different behaviours, shown in Figure~\ref{fig:summary}: the cores where N(CH$_3$OH)/N(C$^{18}$O) peaks at the dust peak, and the cores where the N(CH$_3$OH)/N(C$^{18}$O) ratios peak has a slight offset with respect to the dust peak ($\sim $10000 AU). This trend is particularly clear in the cores where more than 5 offsets have been observed. B213-C1 and C5 for example show a clear peak of the column density ratio towards the dust peak (see Figures~\ref{fig:summary}, ~\ref{fig:B213-C1-results}, and~\ref{fig:B213-C5-results}), while in B213-C10 and C16 the peak is slightly shifted from the dust peak ($\sim $10000 au), (see Figures~\ref{fig:summary}, ~\ref{fig:B213-C10-results} and~\ref{fig:B213-C16-results}). One reason for this behaviour might be the external irradiation on the cores. B213-C1 and C5 are in fact located towards the northern part of B213, that is known to be actively forming stars with $\sim$40 Class 0/I/II sources present in the vicinity of B213-C1, C5, C2, C6 and C7. B213-C10 and C16, on the other side, are located in the central/southern part of B213, which only contains $\sim$15 Class 0/I/II sources around B213-C10, C12 and C16, and 5 around B213-C17. In Figure~\ref{fig:stars} the positions of the young stellar objects in B213 are marked as red triangles on the H$_2$ column density map of the filament. Low-mass protostars can accelerate energetic particles along the outflow, and consequently increase the local cosmic rays flux \citep{padovani16, axen21}. This will in turn increase the destruction of CO, and consequently decrease the production of methanol towards the outer layers of starless cores in the North of B213.

The line-width of the methanol lines in the cores located in the North of B213 are $\sim$25$\%$ broader than the lines of methanol observed towards the central and southern cores of B213, hinting at more prominent effects due to stellar activity in the northern part for B213. This effect has already been observed with ammonia in \cite{seo15}. An additional indication of the environmental effects is the fact that the dust temperature in the southern cores of B213 is lower with respect to the northern cores, at comparable visual extinctions, see Figure~\ref{fig:AV_vs_T}.
In addition to the irradiation, there are also differences in the velocity structure. In B213-C1, C2, and C7, the profiles of C$^{18}$O and CH$_3$OH are very different. In B213-C6, C10 and C16 we have however a better agreement between the profiles of C$^{18}$O and CH$_3$OH. 

The N(CH$_3$OH)/N(C$^{18}$O) column density ratios towards the three cores in TMC-1 can be only calculated towards few offsets close to the dust peak, because methanol was not detected at larger radii, in contrast with B213, likely because the cores in B213 are denser.
Towards TMC-1 NH$_3$ the two different velocities show peaks of the N(CH$_3$OH)/N(C$^{18}$O) ratio at different offsets, maybe hinting at the fact that the gas they are tracing does not match in chemical age. Towards TMC-1 CP, if we exclude the offset 4 for the component at 5.6 km/s because of the large error bar, both velocity components show the same trend, an increase of the N(CH$_3$OH)/N(C$^{18}$O) ratio towards the dust peak. Towards TMC-1 C instead, the N(CH$_3$OH)/N(C$^{18}$O) ratio is constant within error-bars.

\section{Chemical Modelling}

\subsection{Physical structure for spherical 1D static chemical model}
\label{sec_david}

For a proper chemical modeling of the region under study, it is necessary to have a good knowledge of the density and temperature across the region. However, sometimes our knowledge is limited to a few points, and geometrical considerations and simplifications are mandatory. Temperature parameterizations of prestellar cores can be found in, for example, \cite{crapsi07}, in which they choose the spherically symmetric profile

\begin{equation}
    T(r) = T_{\rm out} - \frac{T_{\rm out}-T_{\rm in}}{1+\left(\frac{r}{r_{T0}}\right)^{\alpha}},
\end{equation}

where $T_{\rm out}$ and T$_{\rm in}$ are the temperatures at the outermost and innermost observed positions, respectively. The dependence of the temperature with the radius is controlled by $r_{T0}$, the flat radius of the temperature, and $\alpha$, an asymptotic power index. A similar functional dependence with the radius is found in Plummer-like density profiles. These profiles assume spherical symmetry, and they have become popular density profiles to describe dense cores \citep{tafalla02,priestley18}:

\begin{equation}
    n_{\rm H}(r) = \frac{n_{0}}{1+\left(\frac{r}{r_0}\right)^{\alpha}},
\end{equation}

where $n_{\rm H}$ is the total atomic hydrogen number density across the profile, $n_{0}$ is the central density, $r_{0}$ is the flat radius, and $\alpha$ is the asymptotic power index. These profiles allow us to obtain density and temperature profiles across the B213 cuts in  Table~\ref{table:results}. To have more data to derive physical profiles, we may fit data of different cuts with similar temperatures simultaneously. Therefore, we merge and average the data from the cut B213-C2 with B213-C6 (B213-C2-C6), B213-C10 with B213-C12 (B213-C10-C12), and consider B213-C16 as independent from the rest. The temperature profiles of B213-C2-C6, B213-C10-C12, and B213-C16 can be determined by finding the set of parameters that best fit the experimental data from Table~\ref{table:results}. The results are shown in Table~\ref{tab:tempDavid}.

\begin{table}
	\centering
	\caption{Parameter values for the temperature and density profiles presented in Section~\ref{sec_david}}
	\begin{tabular}{lccc}
		\toprule
		Parameters & B213-C2-C6 & B213-C10-C12 & B213-C16 \\ \midrule
		T$_{\rm in}$ (K) & $10.94 \pm 0.05$ & $10.89 \pm 0.04$ & $10.25 \pm 0.07$ \\
		T$_{\rm out}$ (K) & $14.40 \pm 0.05$ & $14.78 \pm 0.24$ & $13.78 \pm 0.41$ \\
		r$_{T0}$ (au) & $(7.03 \pm 0.20)\times 10^3$ & $(1.46 \pm 0.14)\times 10^4$ & $(1.04 \pm 0.27)\times 10^4$ \\
		$\alpha$ & $2.68 \pm 0.16$ & $1.56 \pm 0.13$ & $1.00 \pm 0.16$ \\ \midrule
		n$_{0}$ (cm$^{-3}$) & $(3.95 \pm 7.22)\times 10^5$ & $(2.06 \pm 3.85)\times 10^5$ & $(1.27 \pm 8.47)\times 10^6$ \\
		r$_{0}$ (au) & $(2.60 \pm 4.37)\times 10^3$ & $(5.00 \pm 9.08)\times 10^3$ & $(4.20 \pm 37.28)\times 10^2$ \\
		$\alpha$ & $2.40 \pm 1.58$ & $2.52 \pm 2.88$ & $1.24 \pm 1.39$ \\
    	\bottomrule
	\end{tabular}
	\label{tab:tempDavid}
\end{table}

We may perform a similar computation to derive the parameters of the density profiles. This time, however, we have information about the total hydrogen column density in the measured visual extinction across B213-C2-C6, B213-C10-C12, and B213-C16 in Table~\ref{table:results}. Given the density profile, we can estimate the total hydrogen column density, assuming spherical symmetry, performing a projection along the line of sight:

\begin{equation*}
    {N_{{\rm H}}}(r) = 2\times\sum_{i}\Delta l_{i}\frac{n_{\rm H}(s_{i})+n_{\rm H}(s_{i+1})}{2},
\end{equation*}

where $r$ is the impact parameter, $\Delta l_{i}= l_{i+1}-l_{i}$, $s_{i}=\sqrt{r^2+l_{i}^{2}}$, $l_{i}$ is a discretization of the segment along the line of sight $l_{\rm max} > \dots > l_{i+1} > l_{i} > \dots > 0$, with $l_{\rm max} = \sqrt{r_{\rm max}^{2}-r^{2}}$, and $r_{\rm max}$ the radius of the density profile. The parameters that best describe the density of the B213 cuts would be those that best fit the visual extinction after the projection along the line of sight. The total hydrogen column density $N_H$ is compared to the visual extinction in Table~\ref{table:results} using the relationship A$_{v}$ = $N_H/1.88\times 10^{21}$ \citep{bohlin78}, where it is assumed all hydrogen is in molecular form. To perform the fitting of the parameters to the extinction values, we have subtracted the background extinction, that is, the extinction at the outermost positions, to each cut. The results of the fitting process are shown in Table~\ref{tab:tempDavid}.

\subsection{Chemical modelling approaches}
\label{sec_modelling}
Two approaches were used to reproduce our observations with chemical models. In the first, we used the physical structure of some of the B213 cores in our sample derived in Section~\ref{sec_david}, and ran a spherical 1D static chemical model to compute the radial abundance profiles of methanol and CO. 
In the second approach, applied to the low-density parts of the filament beyond the starless cores, we ran a 0D chemical model assuming a grid of temperature and A$_V$ to compute the abundances of methanol.

To model the emission of methanol in cold starless cores, it is essential to utilize a so-called three-phase astrochemical model. By three phases, we mean gas phase, surface of icy mantles of 
interstellar grains, and bulk ice of grain mantles. The usage of a three-phase model, a more advanced approach in comparison to two-phase (gas phase + ice without distinction between
surface and bulk), is justified by the importance of the details of chemical processes on interstellar grains for the formation of methanol and its abundance in the gas phase. Since \citealt{geppert06}, the consensus is that observed abundances of interstellar methanol cannot be formed in gas-phase chemical reactions. In contrast, formation of methanol during
hydrogenation of CO molecules on interstellar grains is shown to be very efficient (e.g. \citealt{watanabe02}). The delivery of methanol formed on cold ($\sim$10 K) grains to the gas
phase is most likely due to the process of reactive desorption \citep{garrod07, minissale16}. The efficiency of reactive desorption depends on the details of a
particular chemically reacting system and the composition of the environment, i.e., the underlying ice surface \citep{minissale16, chuang18}. Thus, in order to
model properly a complicated relation between CO and CH$_3$OH in prestellar clouds, we need to utilize the reasonably detailed approach to grain chemistry and gas-grain interaction. 
To model the emission of methanol, we utilized the MONACO model previously applied to studies of chemistry in a number of prestellar cores \citep{vasyunin13, vasyunin17, nagy19, lattanzi20, harju20, scibelli21, jimenez-serra21}. MONACO is a rate equations-based 
numerical code capable of simulation chemistry in interstellar medium under a three-phase approach. As described in details in \cite{vasyunin17}, the code includes treatment of
chemical reactions in the gas phase, as well as diffusive chemistry on surfaces of icy mantles on interstellar grains, and in the bulk of icy mantles. The code includes treatment of 
thermal desorption, cosmic ray-induced desorption \citep{hasegawa93}, photodesorption with photodesorption yield equal to 10$^{-3}$ particles per incident photon \citep{oberg08} and reactive desorption following parametrization by \cite{minissale16}. The efficiency of reactive desorption is further adjusted to the fraction of
ice surface covered with non-water species \citep{vasyunin17}. We assume low diffusion-to-desorption energy ratio E$_{diff}$/E$_{des}$ = 0.3 for thermal diffusion of species on ice 
surface. In addition, quantum tunneling as a source of mobility for atomic and molecular hydrogen is assumed through a rectangular barrier for diffusion with thickness of 1.2 angstroms
\citep{vasyunin17}. For diffusion of bulk species, E$_{diff}$/E$_{des}$ is assumed twice that of for surface, i.e., 0.6 following \cite{garrod13}. 
The results of the spherical 1D static chemical models are presented in Figures~\ref{fig:1D_models} and \ref{fig:david}.
More details on the physical profiles used for the cores can be found in Section~\ref{sec_david}.\\
To convert abundances of species per unit volume to directly observed values of column densities, we assumed that modeled starless cores are spherically symmetric with radial density
profiles derived as described in Section~\ref{sec_david}. Convolution with IRAM 30m beam has also been taken into account. The procedure is described in more details in \cite{jimenez-serra16}.

To explore the abundances of CO and CH$_3$OH in the molecular gas that surrounds starless cores in the filament, we utilized the same chemical model as for the cores. However in contrast to starless cores, the surrounding gas does not exhibit a well-shaped spatial structure. Thus, we utilized simple 0D chemical models with constant values of temperature and density. Since both parameters can vary in the surrounding gas, we calculated grid of 0D models on a parameter space that we believe reasonable covers the ranges of temperature and density in the filament gas. The results of such gridded modeling allows us to explore the behavior of CO and CH$_3$OH abundances beyond the starless cores in B213 filament. The results of the 0D models are presented in Figure~\ref{fig:grids}.

\subsection{Modelling results}

\begin{figure*}
\centering
 \includegraphics [width=1\textwidth]{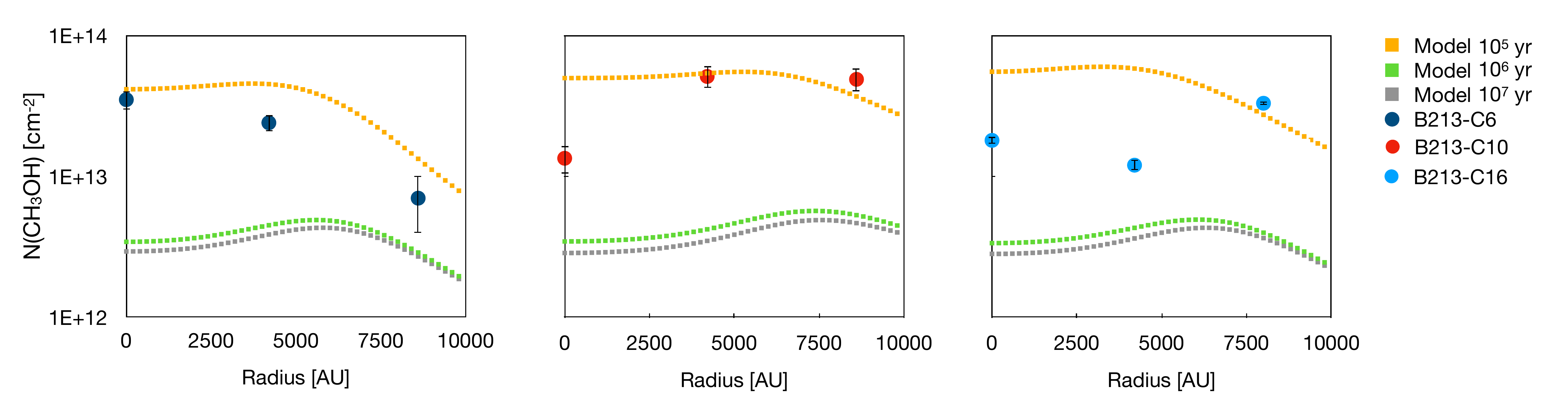}
 \caption{Methanol column densities calculated with the spherical 1D static chemical model using the physical structures derived for the cores B213-C2 and C6 (left panel), C10 (central panel), and C16 (right panel). The yellow, green and grey squares represent the results of the models extracted at 10$^5$, 10$^6$, and 10$^7$ yr. The column densities observed towards B213-C6, C10 and C16 are shown as dark blue, red and light blue in the left, central and right panels, respectively.
}
  \label{fig:1D_models}
\end{figure*}

\begin{sidewaysfigure}
 \centering
 \includegraphics [width=0.9\textheight, height=0.7\textwidth]{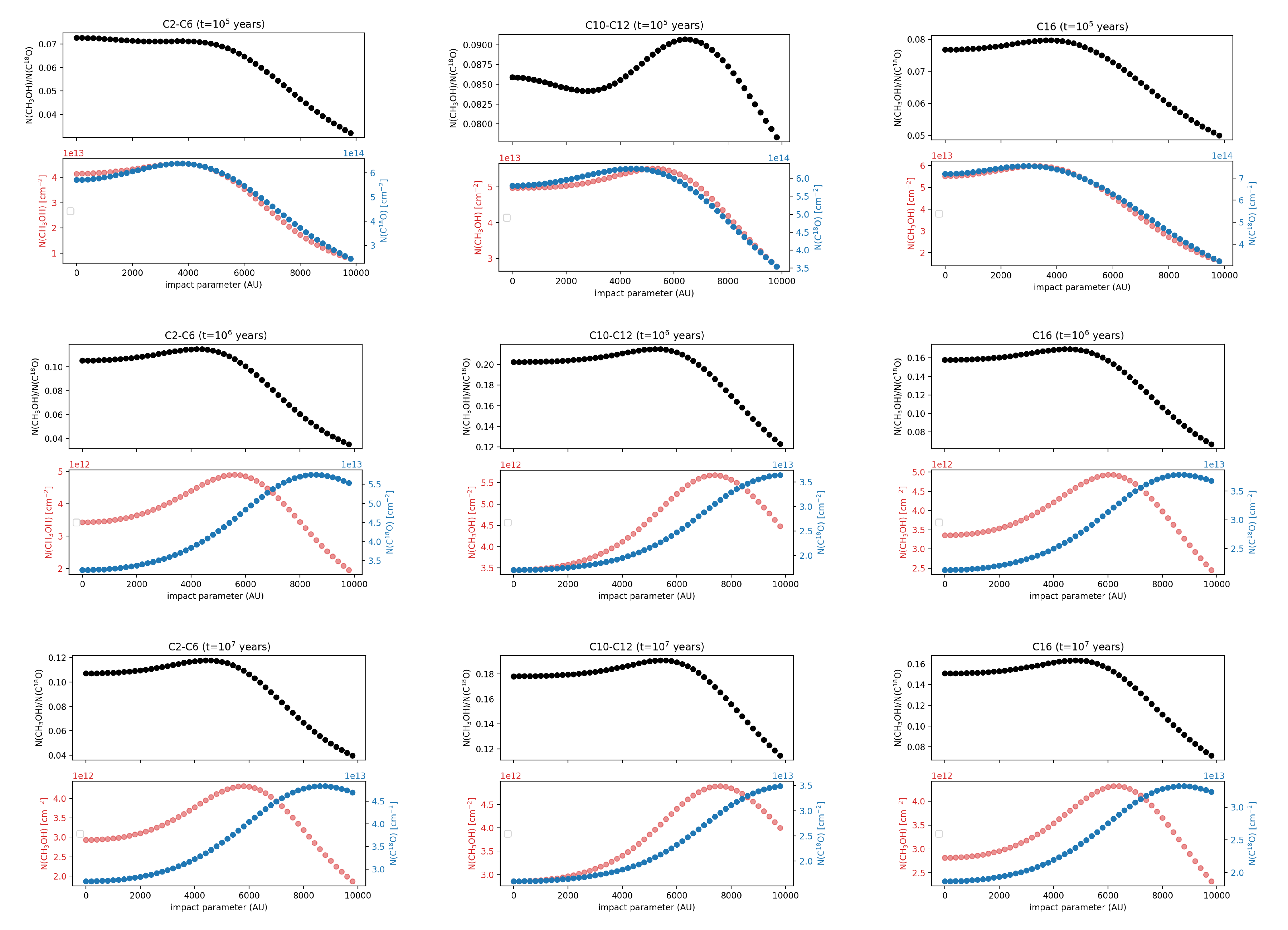}
 \caption{Results of the spherical 1D static chemical model using the physical structure derived for B213 C2 and C6 (left panel), C10 and C12 (central vertical panel) and C16 (right panel). The column density profiles of CH$_3$OH and C$^{18}$O (red and blue points) as well as their ratio (black points) have been computed at 10$^5$ years (upper panel), 10$^6$ (central horizontal panel) and 10$^7$ years (lower panel).
}
  \label{fig:david}
    \end{sidewaysfigure}

Figure~\ref{fig:1D_models} compares the N(CH$_3$OH) derived from the spherical 1D static chemical model used with the physical structures derived for B213 C2-C6, C10-12 and C16 at 10$^5$, 10$^6$ and 10$^7$ years, with the column densities derived in this paper for B213-C6, C10 and C16 and reported in Table~\ref{table:results}. The models at 10$^5$ yr reproduces fairly well (within a factor of a few) the column densities of methanol. The model using  the physical structure derived for the cores C2 and C5 (left panel in Figure~\ref{fig:1D_models}) reproduces very well also the radial profile of observed column densities of methanol towards B213-C5. We decided to plot only the N(CH$_3$OH) computed for C5 and not for C2 because the latter has larger error bars. The radial profiles observed for C10 and C16 (central and right panels in Figure~\ref{fig:1D_models}) are instead not very well reproduced by the models. While it is important to keep in mind the intrinsic large uncertainties of the column densities computed by chemical models \citep{vasyunin04}, we observe that the changes in N(CH$_3$OH) as a function of radii cannot be reproduced by the chemical models for C10 and C16, while it can reproduced fairly well for C6. This is shown even if we look at the radial profiles of methanol and C$^{18}$O in Figures~\ref{fig:TMc1C-results} to \ref{fig:B213-C17-results}, with respect to the one predicted by the models in Figure~\ref{fig:david}.
It is important to note that the physical structures used to run the spherical 1D static chemical model have been derived from $Herschel$ observations, which are not sensitive to the central and densest part of the cores because of the large beam ($\sim$40$"$). The volume density in starless cores could in fact have a steep increase towards the inner 40$"$, especially in the more evolved pre-stellar cores (see for example Figure 5 in \citealt{crapsi07}).
Figure~\ref{fig:david} shows the radial distribution of the N(CH$_3$OH)/N(C$^{18}$O) column density ratio, as well as the column densities of CH$_3$OH and C$^{18}$O (calculated from the main isotopologue assuming a $^{16}$O/$^{18}$O ratio of 500, Solar value in \citealt{penzias81}) in order to facilitate the comparison with our observational results shown in Figures~\ref{fig:TMc1C-results} to \ref{fig:B213-C17-results}.
Independently from the physical structure used, the model predicts a drop of CH$_3$OH column density of about one order of magnitude from 10$^5$ to 10$^6$ years, and an additional, although less significant, drop from 10$^6$ to 10$^7$ years. The column density of C$^{18}$O is underpredicted by the model. However, the model is only considering the core, and not also the cloud where the core is embedded. To check if the difference in C$^{18}$O column densities between model and observation is only due to the lack of the contribution from the cloud, we have calculated the C$^{18}$O column density at offsets outside of the cores, for example offset 7 in B213-C2, and then added it to the column densities predicted by the model, and we saw that the contribution of the cloud can account for the missing CO. 

\begin{figure*}
\centering
 \includegraphics [width=1\textwidth]{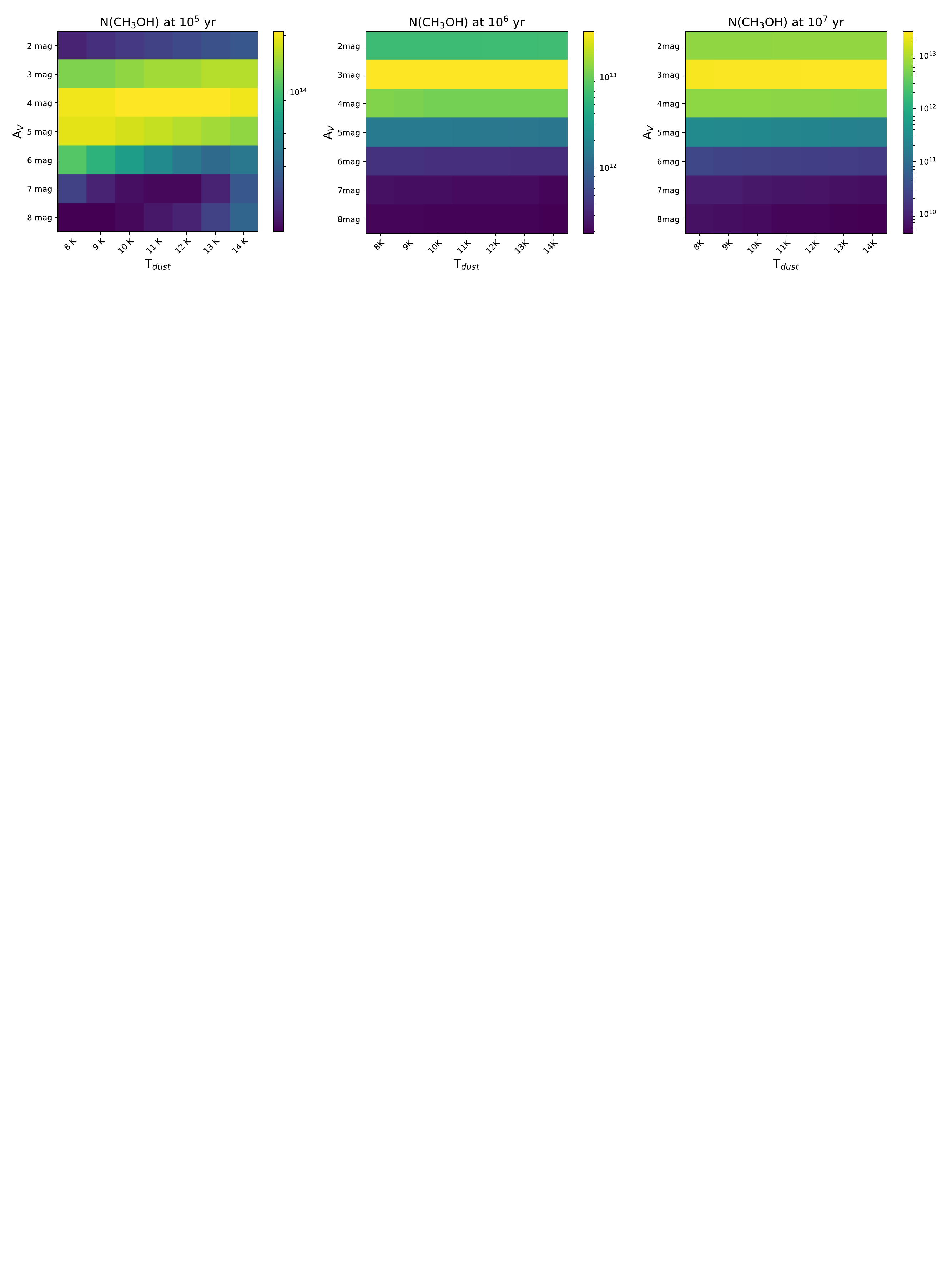}
 \caption{Methanol column densities calculated with the grid 0D chemical model at 10$^5$ years (left panel), at 10$^6$ years (central panel), and at 10$^7$ years (right panel). 
The volume density used is a function of A$_V$ defined by the density profiles of C2-C6 cores used for the spherical 1D static chemical model.
}
  \label{fig:grids}
\end{figure*}

In some of our cores the observed cuts cover over 40000 AU, hence merging into the cloud, while the physical models of the cores are derived assuming a Bonnor-Ebert sphere with a radius of 10000 AU. We therefore computed a grid of models in the temperature range of 8-14 K and A$_V$ range of 2-8 mag. Please note in this case the A$_V$ is the local A$_V$, so it needs to be multiplied by a factor of two to be compared with the A$_V$ tabulated in Table~\ref{table:results}. The column density of methanol resulting from the grid models at 10$^5$, 10$^6$ and 10$^7$ yr are shown in Figure~\ref{fig:grids}. Also in this case, the models cannot reproduce the variety of profiles that we observe. They predict a substantial drop of the methanol in gas-phase between 10$^5$ and 10$^6$ yr because of freeze-out, as well as a defined peak of the methanol column density, independent from the dust temperature, at A$_V$ $\sim$ 4 mag (corresponding to a total visual extinction of $\sim$8 mag along the line of sight). This is in contrast with what we observe, i.e. that the cores where the dust temperature is higher, tend to have the methanol peaking at larger A$_V$.
It is important however to note that protostellar feedback (e.g. higher cosmic-ray ionization rate and/or shocks) is not included in the models.

\section{Conclusions}
\label{sect_conclusions}
Our observations of methanol with the GEMS dataset show that methanol is present and its emission is extended in cold starless cores and in their surrounding clouds. In our dataset in fact, methanol has been detected at visual extinction ranging from 4 to 27 mag. Furthermore, the different behaviours in our sample of starless cores in Taurus show that local environmental differences are significative and have an impact on the distribution of methanol.\\

Previous work on emission maps of methanol towards 4 starless and 2 pre-stellar cores suggested that the asymmetric distribution of methanol is linked to the large scale density structure around the cores as well as different amount of illumination from the external radiation field  \citep{spezzano16, spezzano20}. With this work we can confirm that environmental effects have a strong impact on the methanol distribution. However, we show that the column density and temperature of the medium computed on the line-of-sight are not good enough indicators to identify the effects of the environment on the methanol, and that the level of star formation activity needs to be taken into account. It is clear from our data that low-mass protostars in the surrounding environment have an impact on the distribution of methanol towards starless cores. Low-mass protostars in fact can accelerate energetic particles and increase the local cosmic-ray flux and ionisation rate. This can locally increase the abundance of He$^+$, the main destruction partner of CO in dark clouds, thus reducing its abundance in the gas and solid form (the latter being the first step toward CH$_3$OH production). Increased fluxes of cosmic rays can also increase the CO desorption rate (e.g. \citealt{hasegawa93}), again reducing the amount of available solid CO for its transformation into methanol via successive hydrogenation. The higher CO desorption rate could compensate the gas phase destruction of CO via He$^+$; indeed we do not see any local reduction of the C$^{18}$O column density around the cores in the North.
Dedicated observations targeting molecular ions are needed to map the variation of ionization rate across low-mass star forming regions in order to test the effect on the chemistry with the chemical models. 
Another consequence of the presence of low-mass stars nearby is that they contribute to the clumpiness of the medium, escavated by the cones of the outflows, which could lead to a more efficient penetration of the interstellar radiation field in the molecular cloud. In these conditions, only the densest regions (i.e. the centers of the starless cores) will be screened enough from UV photons to allow CO to accumulate on the surface of dust grains. The higher dust temperature measured by $Herschel$ in the northern region of B213 could indeed hint to this scenario, but higher resolution mapping of the dust continuum emission and gas tracers are needed to make quantitative conclusions.\\

The comparison with state-of-the-art chemical models shows that overall the agreement of the modelled column densities with the observed ones is quite good, and that reactive desorption is a very efficient method to release the methanol formed on the surface of dust grains into the gas-phase, reproducing the observations. The discrepancies among models and observations in the reproduction of the radial profiles might be due to incomplete formation/destruction/desorption mechanisms in the chemical models, to the fact that the protostellar feedback is not included in the models, as well as to the intrinsic limit of using spherical models to reproduce an asymmetric emission.
Additional experimental and theoretical work on, for example, the chemical desorption and cosmic ray sputtering on different ice mixtures would be helpful for a better match between observations and models. Collisional rates for methanol with molecular hydrogen at temperatures lower than 10 K would also be beneficial.


\begin{acknowledgements}
The authors are grateful to the anonymous referee for insightful comments.\\
A large part of the data analysis described in this paper was performed during the spring of 2020, in the beginning of the COVID pandemic and during a hard lockdown. S.S. wishes to thank the Max Planck Society for the flexibility that was allowed during the pandemic, because it contributed to maintaining a clear and focus mind during the hours that she could dedicate to her work, and overall to keep calm, while waiting for the "storm" to pass.\\
Based on analysis carried out with the CASSIS software (http://cassis.irap.omp.eu;) and CDMS and JPL spectroscopic databases and LAMDA molecular databases. CASSIS has been developed by IRAP-UPS/CNRS.
S.S. wishes to thank the Max Planck Society for the Independent Max Planck Research Group funding.
AF,  DNA and MRB are funded by Spanish MICINN through PID2010-106235GB-I00 national research project. 
V.W. acknowledges the CNRS program "Physique et Chimie du Milieu Interstellaire" (PCMI) co-funded by the Centre National d’Etudes Spatiales (CNES).
A.V. and A.P. are the members of the Max Planck Partner Group at the Ural Federal University. A.V. and A.P. acknowledge the support of the Russian Ministry of Science and Education via the State Assignment Contract no. FEUZ-2020-0038.
\end{acknowledgements}

{}

\begin{appendix}

\section{Additional information on the single cores}
\label{sec_app_dataset}

\subsection{TMC-1 C}
\begin{figure*}
\centering
 \includegraphics [width=0.8\textwidth]{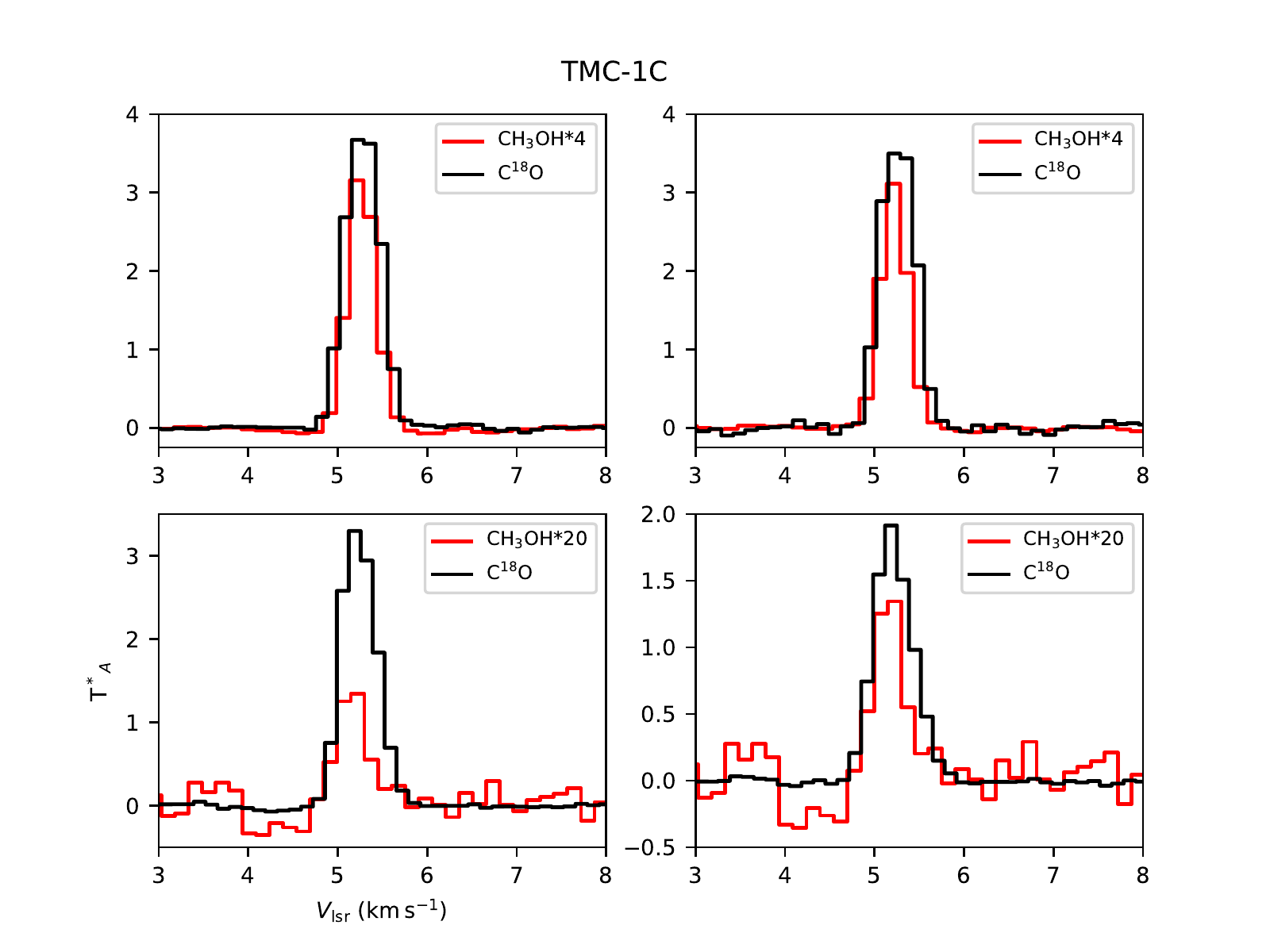}
 \caption{Spectra of the 2$_{1,2}$-1$_{1,1}$ ($E_2$) transition of methanol overlaid with the $J$ = 1-0 transition of C$^{18}$O in TMC-1 C.
}
  \label{fig:TMC-1C-co}
\end{figure*}

We have observed CH$_3$OH and C$^{18}$O (both lines) towards four positions in TMC-1 C. The  $J_{K_a,K_c}$ = 2$_{1,2}$-1$_{1,1}$ ($E_2$), 2$_{0,2}$-1$_{0,1}$ ($A^+$),  2$_{0,2}$-1$_{0,1}$ ($E_1$), and 2$_{0,2}$-1$_{0,1}$ ($E_1$-$E_2$) transitions of methanol have been observed towards all four positions, while the 1$_{0,1}$-0$_{0,0}$ ($A^+$) transition has been observed only towards the offsets 3 and 4. All lines have been detected with a signal to noise larger than 3 towards all positions. 
The spectra of the 2$_{1,2}$-1$_{1,1}$ ($E_2$) transition of methanol overlaid with the $J$ = 1-0 transition of C$^{18}$O are shown in Figure~\ref{fig:TMC-1C-co}, and they show a very good match both in $v_{LSR}$ and in line-width. \\
Figure~\ref{fig:TMc1C-results} shows in the upper panel the observed offsets within TMC-1 C on the H$_2$ column density map derived from $Herschel$ and $Planck$ data \citep{rodriguez21} data, and the variation of the CH$_3$OH and C$^{18}$O column density ratio, as well as the single column densities N(CH$_3$OH) and N(C$^{18}$O) in the observed cut across TMC-1 C in the lower panel. Both N(CH$_3$OH) and N(C$^{18}$O) increase when going towards dust peak of TMC-1 C. The column density ratio is constant, within error-bars, across the observed cut.

\subsection{TMC-1 CP}
\begin{figure*}
\centering
 \includegraphics [width=0.8\textwidth]{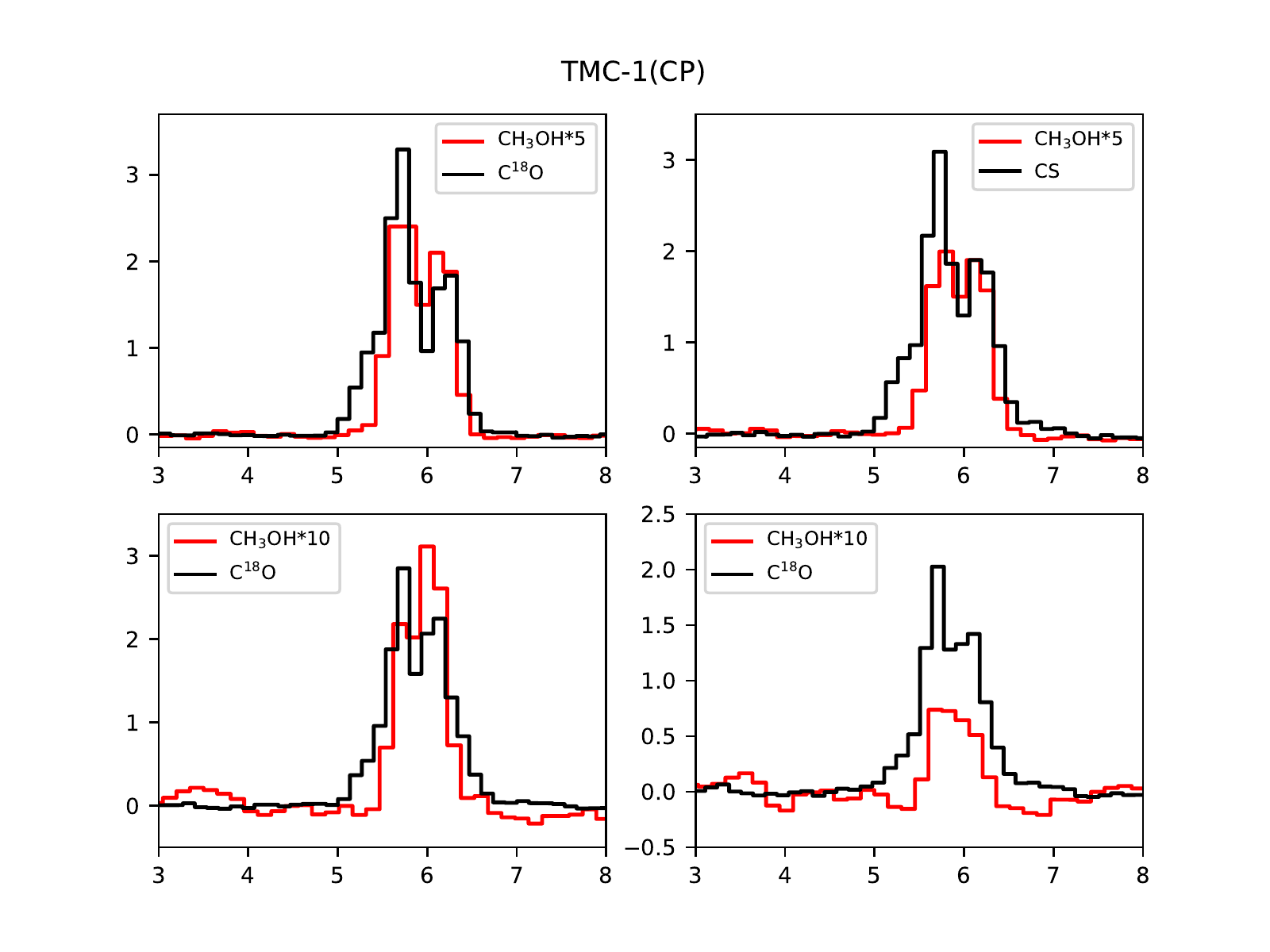}
 \caption{Spectra of the 2$_{1,2}$-1$_{1,1}$ ($E_2$) transition of methanol overlaid with the $J$ = 1-0 transition of C$^{18}$O in TMC-1 CP.
}
  \label{fig:TMC-1CP-co}
\end{figure*}

We have observed CH$_3$OH and C$^{18}$O (both lines) towards four positions in TMC-1 CP. The  $J_{K_a,K_c}$ = 2$_{1,2}$-1$_{1,1}$ ($E_2$), 2$_{0,2}$-1$_{0,1}$ ($A^+$),  2$_{0,2}$-1$_{0,1}$ ($E_1$), and 2$_{0,2}$-1$_{0,1}$ ($E_1$-$E_2$) transitions of methanol have been observed towards all four positions, while the 1$_{0,1}$-0$_{0,0}$ ($A^+$) transition has been observed only towards the offset 4. All lines have been detected with a signal to noise larger than 3 towards all positions. 
The spectra of the 2$_{1,2}$-1$_{1,1}$ ($E_2$) transition of methanol overlaid with the $J$ = 1-0 transition of C$^{18}$O are shown in Figure~\ref{fig:TMC-1CP-co}. TMC-1 CP is known to have multiple velocity components (Fuente 2019 and references therein). From the spectra shown in Figure~\ref{fig:TMC-1CP-co} we see that methanol is tracing two of the many velocity components seen in C$^{18}$O.\\	
Two velocity components are present towards TMC-1 CP, one at 5.6 km/s and one at 6.0 km/s, with the latter having column densities of both CH$_3$OH and C$^{18}$O larger by about a factor of 2 with respect to the lower velocity component.
Figure~\ref{fig:TMc1CP-results} shows the observed offsets on the H$_2$ column density map in the upper panel, and the variation of the CH$_3$OH and C$^{18}$O column density ratio, as well as the single column densities N(CH$_3$OH) and N(C$^{18}$O) in the observed cut across the core for both velocity components, in the lower panel. In the lower velocity component the column density of both methanol and C$^{18}$O increase towards the dust peak. Excluding the offset 4, that has a larger error in comparison with the other positions, also the column density ratio increases towards the dust peak.

\subsection{TMC-1 NH$_3$}
\begin{figure*}
\centering
 \includegraphics [width=0.8\textwidth]{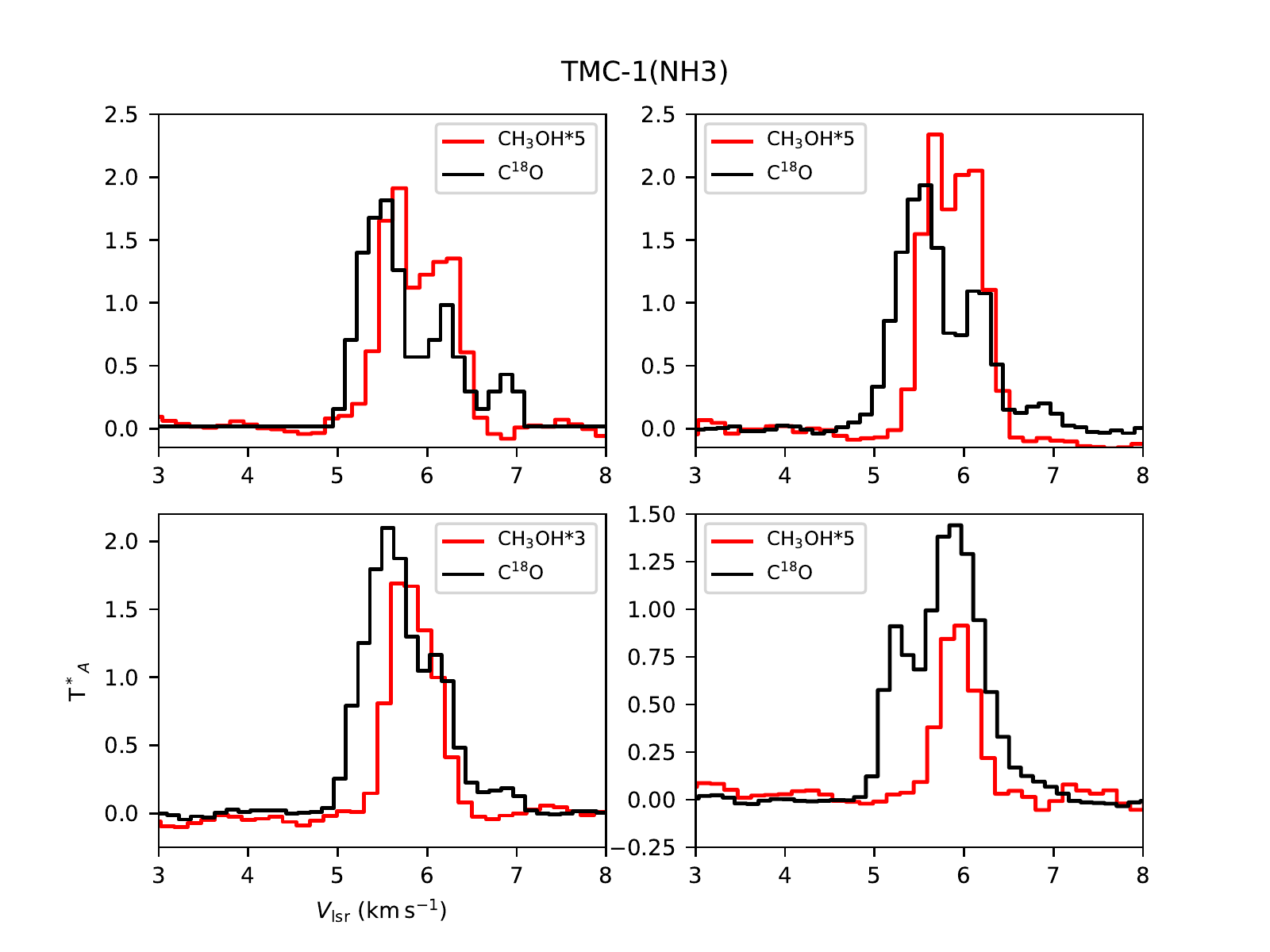}
 \caption{Spectra of the 2$_{1,2}$-1$_{1,1}$ ($E_2$) transition of methanol overlaid with the $J$ = 1-0 transition of C$^{18}$O in TMC-1 NH$_3$.
}
  \label{fig:TMC-1NH3-co}
\end{figure*}

We have observed CH$_3$OH and C$^{18}$O (both lines) towards four positions in TMC-1 NH$_3$. The  $J_{K_a,K_c}$ = 2$_{1,2}$-1$_{1,1}$ ($E_2$), 2$_{0,2}$-1$_{0,1}$ ($A^+$),  2$_{0,2}$-1$_{0,1}$ ($E_1$), and 2$_{0,2}$-1$_{0,1}$ ($E_1$-$E_2$) transitions of methanol have been observed towards all four positions, while the 1$_{0,1}$-0$_{0,0}$ ($A^+$) transition has been observed only towards the offset 4. All lines have been detected with a signal to noise larger than 3 towards all positions. 
The spectra of the 2$_{1,2}$-1$_{1,1}$ ($E_2$) transition of methanol overlaid with the $J$ = 1-0 transition of C$^{18}$O are shown in Figure~\ref{fig:TMC-1NH3-co}. The velocity structure that we can see in the C$^{18}$O lines is quite complex, and the mismatch with the peaks of the methanol lines suggests the presence of more velocity components that we can actually resolve, at least in C$^{18}$O.\\
TMC-1 NH$_3$ is the ammonia peak in TMC-1 and supposed to be the most evolved core, with respect to TMC-1 CP for example \citep{hirahara92}. Two velocity components are present for methanol, while a more complex structure is present in C$^{18}$O, see Figure~\ref{fig:TMC-1NH3-co}. 
Figure~\ref{fig:TMc1NH3-results} shows in the upper panel the observed offsets within TMC-1 NH$_3$ on the H$_2$ column density map derived from $Herschel$ and $Planck$ data \citep{rodriguez21}, and the variation of the CH$_3$OH and C$^{18}$O column density ratio, as well as the single column densities N(CH$_3$OH) and N(C$^{18}$O) in the observed cut across the core for both velocity components, in the lower panel. In the component at 5.7 km/s, the column densities of both molecules have a peak at the offset 3, and then decreases moving towards the dust peak. The column density ratio shows the same behaviour. 
In the component at 6.0 km/s instead, while the column density of C$^{18}$O decreases towards the dust peak, the methanol column density has a peak at the offset 2, and so does the column density ratio.

\subsection{B213 C1}
We have observed all lines of CH$_3$OH reported in Table~\ref{table:parameters}, with the exception of the 1$_{0,1}$-0$_{0,0}$ ($A^+$) transition, and the 1-0 transition of C$^{18}$O towards nine positions in B213-C1. The observed lines have been detected with a signal to noise larger than 3 towards all positions. 
The spectra of the 2$_{1,2}$-1$_{1,1}$ ($E_2$) transition of methanol overlaid with the $J$ = 1-0 transition of C$^{18}$O are shown in Figure~\ref{fig:figure1}.
The velocity structure is quite complex, with both molecules showing multiple velocity components. A shift in the v$_{LSR}$ of the main velocity component across the observed cut, is also clearly seen.\\
The upper panel in Figure~\ref{fig:B213-C1-results} shows the H$_2$ column density map of B213-C1 as well as the offsets of our observations. The observed stripe is centred at the denser peak, labeled as offset 1, but also passes through a less dense tail towards the East (offsets 6 and 7).
In the lower panel of Figure~\ref{fig:B213-C1-results} are shown both the column density ratio of CH$_3$OH/C$^{18}$O and the single column densities across the observed stripe, which all decrease rather sharply towards the East after the offset 3 and peak towards the dust peak.

\subsection{B213 C2}
\begin{figure*}
\centering
 \includegraphics [width=0.8\textwidth]{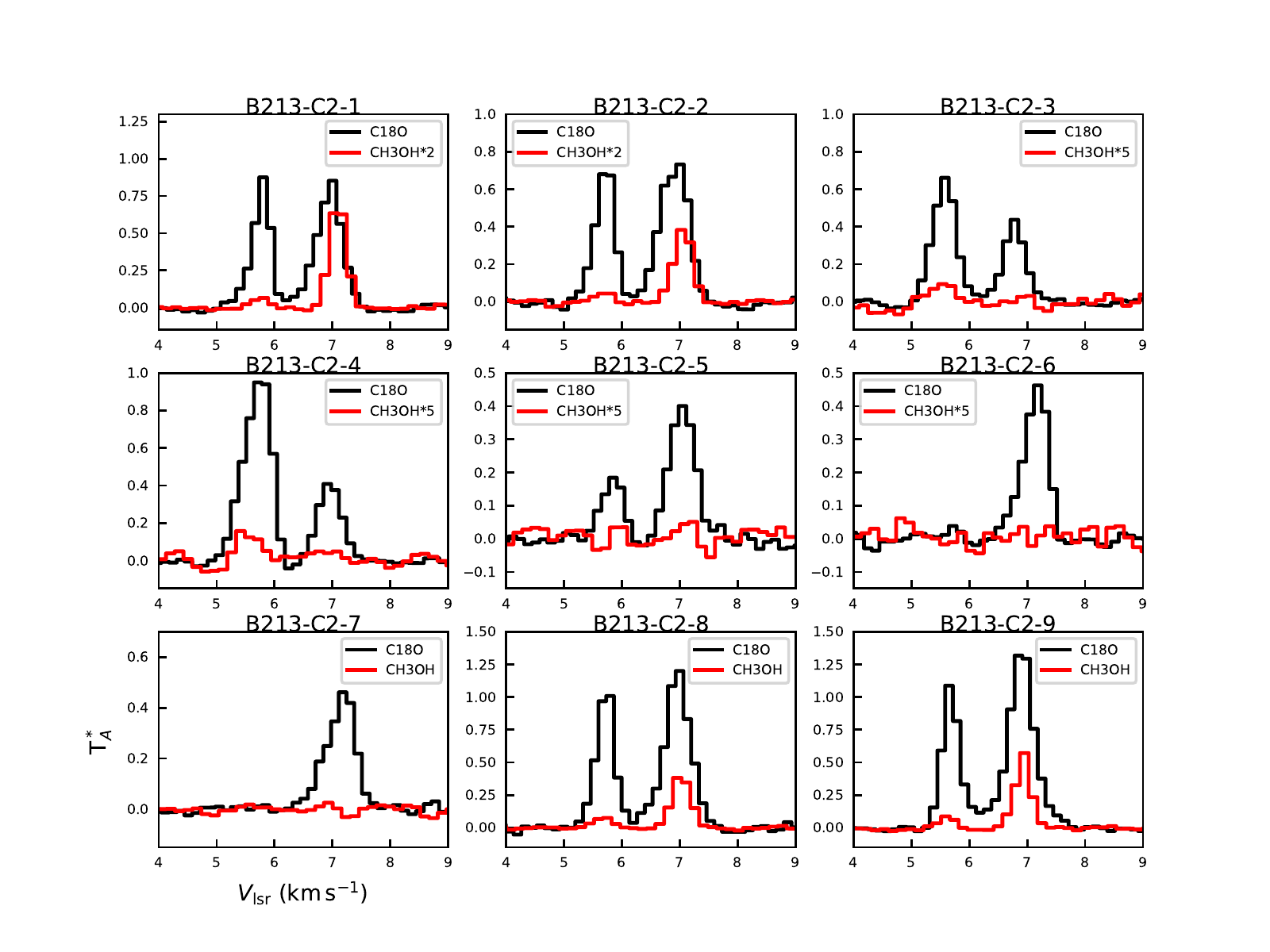}
 \caption{Spectra of the 2$_{1,2}$-1$_{1,1}$ ($E_2$) transition of methanol overlaid with the $J$ = 1-0 transition of C$^{18}$O in B213-C2
}
  \label{fig:B213-C2-all}
\end{figure*}

We have observed all lines of CH$_3$OH reported in Table~\ref{table:parameters}, with the exception of the 1$_{0,1}$-0$_{0,0}$ ($A^+$) transition, and the 1-0 transition of C$^{18}$O towards nine positions in B213-C2. The observed lines have been detected with a signal to noise larger than 3 towards positions 1, 2, 8 and 9. 
The spectra of the 2$_{1,2}$-1$_{1,1}$ ($E_2$) transition of methanol overlaid with the $J$ = 1-0 transition of C$^{18}$O are shown in Figure~\ref{fig:B213-C2-all}.
The C$^{18}$O transition shows two velocity components in all position with the exception of positions 6 and 7. The methanol transition instead shows a brighter velocity component at $\sim$7 km/s.\\
The H$_2$ column density map of B213-C2 is shown in the upper panel of Figure~\ref{fig:B213-C2-results} with all the offsets observed within B213-C2 in the GEMS large project, however methanol was only detected towards the densest offsets (1,2 8 and 9).
Both the variation of the CH$_3$OH and C$^{18}$O column density ratio, as well as the single column densities N(CH$_3$OH) and N(C$^{18}$O) are shown in the lower panel. The column densities of CH$_3$OH and C$^{18}$O show different behaviours. While C$^{18}$O is at its lowest at the dust peak and increases towards the West (offsets 8 and 9), methanol peaks at the dust peak and decreases towards offsets 8 and 9, although the error bars on the methanol column density at the dust peak are rather large so it might be only a slight increase. The column density ratio also peaks towards the dust peak of B213-C2.

\subsection{B213 C5}
\begin{figure*}
\centering
 \includegraphics [width=0.8\textwidth]{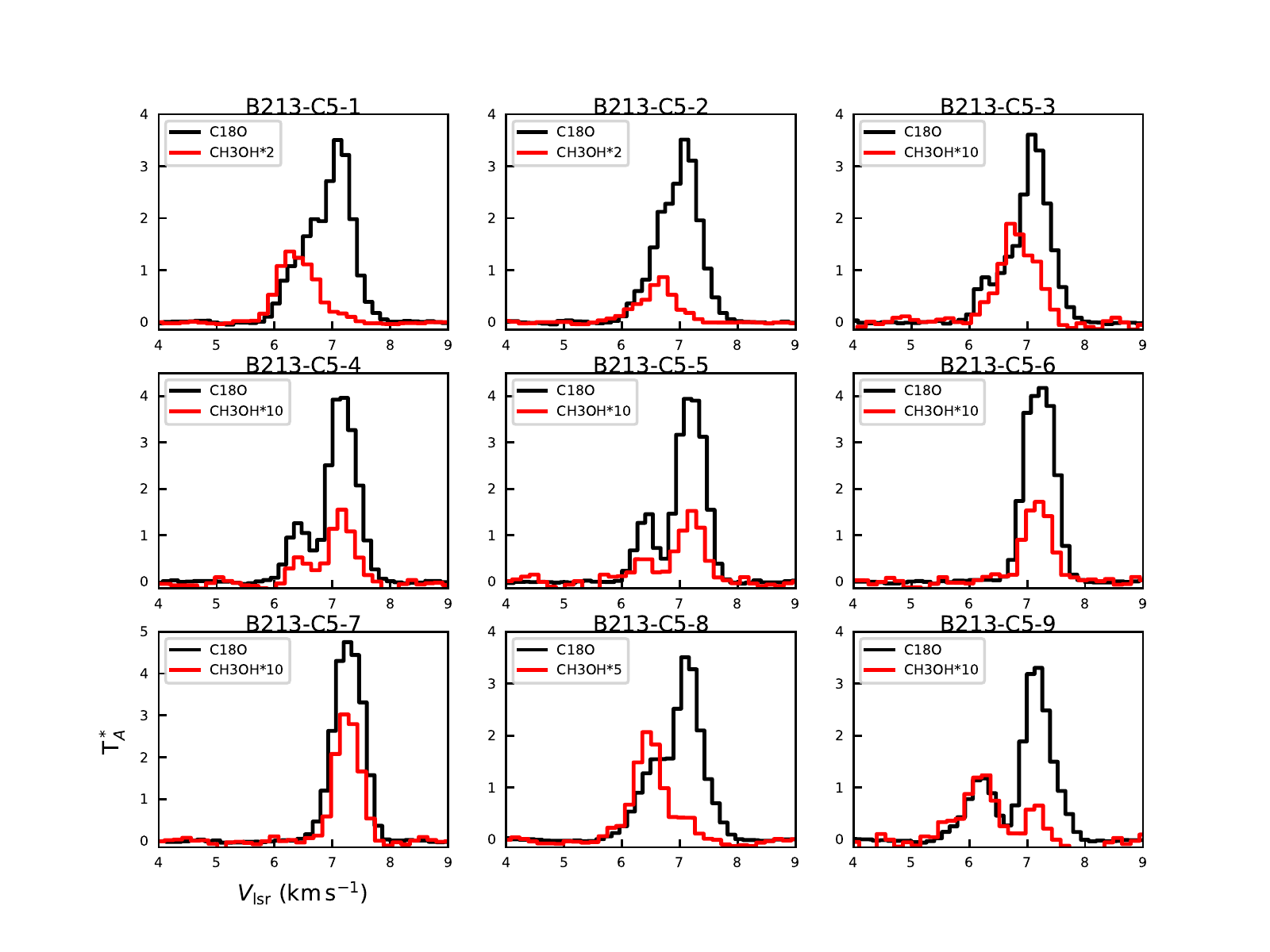}
 \caption{Spectra of the 2$_{1,2}$-1$_{1,1}$ ($E_2$) transition of methanol overlaid with the $J$ = 1-0 transition of C$^{18}$O in B213-C5
}
  \label{fig:B213-C5-all}
\end{figure*}

We have observed all lines of CH$_3$OH reported in Table~\ref{table:parameters}, with the exception of the 1$_{0,1}$-0$_{0,0}$ ($A^+$) transition, and the 1-0 transition of C$^{18}$O towards nine positions in B213-C5. The observed lines have been detected with a signal to noise larger than 3 towards all positions. 
The spectra of the 2$_{1,2}$-1$_{1,1}$ ($E_2$) transition of methanol overlaid with the $J$ = 1-0 transition of C$^{18}$O are shown in Figure~\ref{fig:B213-C5-all}.
The velocity structure is rather complex, with the methanol line tracing a portion of a lower velocity wing in the C$^{18}$O line towards the positions 1, 2 3, and 8. Two main velocity components are visible in both lines towards the positions 4, 5, and 9. While a single velocity component in both lines is present at the positions 6 and 7. \\
The upper panel in Figure~\ref{fig:B213-C5-results} shows the H$_2$ column density map of B213-C5 as well as the offsets of our observations. The observed stripe is centred at a dense peak towards the West of the map, labeled as offset 1, and also passes through an another dense clump towards the East (offsets 6 and 7).
The column densities and column density ratio plots in the lower panel of Figure~\ref{fig:B213-C5-results} show a very clear behaviour: the C$^{18}$O column density decreases when moving from East to the West in the observed offsets, and both the methanol column density and the column density ratio have a very well defined peak at the dust peak of B213-C5 (offset 1).

\subsection{B213 C6}
\begin{figure*}
\centering
 \includegraphics [width=0.8\textwidth]{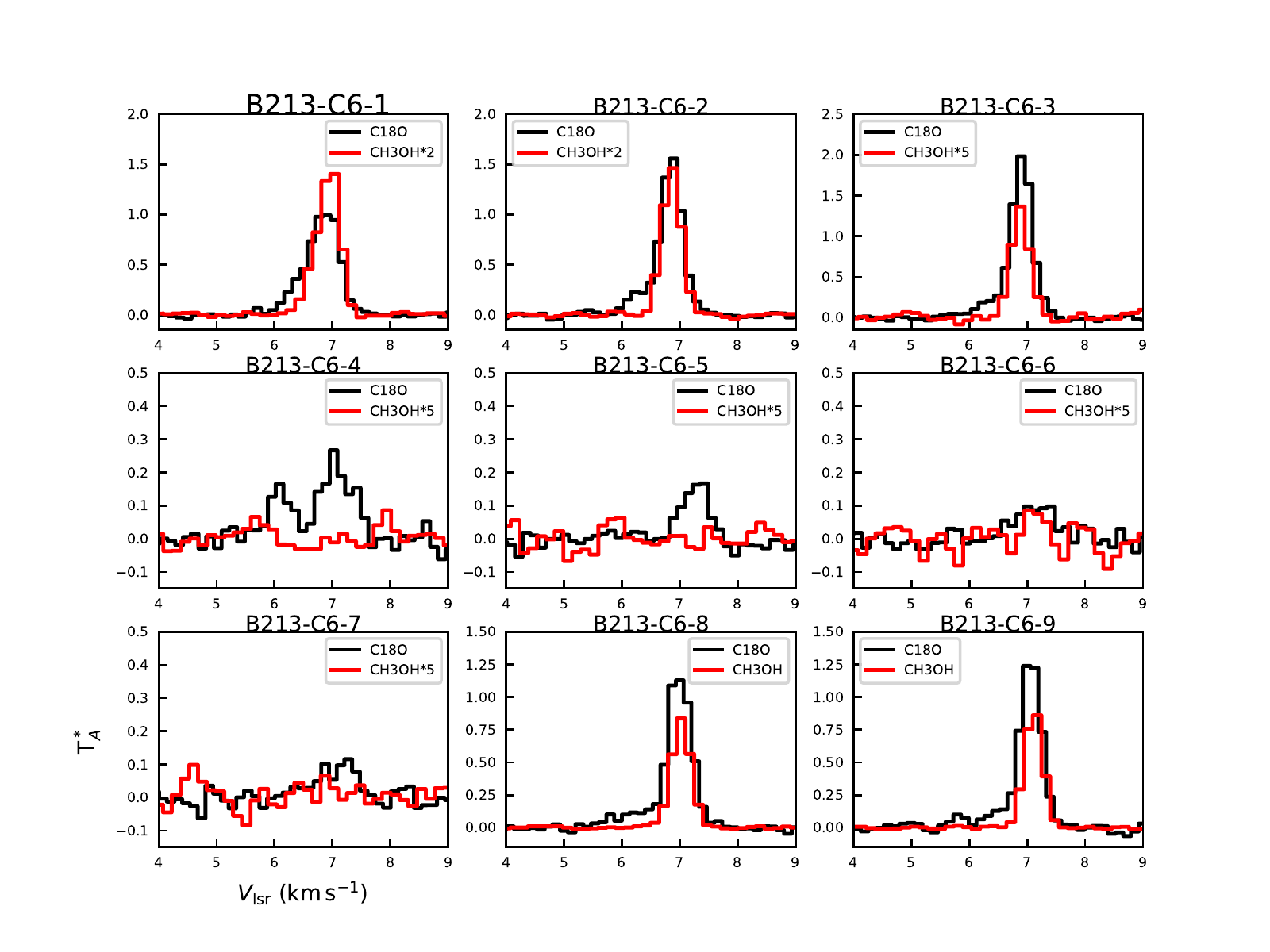}
 \caption{Spectra of the 2$_{1,2}$-1$_{1,1}$ ($E_2$) transition of methanol overlaid with the $J$ = 1-0 transition of C$^{18}$O in B213-C6
}
  \label{fig:B213-C6-all}
\end{figure*}

We have observed all lines of CH$_3$OH reported in Table~\ref{table:parameters}, with the exception of the 1$_{0,1}$-0$_{0,0}$ ($A^+$) transition, and the 1-0 transition of C$^{18}$O towards nine positions in B213-C6. The observed lines have been detected with a signal to noise larger than 3 towards positions 1, 2, 3, 8 and 9. 
The spectra of the 2$_{1,2}$-1$_{1,1}$ ($E_2$) transition of methanol overlaid with the $J$ = 1-0 transition of C$^{18}$O are shown in Figure~\ref{fig:B213-C6-all}.
In this case, the velocity structure is quite simple compared to the others cores in B213, with only one prominent velocity component. Both v$_{LSR}$ and line widths match very well for C$^{18}$O and CH$_3$OH.\\
The H$_2$ column density map of B213-C6 is shown in the upper panel of Figure~\ref{fig:B213-C6-results} with all the offsets observed within B213-C2 in the GEMS large project, however methanol was only detected towards the densest offsets (1,2, 3, 8 and 9).
The column densities and column density ratio plots are shown in the lower panel: the C$^{18}$O column density decreases towards the denser offsets, while  both the methanol column density and the column density ratio increase towards the dust peak of B213-C6.

\subsection{B213 C7}
\begin{figure*}
\centering
 \includegraphics [width=0.8\textwidth]{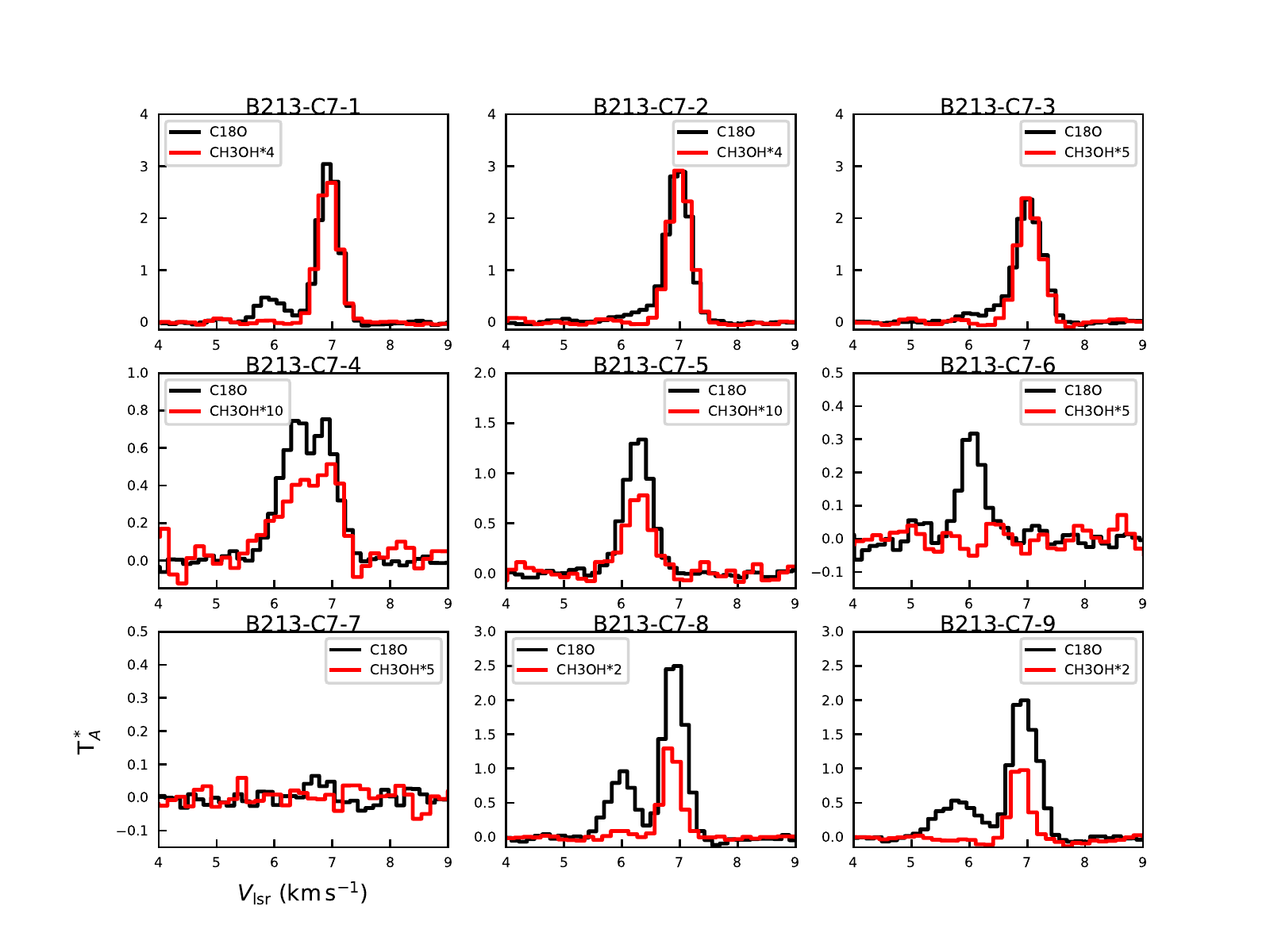}
 \caption{Spectra of the 2$_{1,2}$-1$_{1,1}$ ($E_2$) transition of methanol overlaid with the $J$ = 1-0 transition of C$^{18}$O in B213-C7
}
  \label{fig:B213-C7-all}
\end{figure*}

We have observed all lines of CH$_3$OH reported in Table~\ref{table:parameters}, with the exception of the 1$_{0,1}$-0$_{0,0}$ ($A^+$) transition, and the 1-0 transition of C$^{18}$O towards nine positions in B213-C7. The observed lines have been detected with a signal to noise larger than 3 towards positions 1, 2, 3, 4, 5, 8 and 9. Only C$^{18}$O has been observed towards the position 6, and neither C$^{18}$O nor CH$_3$OH have been observed towards position 7.
The spectra of the 2$_{1,2}$-1$_{1,1}$ ($E_2$) transition of methanol overlaid with the $J$ = 1-0 transition of C$^{18}$O are shown in Figure~\ref{fig:B213-C7-all}.
Different velocity components are present in the different positions. Two velocity components (at $\sim$6 and 7 km/s) are visible in positions 1, 8 and 9 for C$^{18}$O, with only the brightest being observed also in methanol. Only the component at 7 km/s is visible in both molecules towards the positions 2 and 3. Positions 5 and 6 show only one velocity component at $\sim$6 km/s. In positions 4 two velocity components are also present, but with a smaller separation in velocity with respect to positions 1, 8 and 9.\\
The H$_2$ column density map of B213-C7 is shown in the upper panel of Figure~\ref{fig:B213-C7-results} with all the offsets observed within B213-C2 in the GEMS large project, however methanol was only detected towards the densest offsets (1,2, 3, 8 and 9). The column densities and column density ratio plots are shown in the lower panel of Figure~\ref{fig:B213-C7-results}.
The column density of C$^{18}$O has a peak at the offset 2 and then decreases towards the West. The methanol column density, as well as the column density ratio have an oscillating pattern.

\subsection{B213 C10}
\begin{figure*}
\centering
 \includegraphics [width=0.8\textwidth]{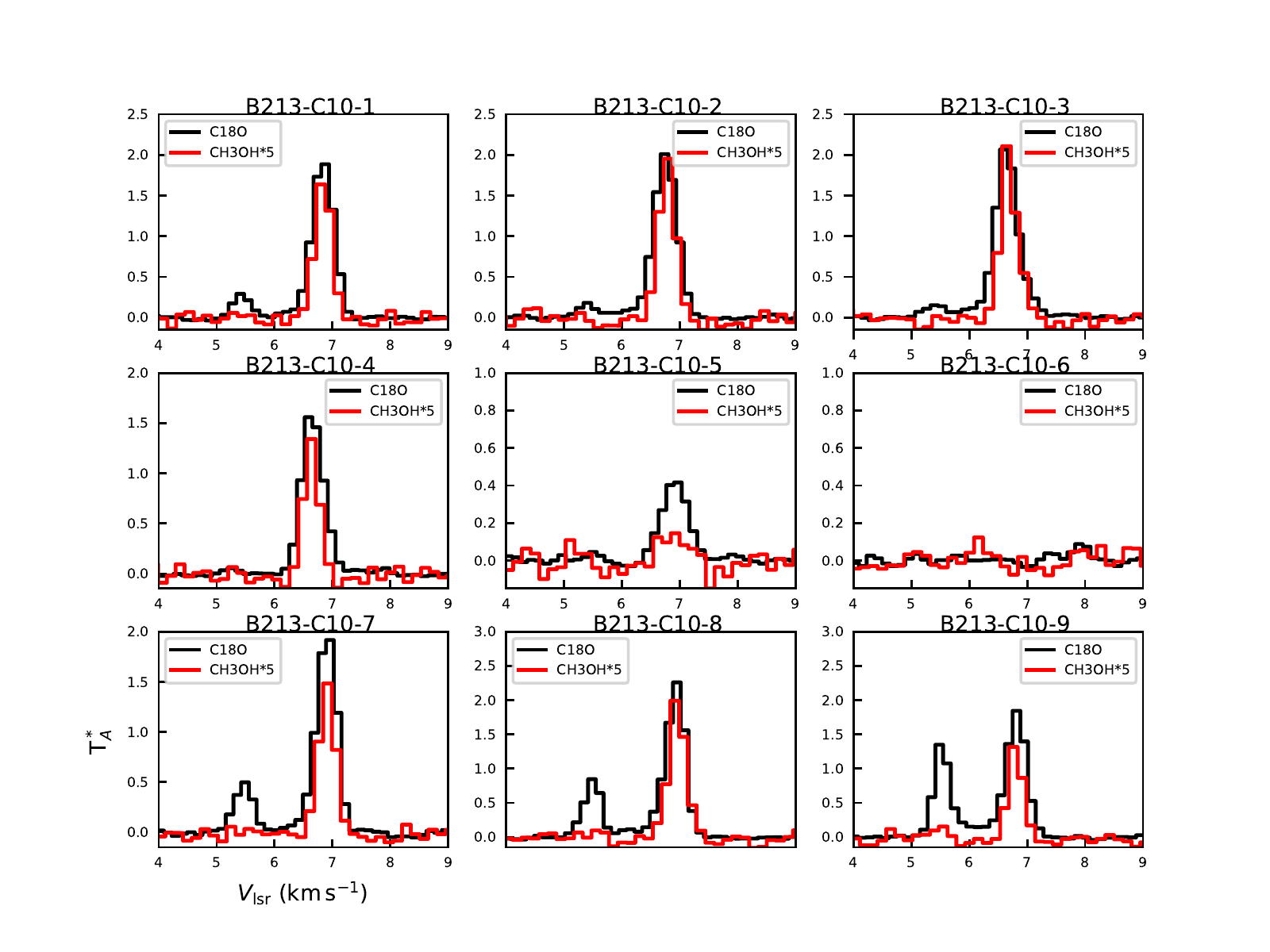}
 \caption{Spectra of the 2$_{1,2}$-1$_{1,1}$ ($E_2$) transition of methanol overlaid with the $J$ = 1-0 transition of C$^{18}$O in B213-C10
}
  \label{fig:B213-C10-all}
\end{figure*}

We have observed all lines of CH$_3$OH reported in Table~\ref{table:parameters}, with the exception of the 1$_{0,1}$-0$_{0,0}$ ($A^+$) transition, and the 1-0 transition of C$^{18}$O towards nine positions in B213-C10. The observed lines have been detected with a signal to noise larger than 3 towards all positions except positions 5 and 6. Only C$^{18}$O has been observed towards the position 5.
The spectra of the 2$_{1,2}$-1$_{1,1}$ ($E_2$) transition of methanol overlaid with the $J$ = 1-0 transition of C$^{18}$O are shown in Figure~\ref{fig:B213-C10-all}.
Two velocity components at $\sim$5.5 and $\sim$7 km/s are present for C$^{18}$O towards the positions 1, 2, 3, 7, 8, and 9. Only the brightest velocity component is observed in methanol.\\
The upper panel in Figure~\ref{fig:B213-C10-results} shows the H$_2$ column density map of B213-C10 as well as the offsets of our observations. Towards B213-C10, methanol was observed in all offset with the exception of offsets 5 and 6.
The column densities and column density ratio plots are shown in the lower panel of Figure~\ref{fig:B213-C10-results}. The C$^{18}$O column density shows two peaks, one around offsets 2 and 3, and another at offset 8. The methanol column density, as well as the column density ratio instead have a clear peak only at offsets 2 and 3, towards a tail of in the H$_2$ column density map.

\subsection{B213 C12}
\begin{figure*}
\centering
 \includegraphics [width=0.8\textwidth]{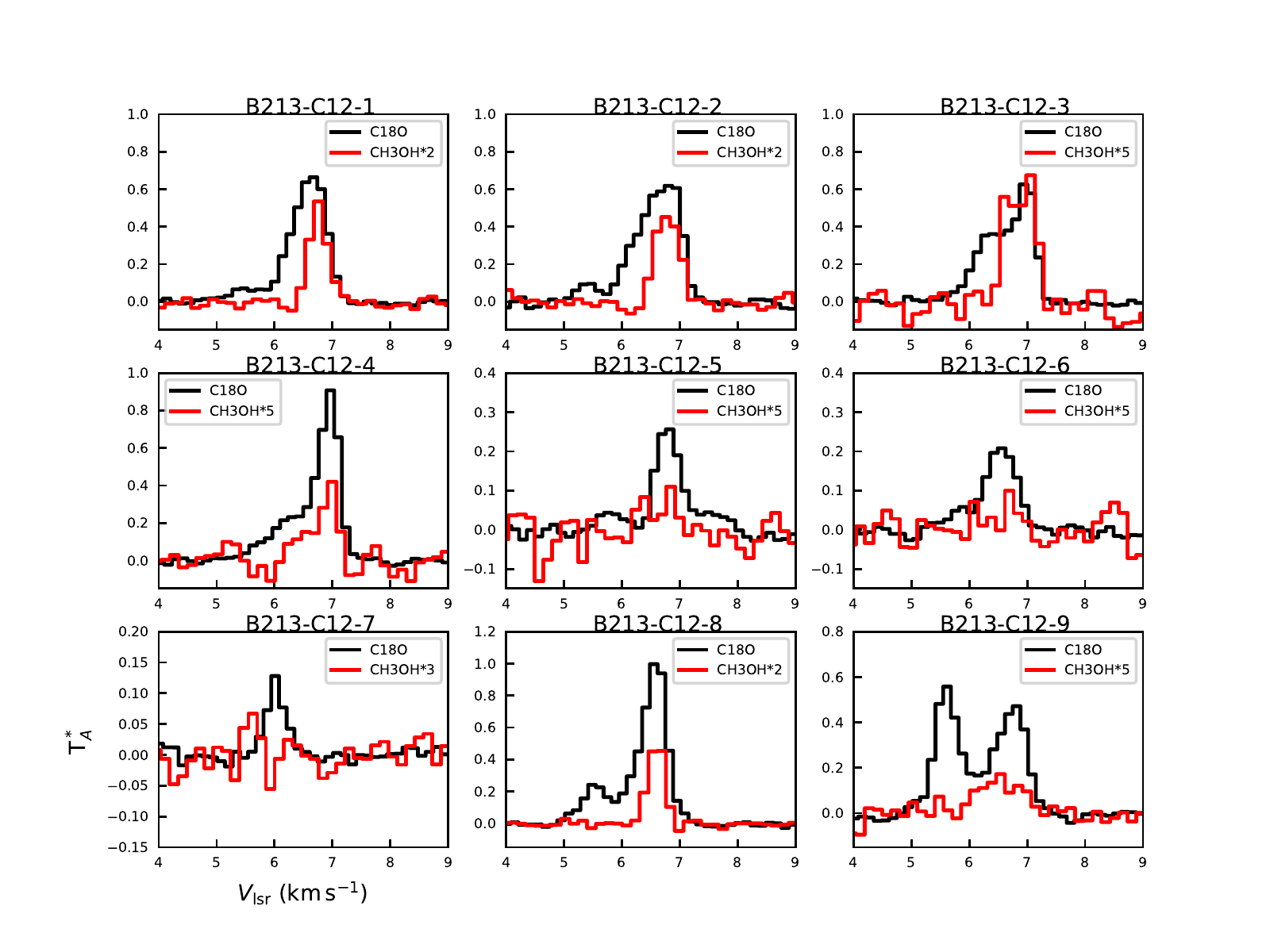}
 \caption{Spectra of the 2$_{1,2}$-1$_{1,1}$ ($E_2$) transition of methanol overlaid with the $J$ = 1-0 transition of C$^{18}$O in B213-C12
}
  \label{fig:B213-C12-all}
\end{figure*}

We have observed all lines of CH$_3$OH reported in Table~\ref{table:parameters}, with the exception of the 1$_{0,1}$-0$_{0,0}$ ($A^+$) transition, and the 1-0 transition of C$^{18}$O towards nine positions in B213-C12. The observed lines have been detected with a signal to noise larger than 3 towards positions 1, 2, 3, 4, 8 and 9.
The spectra of the 2$_{1,2}$-1$_{1,1}$ ($E_2$) transition of methanol overlaid with the $J$ = 1-0 transition of C$^{18}$O are shown in Figure~\ref{fig:B213-C12-all}.
The C$^{18}$O lines show a complex velocity structure, with multiple velocity components unresolved towards the positions 1, 2, 3, and 4, and two main velocity components towards positions 8 and 9. The methanol line shows only one velocity components at $\sim$7 km/s. \\
The H$_2$ column density map of B213-C12 is shown in the upper panel of Figure~\ref{fig:B213-C12-results} with all the offsets observed within B213-C12 in the GEMS large project, however methanol was only detected towards the densest offsets (1,2, 3 and 8).
Both the variation of the CH$_3$OH and C$^{18}$O column density ratio, as well as the single column densities N(CH$_3$OH) and N(C$^{18}$O) are shown in the lower panel. The column densities of CH$_3$OH and C$^{18}$O show different behaviours: C$^{18}$O increases from offsets 3 to offset 8, passing through the dust peak at offset 1, while methanol shows the opposite behaviour. The column density shows the same trend as the methanol column density.

\subsection{B213 C16}
\begin{figure*}
\centering
 \includegraphics [width=0.8\textwidth]{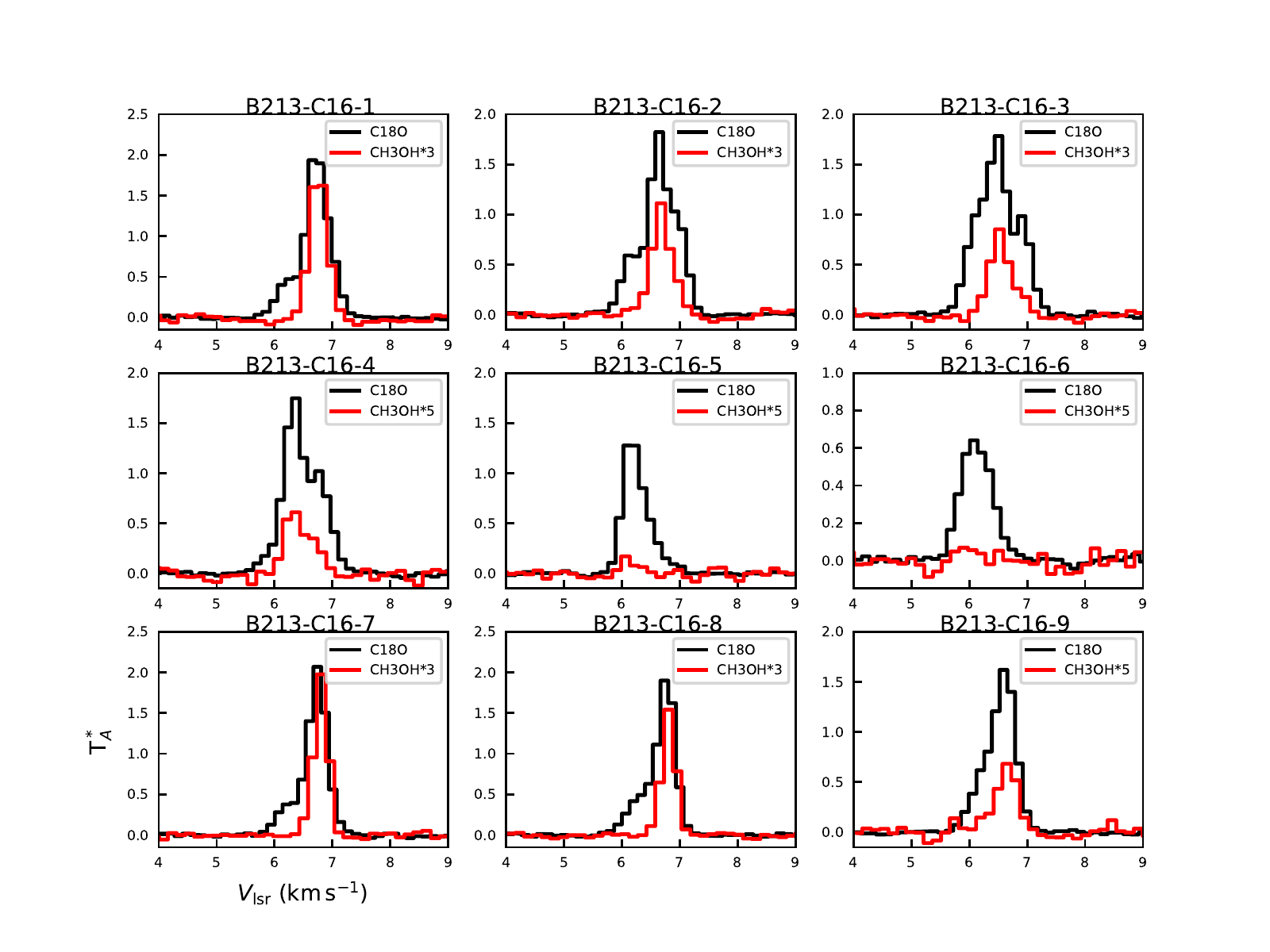}
 \caption{Spectra of the 2$_{1,2}$-1$_{1,1}$ ($E_2$) transition of methanol overlaid with the $J$ = 1-0 transition of C$^{18}$O in B213-C16
}
  \label{fig:B213-C16-all}
\end{figure*}

We have observed all lines of CH$_3$OH reported in Table~\ref{table:parameters}, with the exception of the 1$_{0,1}$-0$_{0,0}$ ($A^+$) transition, and the 1-0 transition of C$^{18}$O towards nine positions in B213-C16. The observed lines have been detected with a signal to noise larger than 3 towards all positions except positions 5 and 6, where only the line of C$^{18}$O have been observed.
The spectra of the 2$_{1,2}$-1$_{1,1}$ ($E_2$) transition of methanol overlaid with the $J$ = 1-0 transition of C$^{18}$O are shown in Figure~\ref{fig:B213-C16-all}.
The C$^{18}$O lines show wings in most positions, hinting at the presence of more than one velocity component on the line of sight. The methanol lines do not show wings, with the exception of the position 4.\\
The upper panel in Figure~\ref{fig:B213-C16-results} shows the H$_2$ column density map of B213-C16 as well as the offsets of our observations. Towards B213-C16, methanol was observed in all offset with the exception of offsets 5 and 6.
The column densities and column density ratio plots are shown in the lower panel of Figure~\ref{fig:B213-C16-results}. The C$^{18}$O column density shows a very shallow peak around offsets 2 and 3. Both the  methanol column density, as well as the column density ratio show a steep increase moving from offset 2 towards offset 4 (opposite with respect to the direction of the dust peak), and have a shallower at offsets 1 and 7.

\subsection{B213 C17}
\begin{figure*}
\centering
 \includegraphics [width=0.8\textwidth]{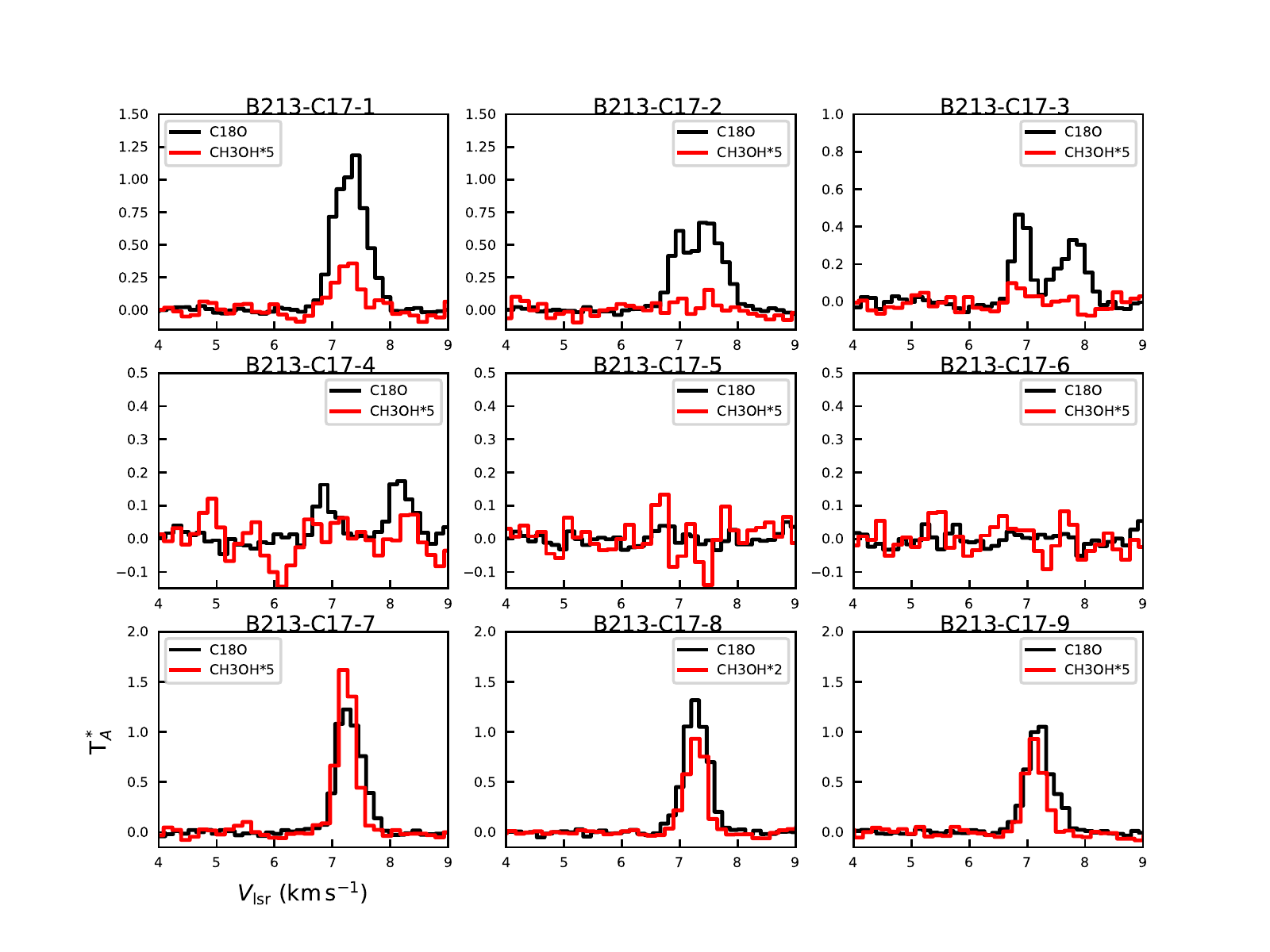}
 \caption{Spectra of the 2$_{1,2}$-1$_{1,1}$ ($E_2$) transition of methanol overlaid with the $J$ = 1-0 transition of C$^{18}$O in B213-C17
}
  \label{fig:B213-C17-all}
\end{figure*}

\begin{figure*}
\centering
 \includegraphics [width=1.0\textwidth]{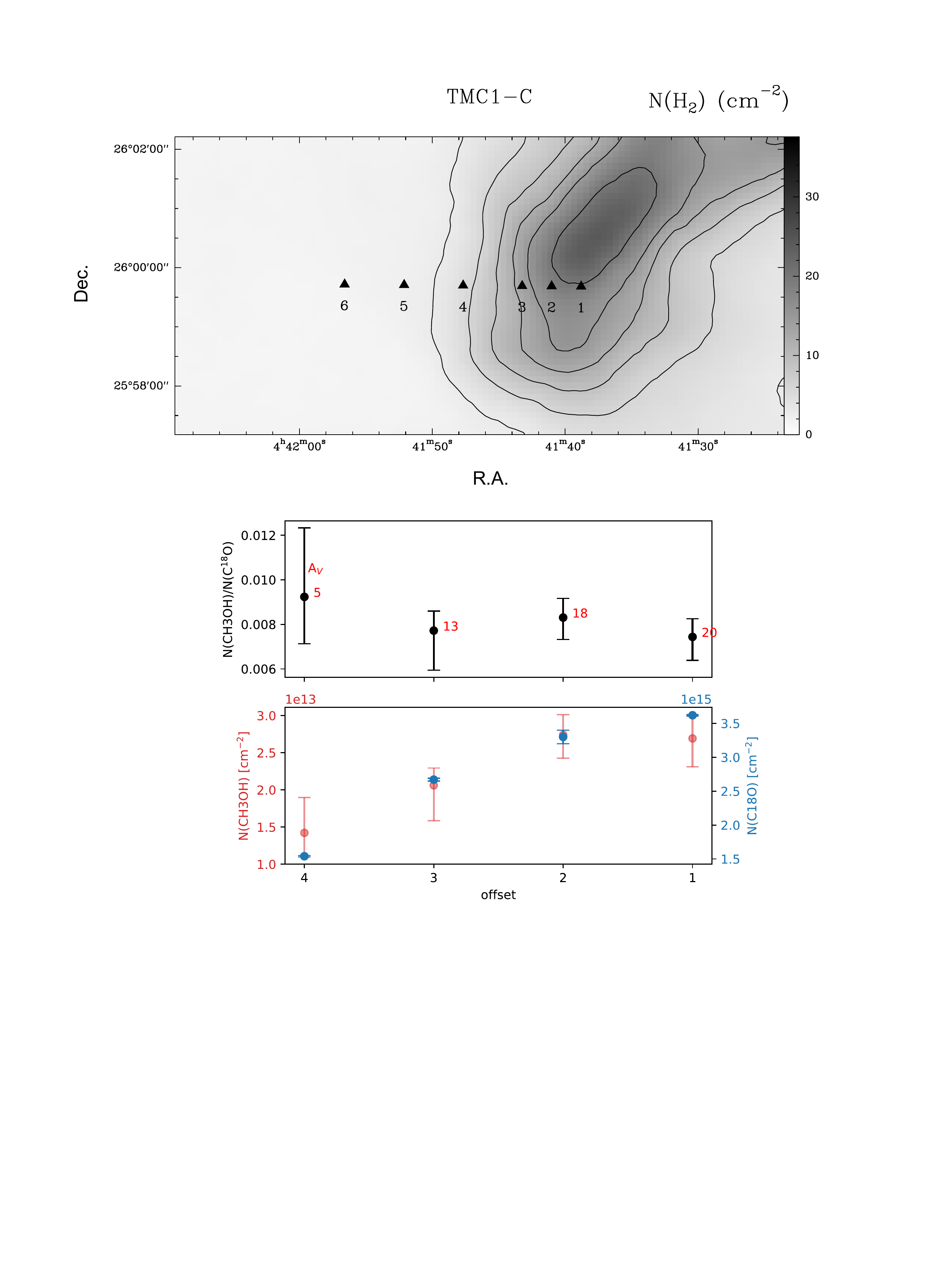}
 \caption{Upper panel: H$_2$ column density of TMC-1 C derived from $Herschel$ and $Planck$ data \citep{rodriguez21}. The triangles mark the positions observed with the GEMS large project. Lower panel: CH$_3$OH and C$^{18}$O column densities, and column density ratios computed in the different offsets in TMC-1 C. In the plot of the CH$_3$OH and C$^{18}$O column density ratio, the A$_V$ in each offset is marked in red.
}
  \label{fig:TMc1C-results}
\end{figure*}

\begin{figure*}
\centering
 \includegraphics [width=1.0\textwidth]{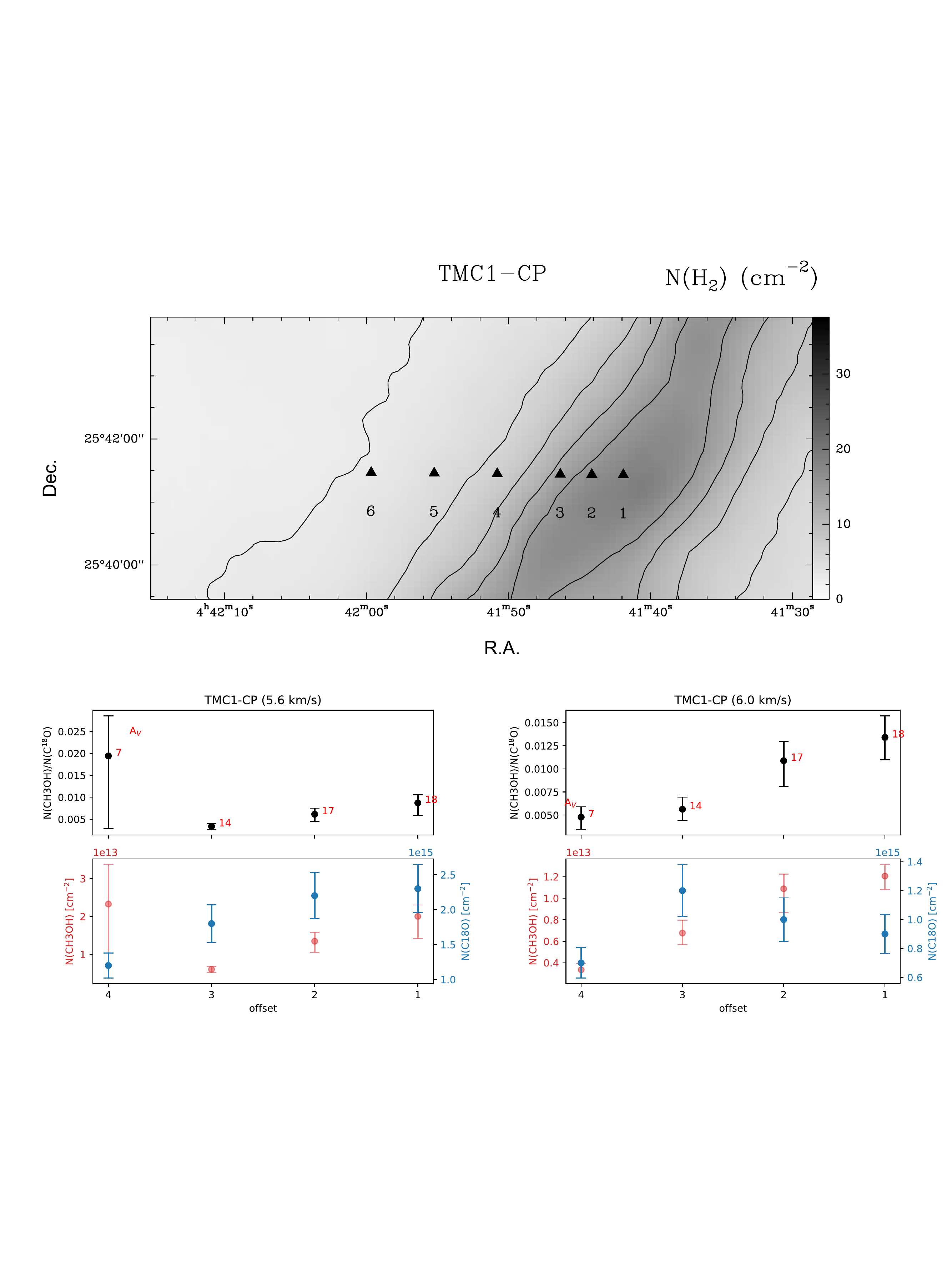}
 \caption{Upper panel: H$_2$ column density of TMC-1 CP derived from $Herschel$ and $Planck$ data \citep{rodriguez21}. The triangles mark the positions observed with the GEMS large project. Lower panel: CH$_3$OH and C$^{18}$O column densities, and column density ratios computed in the different offsets in TMC-1 CP in the two velocity components observed for methanol. In the plot of the CH$_3$OH and C$^{18}$O column density ratio, the A$_V$ in each offset is marked in red.
}
  \label{fig:TMc1CP-results}
\end{figure*}

\begin{figure*}
\centering
 \includegraphics [width=1.0\textwidth]{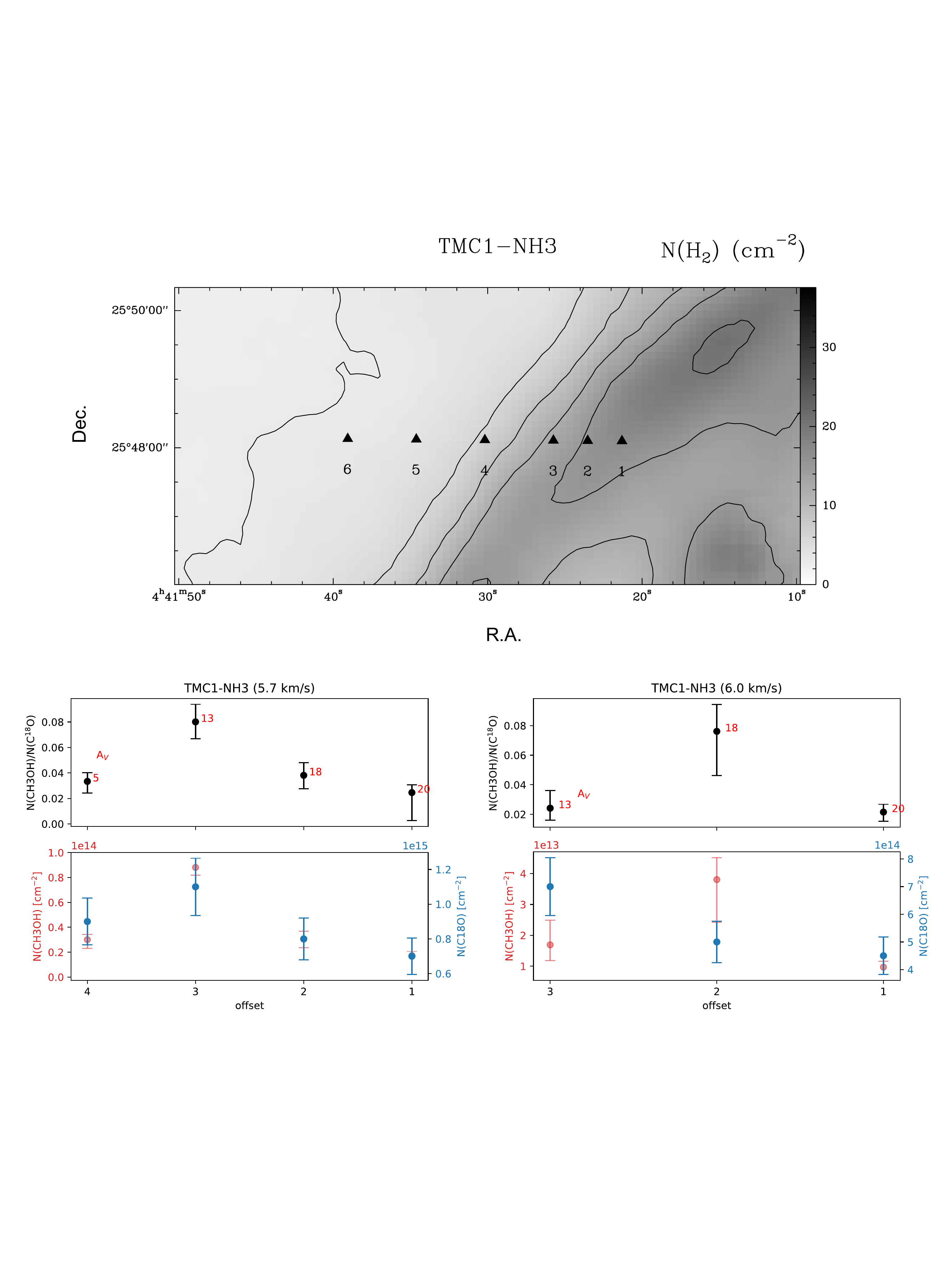}
 \caption{Upper panel: H$_2$ column density of TMC-1 NH$_3$ derived from $Herschel$ and $Planck$ data \citep{rodriguez21}. The triangles mark the positions observed with the GEMS large project. Lower panel: CH$_3$OH and C$^{18}$O column densities, and column density ratios computed in the different offsets in TMC-1 NH$_3$ in the two velocity components observed for methanol. In the plot of the CH$_3$OH and C$^{18}$O column density ratio, the A$_V$ in each offset is marked in red.
}
  \label{fig:TMc1NH3-results}
\end{figure*}

\begin{figure*}
\centering
 \includegraphics [width=1.0\textwidth]{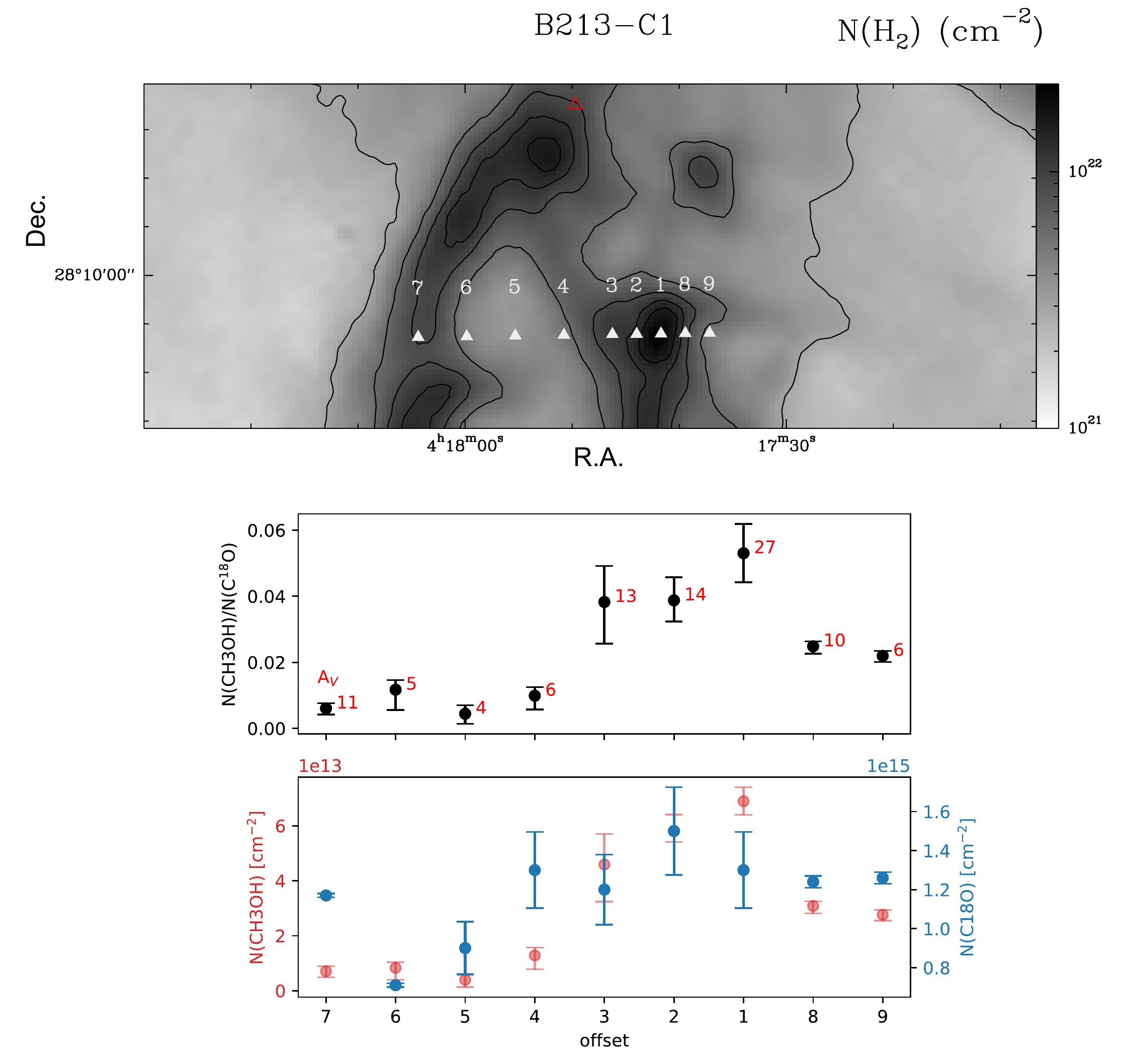}
 \caption{Upper panel: H$_2$ column density of B213-C1 derived from $Herschel$ and $Planck$ data \citep{palmeirim13}. The triangles mark the positions observed with the GEMS large project. The red triangle shows the position of a Class II/III protostellar core \citep{rebull10}.Lower panel: CH$_3$OH and C$^{18}$O column densities, and column density ratios computed in the different offsets in B213-C1. In the plot of the CH$_3$OH and C$^{18}$O column density ratio, the A$_V$ in each offset is marked in red.
}
  \label{fig:B213-C1-results}
\end{figure*}

\begin{figure*}
\centering
 \includegraphics [width=1.0\textwidth]{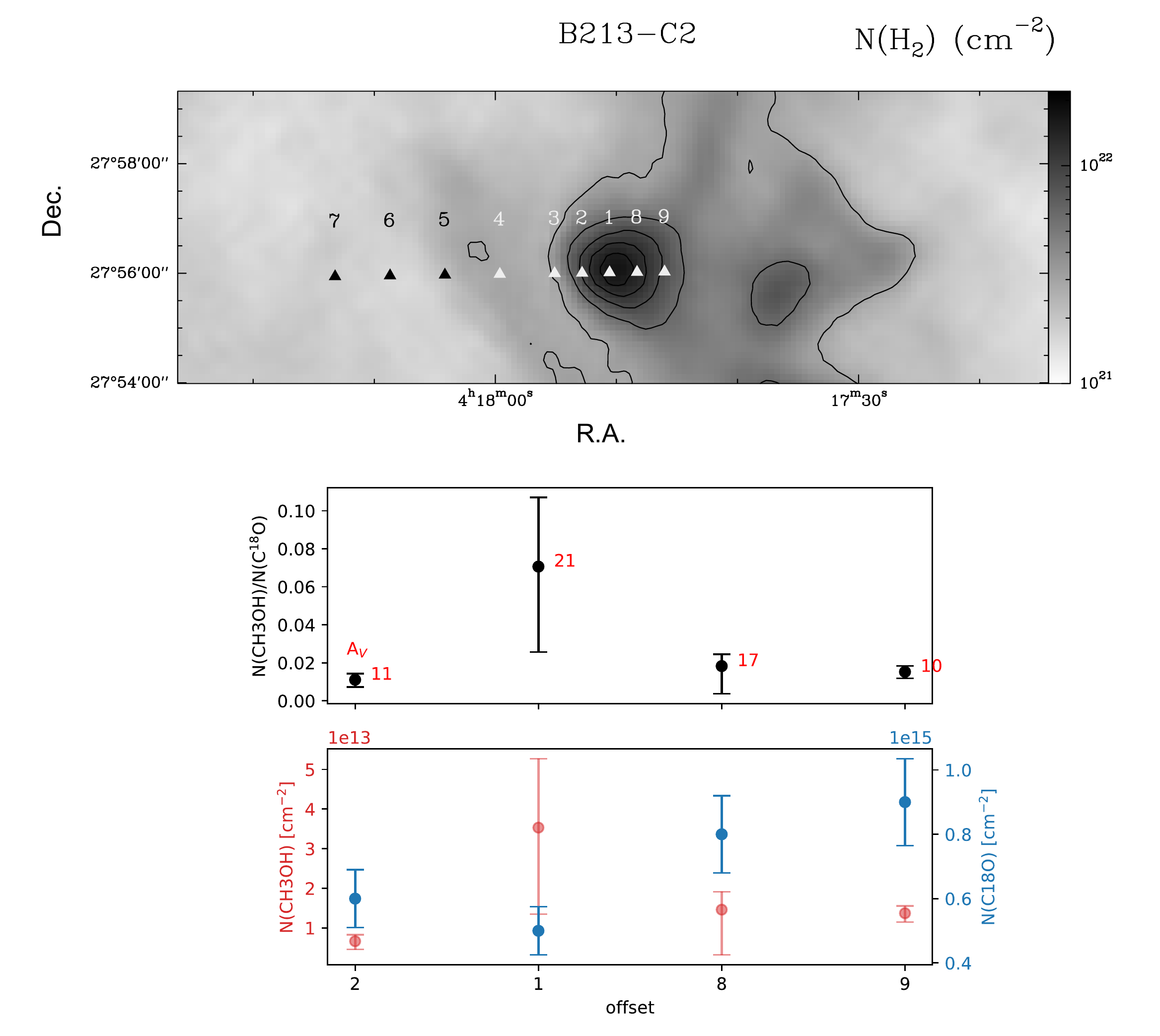}
 \caption{Upper panel: H$_2$ column density of B213-C2 derived from $Herschel$ and $Planck$ data \citep{palmeirim13}. The triangles mark the positions observed with the GEMS large project. Lower panel: CH$_3$OH and C$^{18}$O column densities, and column density ratios computed in the different offsets in B213-C2. In the plot of the CH$_3$OH and C$^{18}$O column density ratio, the A$_V$ in each offset is marked in red.
}
  \label{fig:B213-C2-results}
\end{figure*}

\begin{figure*}
\centering
 \includegraphics [width=1.0\textwidth]{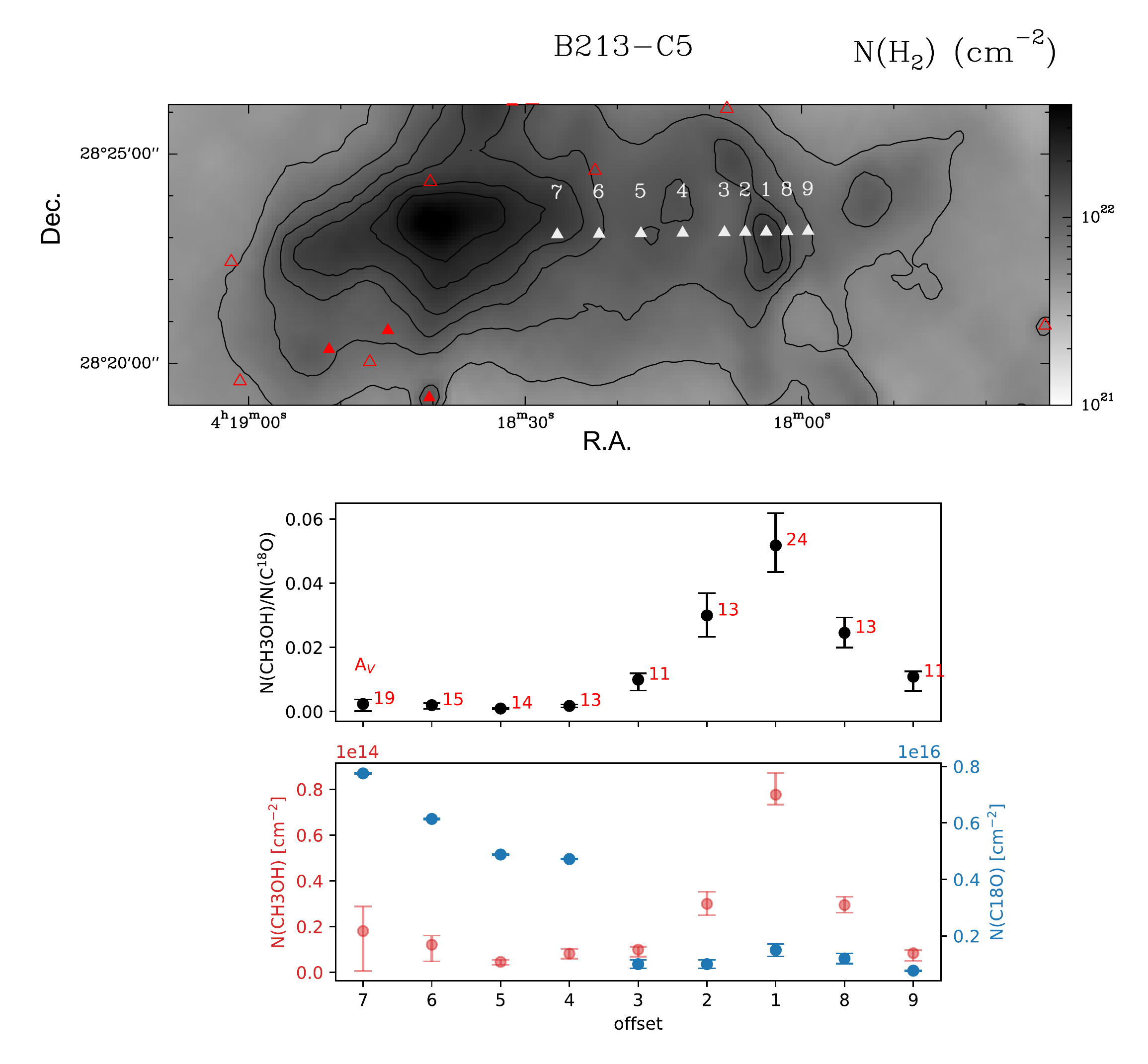}
 \caption{Upper panel: H$_2$ column density of B213-C5 derived from $Herschel$ and $Planck$ data \citep{palmeirim13}. The triangles mark the positions observed with the GEMS large project. The red triangles show the positions of Class I/flat (full triangle) or Class II/III (empty triangle) protostellar cores \citep{rebull10}. Lower panel: CH$_3$OH and C$^{18}$O column densities, and column density ratios computed in the different offsets in B213-C5. In the plot of the CH$_3$OH and C$^{18}$O column density ratio, the A$_V$ in each offset is marked in red.
}
  \label{fig:B213-C5-results}
\end{figure*}

\begin{figure*}
\centering
 \includegraphics [width=1.0\textwidth]{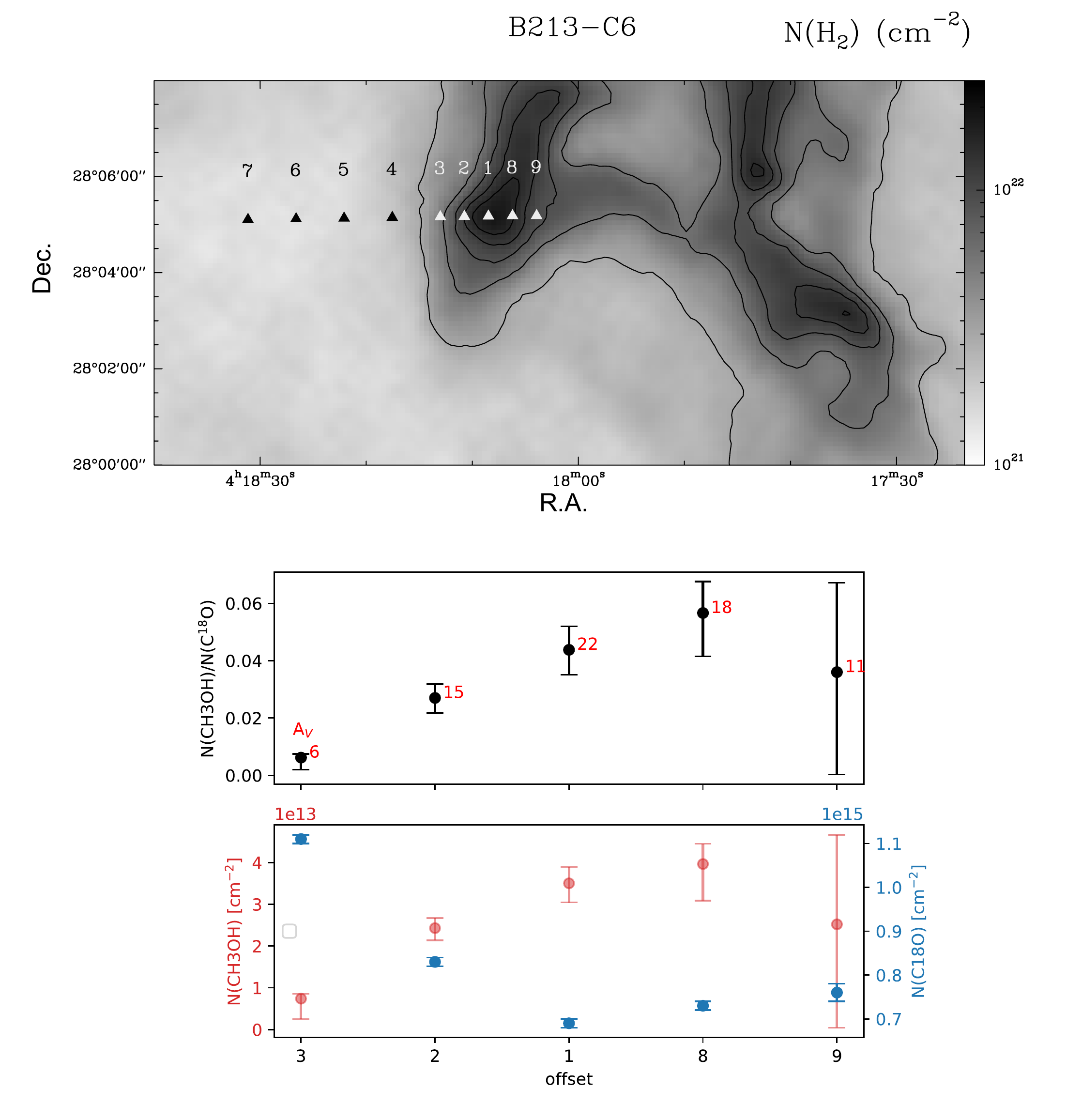}
 \caption{Upper panel: H$_2$ column density of B213-C6 derived from $Herschel$ and $Planck$ data \citep{palmeirim13}. The triangles mark the positions observed with the GEMS large project. Lower panel: CH$_3$OH and C$^{18}$O column densities, and column density ratios computed in the different offsets in B213-C6. In the plot of the CH$_3$OH and C$^{18}$O column density ratio, the A$_V$ in each offset is marked in red.
}
  \label{fig:B213-C6-results}
\end{figure*}

\begin{figure*}
\centering
 \includegraphics [width=1.0\textwidth]{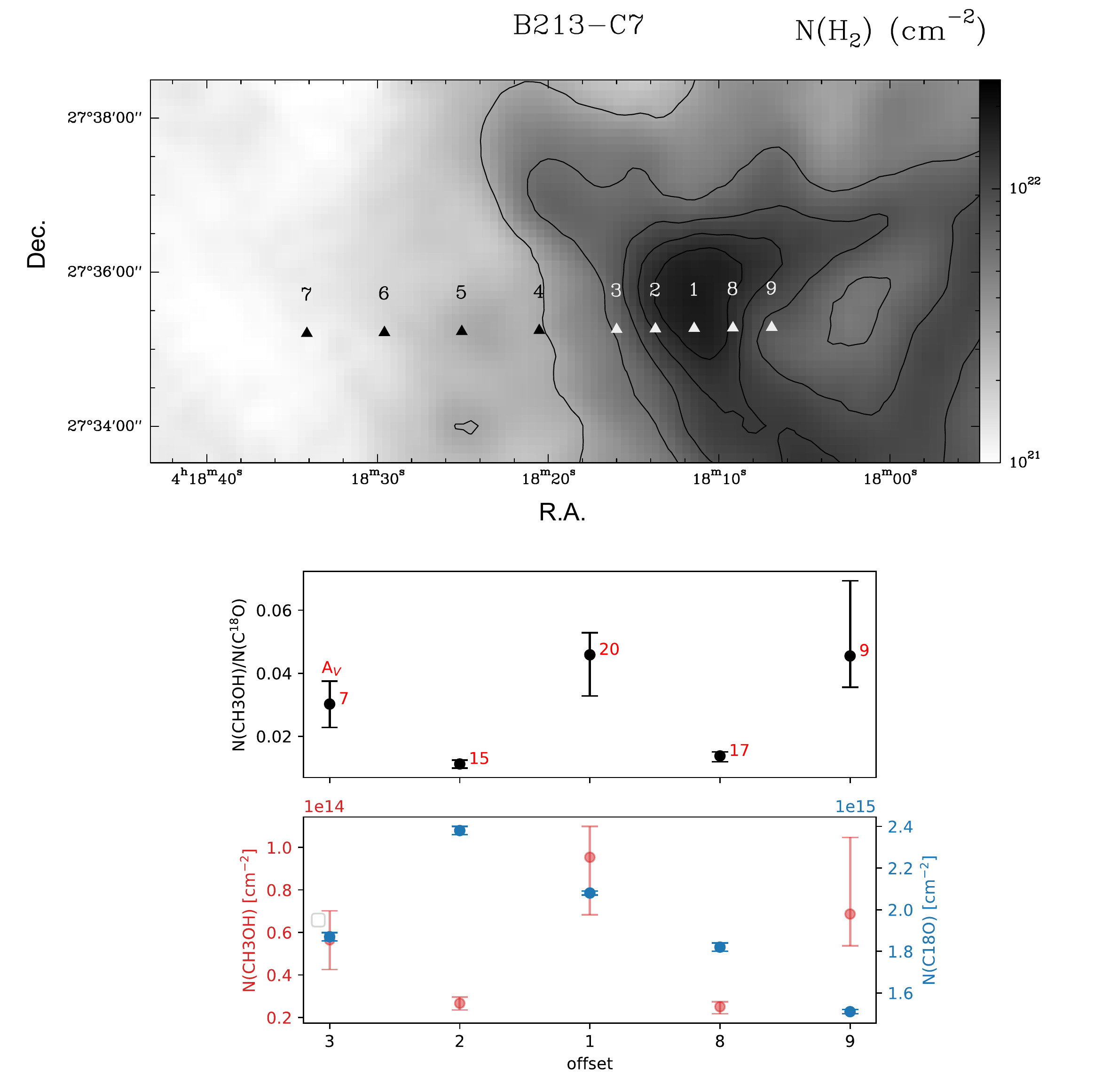}
 \caption{Upper panel: H$_2$ column density of B213-C7 derived from $Herschel$ and $Planck$ data \citep{palmeirim13}. The triangles mark the positions observed with the GEMS large project. Lower panel: CH$_3$OH and C$^{18}$O column densities, and column density ratios computed in the different offsets in B213-C7. In the plot of the CH$_3$OH and C$^{18}$O column density ratio, the A$_V$ in each offset is marked in red.
}
  \label{fig:B213-C7-results}
\end{figure*}

\begin{figure*}
\centering
 \includegraphics [width=0.8\textwidth]{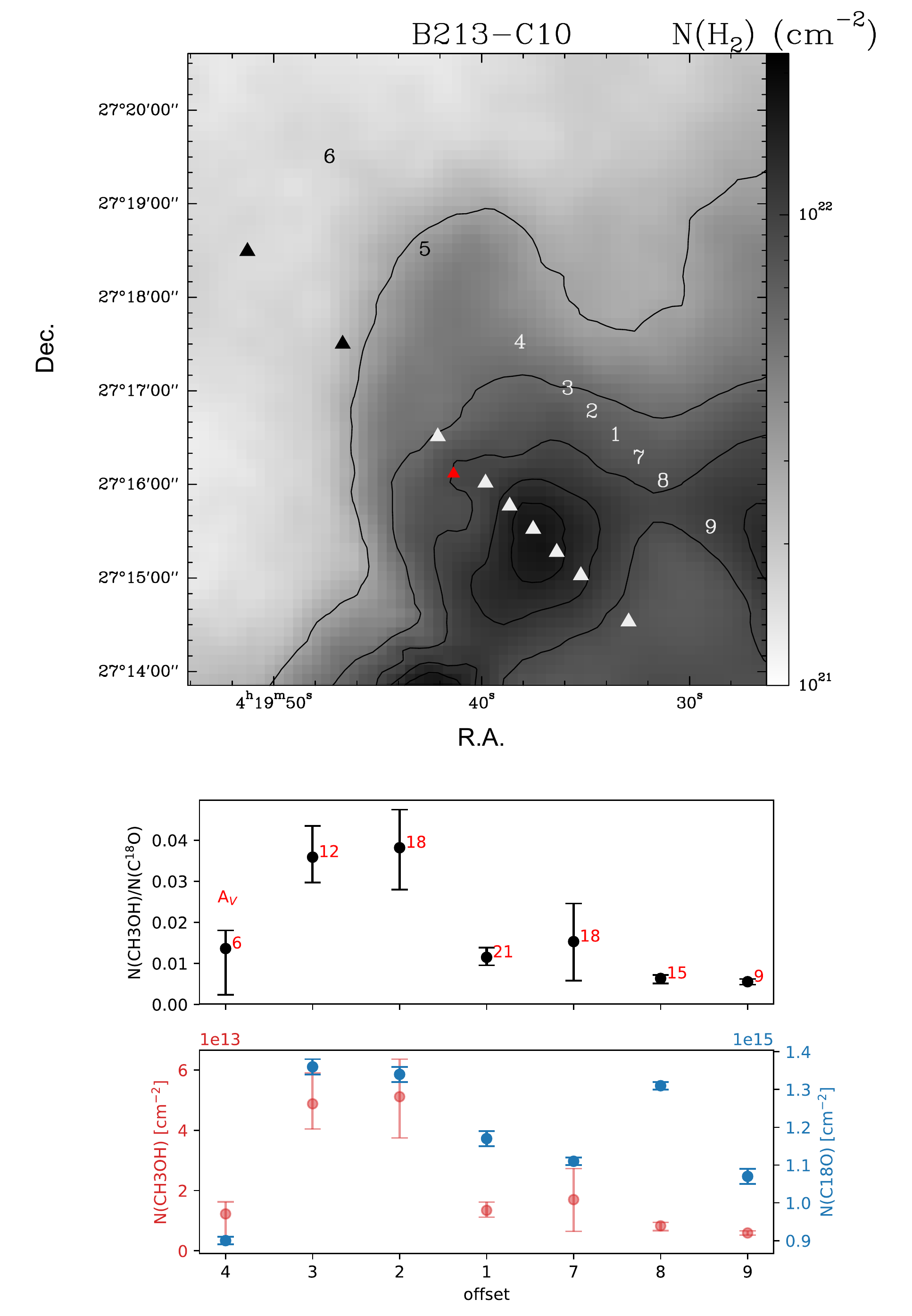}
 \caption{Upper panel: H$_2$ column density of B213-C10 derived from $Herschel$ and $Planck$ data \citep{palmeirim13}. The triangles mark the positions observed with the GEMS large project. The red triangle shows the position of a Class I/flat protostellar core \citep{rebull10}. Lower panel: CH$_3$OH and C$^{18}$O column densities, and column density ratios computed in the different offsets in B213-C10. In the plot of the CH$_3$OH and C$^{18}$O column density ratio, the A$_V$ in each offset is marked in red.
}
  \label{fig:B213-C10-results}
\end{figure*}

\begin{figure*}
\centering
 \includegraphics [width=0.8\textwidth]{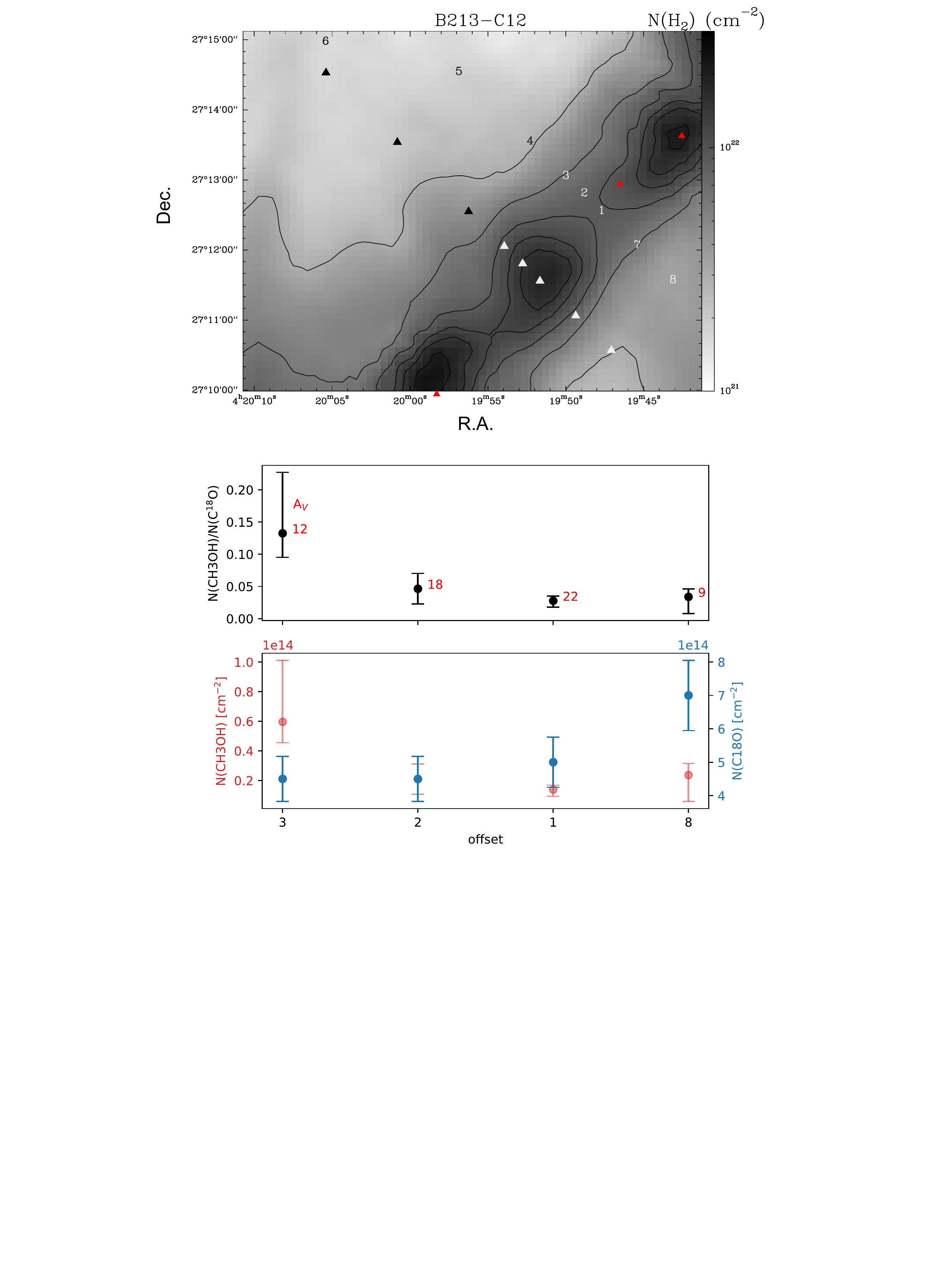}
 \caption{Upper panel: H$_2$ column density of B213-C12 derived from $Herschel$ and $Planck$ data \citep{palmeirim13}. The triangles mark the positions observed with the GEMS large project. The red triangles show the positions of Class I/flat protostellar cores \citep{rebull10}. Lower panel: CH$_3$OH and C$^{18}$O column densities, and column density ratios computed in the different offsets in B213-C12. In the plot of the CH$_3$OH and C$^{18}$O column density ratio, the A$_V$ in each offset is marked in red.
}
  \label{fig:B213-C12-results}
\end{figure*}

\begin{figure*}
\centering
 \includegraphics [width=0.8\textwidth]{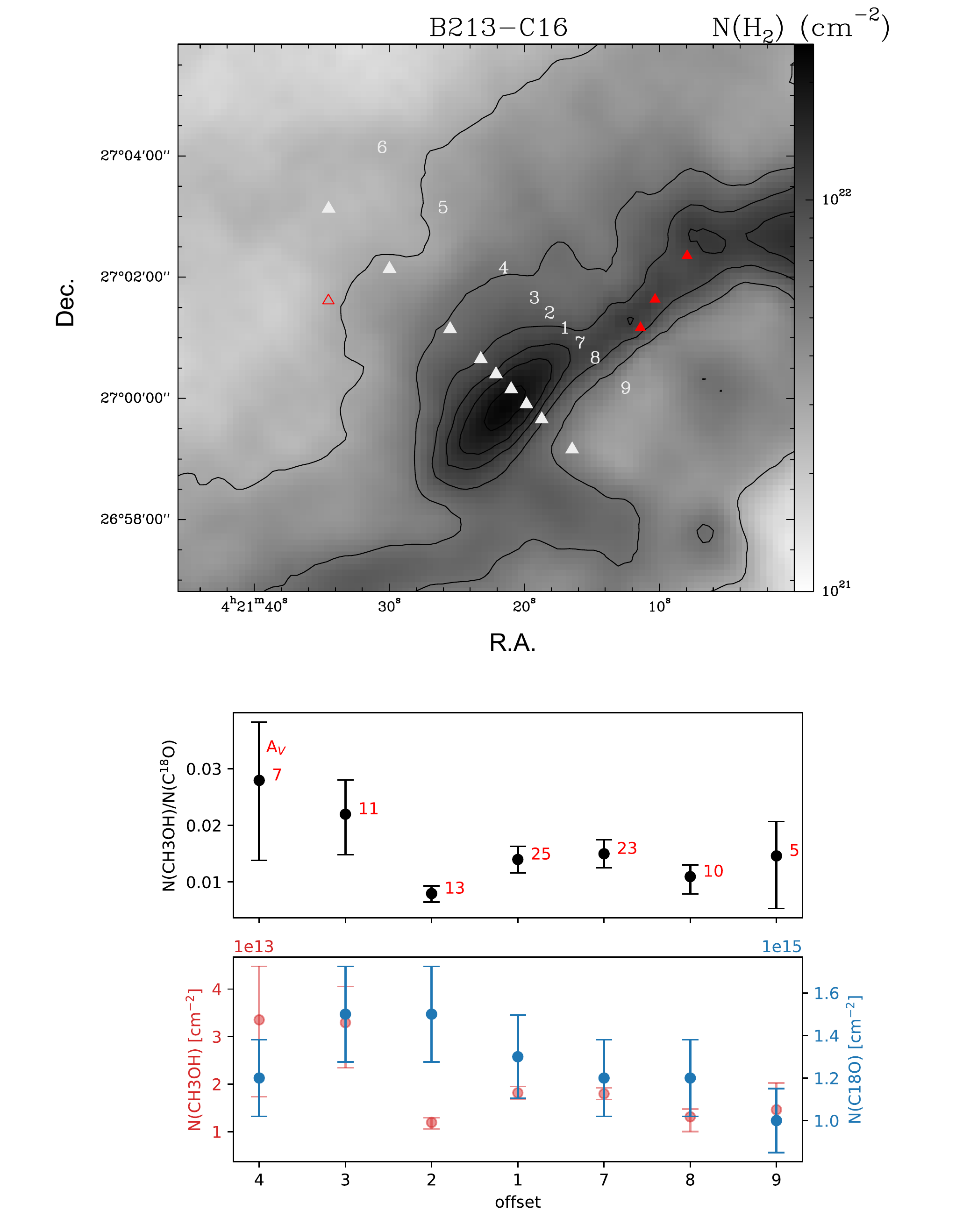}
 \caption{Upper panel: H$_2$ column density of B213-C16 derived from $Herschel$ and $Planck$ data \citep{palmeirim13}. The triangles mark the positions observed with the GEMS large project. The red triangles show the positions of Class I/flat (full triangle) or Class II/III (empty triangle) protostellar cores \citep{rebull10}. Lower panel: CH$_3$OH and C$^{18}$O column densities, and column density ratios computed in the different offsets in B213-C16. In the plot of the CH$_3$OH and C$^{18}$O column density ratio, the A$_V$ in each offset is marked in red.
}
  \label{fig:B213-C16-results}
\end{figure*}

\begin{figure*}
\centering
 \includegraphics [width=0.8\textwidth]{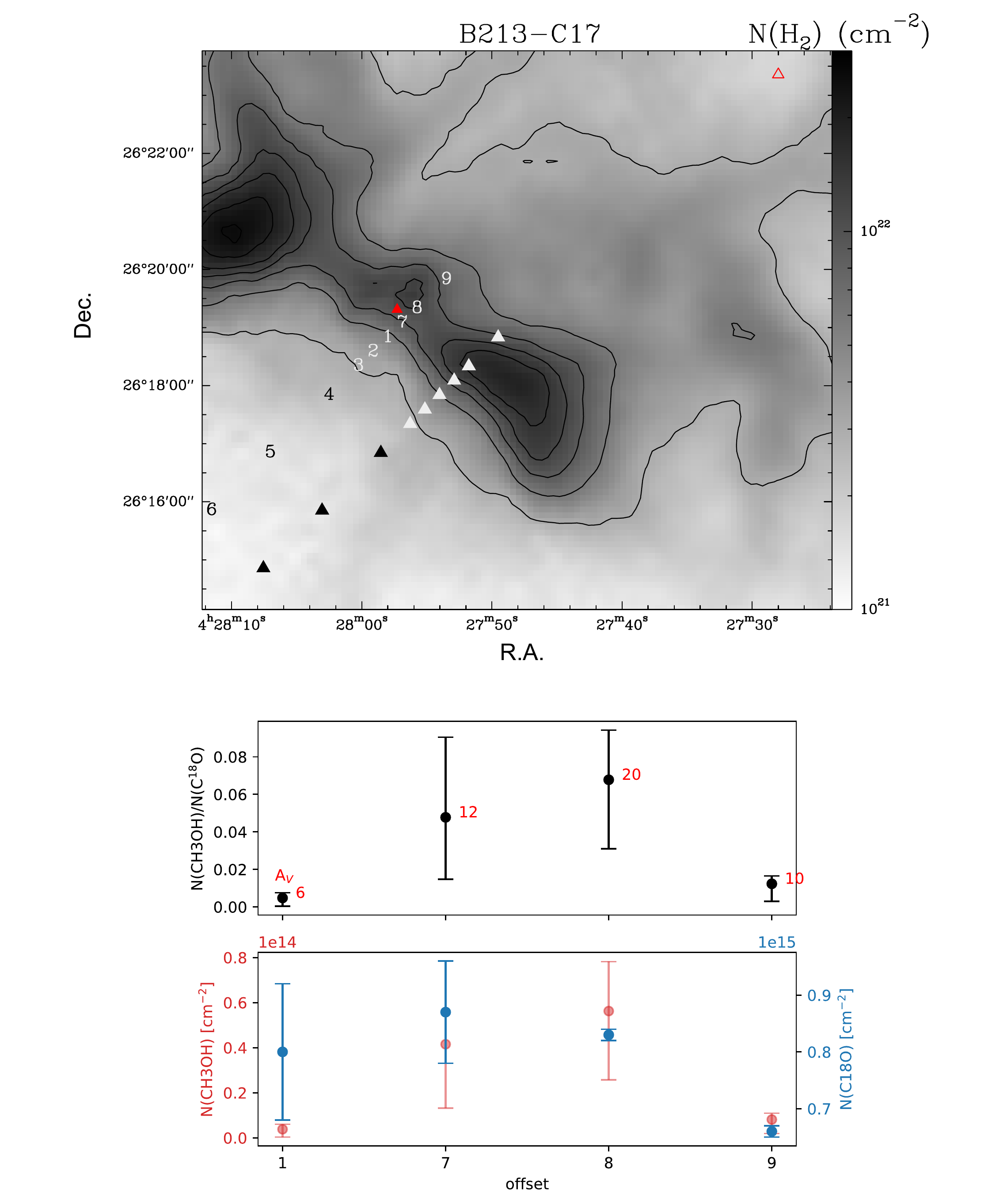}
 \caption{Upper panel: H$_2$ column density of B213-C17 derived from $Herschel$ and $Planck$ data \citep{palmeirim13}. The triangles mark the positions observed with the GEMS large project. The red triangle shows the position of a Class I/flat protostellar core \citep{rebull10}. Lower panel: CH$_3$OH and C$^{18}$O column densities, and column density ratios computed in the different offsets in B213-C17. In the plot of the CH$_3$OH and C$^{18}$O column density ratio, the A$_V$ in each offset is marked in red. 
}
  \label{fig:B213-C17-results}
\end{figure*}

We have observed all lines of CH$_3$OH reported in Table~\ref{table:parameters}, with the exception of the 1$_{0,1}$-0$_{0,0}$ ($A^+$) transition, and the 1-0 transition of C$^{18}$O towards nine positions in B213-C17. The observed lines have been detected with a signal to noise larger than 3 towards positions 1, 2, 3, 7, 8, and 9 for C$^{18}$O, and positions 1, 7, 8, and 9 for methanol.
The spectra of the 2$_{1,2}$-1$_{1,1}$ ($E_2$) transition of methanol overlaid with the $J$ = 1-0 transition of C$^{18}$O are shown in Figure~\ref{fig:B213-C17-all}.
The lines show a single velocity component at the positions 1, 7, 8 and 9; while the C$^{18}$O line shows a double peak structure at the positions 2 and 3.\\
The H$_2$ column density map of B213-C17 is shown in the upper panel of Figure~\ref{fig:B213-C17-results} with all the offsets observed within B213-C17 in the GEMS large project, however methanol was only towards few offsets (1, 7, 8 and 9).
Both the variation of the CH$_3$OH and C$^{18}$O column density ratio, as well as the single column densities N(CH$_3$OH) and N(C$^{18}$O) are shown in the lower panel and show a shallow peak towards the denser offsets observed (7 and 8).

\begin{figure*}
\centering
 \includegraphics [width=0.5\textwidth]{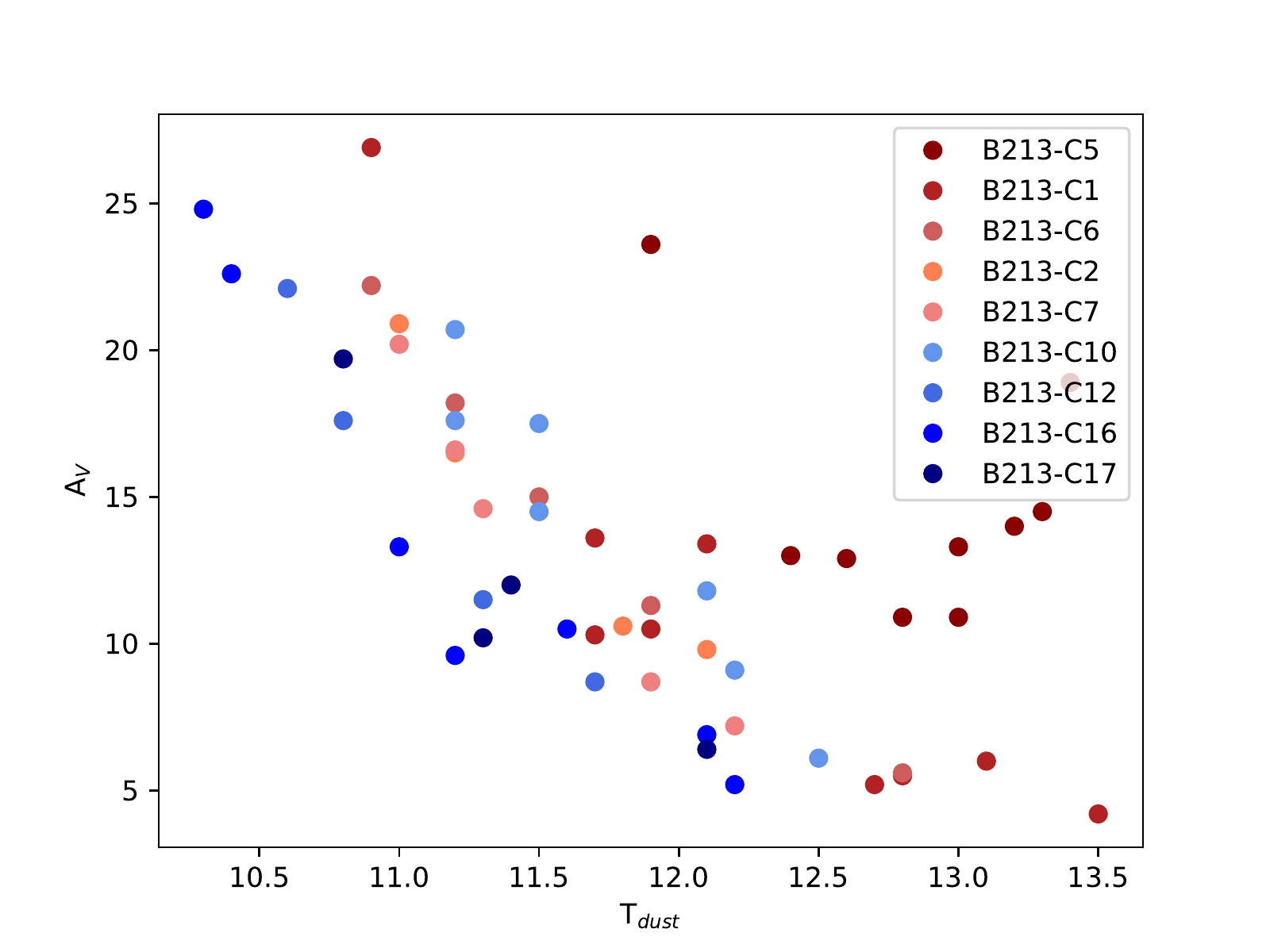}
 \caption{A$_V$ and T$_{dust}$ in the B213 cores within our sample are plotted against each other to show the effect of the larger stellar activity in the northern part of B213. In this Figure, the cores in the North of B213 are plotted in different shades of red that go from darker to lighter when moving from North to South. The cores in the central and southern part of B213 are plotted in different shades of blue, with the darkest being the core located more to the South.
}
  \label{fig:AV_vs_T}
\end{figure*}


\begin{figure*}
\centering
 \includegraphics [width=1\textwidth]{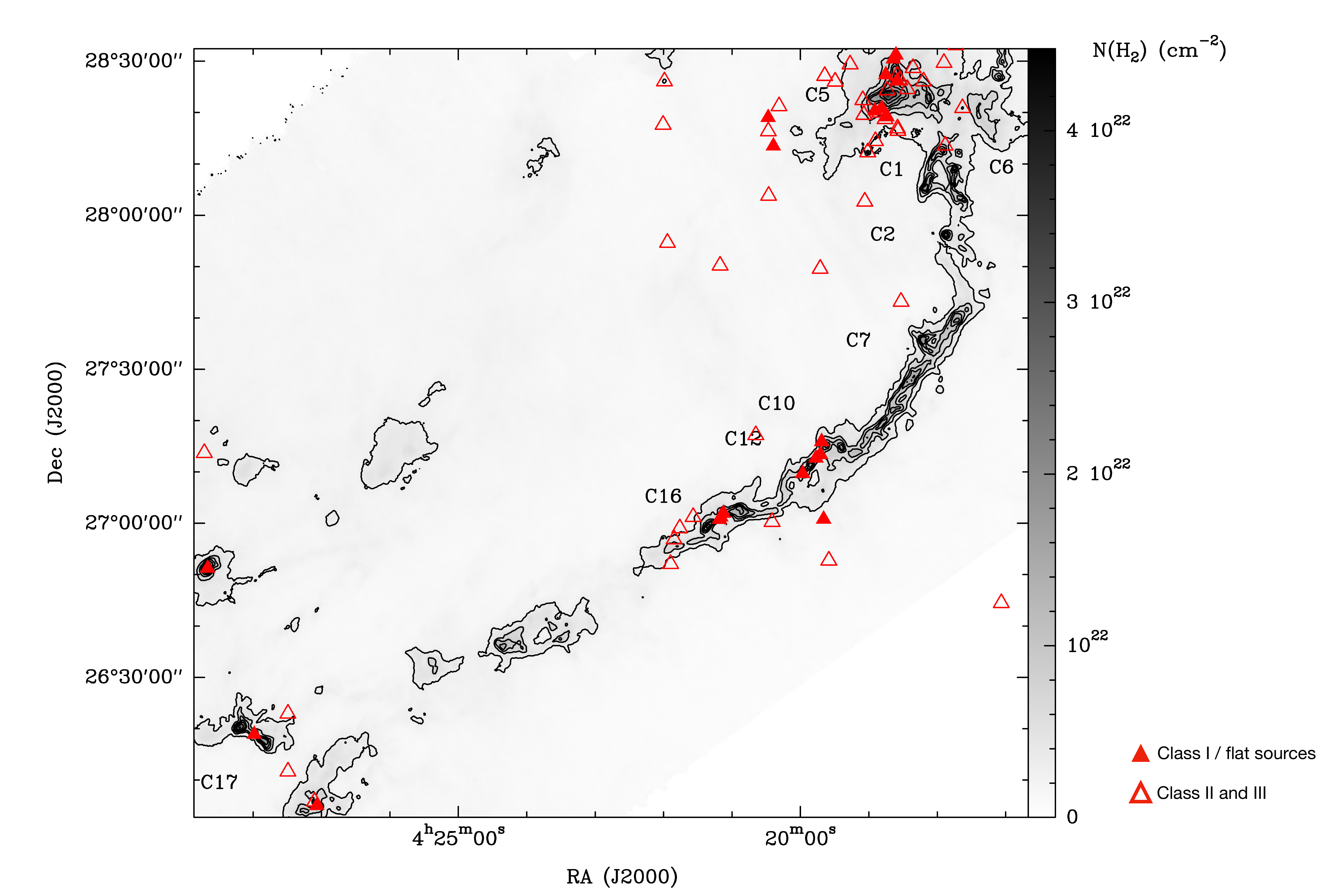}
 \caption{H$_2$ column density map of the B213 filament derived from $Herschel$ and $Planck$ data \citep{palmeirim13}. The young stellar objects reported in \cite{rebull10} are shown as triangles.
}
  \label{fig:stars}
\end{figure*}

\end{appendix}

\end{document}